\documentclass[prd, onecolumn, superscriptaddress, nofootinbib, notitlepage, floatfix, preprintnumbers]{revtex4-1}
\usepackage[dvipsnames]{xcolor}
\usepackage{graphicx}
\usepackage{wrapfig}
\usepackage{amsmath}
\usepackage{amsfonts}
\usepackage{amssymb}
\usepackage{multirow}
\usepackage{slashed}
\usepackage{physics}

\usepackage{soul}

\usepackage{ulem}
\usepackage[colorlinks=true, linkcolor=blue, citecolor=blue, urlcolor=blue]{hyperref}
\newcommand{\msbar}{{\overline{\mathrm{MS}}}}
\def\rpitd{\mathfrak{Y}}

\def\be{\begin{equation}}
\def\ee{\end{equation}}
\newcommand\suchthat[1]{\left|\vphantom{#1}\right.}
\newcommand\befs{\begin{figure*}}
\newcommand\eefs[1]{\label{fig:#1}\end{figure*}}
\newcommand\bef{\begin{figure}}
\newcommand\eef[1]{\label{fig:#1}\end{figure}}
\newcommand\beq{\begin{equation}}
\newcommand\eeq[1]{\label{#1}\end{equation}}
\newcommand\beqa{\begin{eqnarray}&&\quad\cr}
\newcommand\eeqa[1]{\cr &&\quad \label{#1}\end{eqnarray}}
\newcommand\bet{\begin{table}}
\newcommand\eet[1]{\label{tb:#1}\end{table}}
\newcommand\bets{\begin{table*}}
\newcommand\eets[1]{\label{tb:#1}\end{table*}}

\newcommand{\ensem}{{\tt a094m358}}

\begin{document}
\setstcolor{red}
\widetext
\title{
   Non-singlet quark helicity PDFs of the nucleon from pseudo-distributions
}

\newcommand*{\Jlab}{Thomas Jefferson National Accelerator Facility, Newport News, Virginia, USA.}\affiliation{\Jlab}    %1
\newcommand*{\WM}{Department of Physics, William and Mary, Williamsburg, Virginia, USA.}\affiliation{\WM}       %2
\newcommand*{\ODU}{Department of Physics, Old Dominion University, Norfolk, Virginia, USA.}\affiliation{\ODU}    %4
\newcommand*{\CNRS}{Aix Marseille Univ, Universit\'e de Toulon, CNRS, CPT, Marseille, France.}\affiliation{\CNRS}    %4

%%%%%%%%%%%%%%%%%%%%%%%%%%%
% AUTH BLOCK
%%%%%%%%%%%%%%%%%%%%%%%%%%%
\author{Robert G. Edwards}\affiliation{\Jlab}
\author{Colin Egerer}\affiliation{\Jlab}
\author{Joseph Karpie}\affiliation{\Jlab}
\author{Nikhil Karthik}\affiliation{\Jlab}\affiliation{\WM}
\author{Christopher J. Monahan}\affiliation{\WM}\affiliation{\Jlab}
\author{Wayne Morris}\affiliation{\ODU}\affiliation{\Jlab}
\author{Kostas Orginos}\affiliation{\WM}\affiliation{\Jlab}
\author{Anatoly  Radyushkin}\affiliation{\ODU}\affiliation{\Jlab}
\author{David Richards}\affiliation{\Jlab}
\author{Eloy Romero}\affiliation{\Jlab}
\author{Raza Sabbir Sufian}\affiliation{\Jlab}\affiliation{\WM}
\author{Savvas Zafeiropoulos}\affiliation{\CNRS}
\collaboration{On behalf of the \textit{HadStruc Collaboration}}
\begin{abstract}
The non-singlet helicity quark parton distribution functions (PDFs) of the nucleon are determined from lattice QCD, by jointly leveraging pseudo-distributions and the distillation spatial smearing paradigm. A Lorentz decomposition of appropriately isolated space-like matrix elements reveals pseudo-distributions that contain information on the leading-twist helicity PDFs, as well as an invariant amplitude that induces an additional $z^2$ contamination of the leading-twist signal. An analysis of the short-distance behavior of the space-like matrix elements using matching coefficients computed to next-to-leading order (NLO) exposes the desired PDF up to this additional $z^2$ contamination. Due to the non-conservation of the axial current, we elect to isolate the helicity PDFs normalized by the nucleon axial charge at the same scale $\mu^2$. The leading-twist helicity PDFs as well as several sources of systematic error, including higher-twist effects, discretization errors, and the aforementioned $z^2$ contaminating amplitude are jointly determined by characterizing the computed pseudo-distribution in a basis of Jacobi polynomials.  The Akaike Information Criterion is exploited to effectively average over distinct model parameterizations and cuts on the pseudo-distribution. Encouraging agreement is observed with recent global analyses of each non-singlet quark helicity PDF, notably a rather small non-singlet anti-quark helicity PDF for all quark momentum fractions.
\end{abstract}

\preprint{JLAB-THY-22-3751}
\date{\today}
\maketitle

\section{Introduction}
\label{sec:intro}
One of the primary foci of the nuclear physics community in the last three decades has been to discern the origin of the proton's spin in terms of its quark and gluon, collectively partonic, constituents. Efforts to resolve the dynamical origin of the proton's spin date to the earliest measurements of the spin-dependent structure functions describing polarized inclusive deep-inelastic scattering (DIS) cross sections~\cite{EuropeanMuon:1987isl}, which, remarkably, found that the total quark spin contributes only marginally to the total spin of the nucleon. 

The bulk of the information concerning the quark helicity parton distribution functions (PDFs) is deduced from polarized DIS data, where, for a parity-conserving interaction the hadronic tensor is described in terms of the polarized structure functions $g_1\left(x,Q^2\right)$ and $g_2\left(x,Q^2\right)$~\cite{Collins:2011zzd,Anselmino:1993tc,Blumlein:1996vs,ParticleDataGroup:2020ssz}. Within the parton model~\cite{Feynman:1973xc,Bjorken:1969ja,Drell:1969jm}, the $g_1\left(x,Q^2\right)$ structure function is readily interpreted as a linear combination of the quark and anti-quark helicity PDFs, appearing often in the literature as $g_{q/\bar{q}}\left(x\right)$ or $\Delta q/\bar{q}\left(x\right)$. These PDFs quantify the helicity asymmetry of quarks and anti-quarks, respectively, within a hadronic state of definite helicity.

The first determination of the polarized PDFs of the nucleon from a global analysis of longitudinally polarized DIS data was presented by the NNPDF collaboration in Ref.~\cite{Ball:2013lla}, and extended shortly thereafter in Ref.~\cite{Nocera:2014gqa} to include contemporary polarized hadron collider data for inclusive jet and $W$-production from the STAR~\cite{STAR:2010xwx,STAR:2012hth,STAR:2014afm,STAR:2014wox} and PHENIX~\cite{PHENIX:2010aru,PHENIX:2010rkr} experiments at RHIC. Although the inclusion of the STAR and PHENIX data in NNPDF's updated helicity PDF set, called {\tt NNPDFpol1.1}, did not extend the kinematic coverage of the helicity PDFs exposed by the polarized DIS data, the data did further constrain helicity PDF combinations accessible from polarized DIS alone and provided insight on novel helicity PDF combinations, such as $\Delta\bar{u}$.
In particular, the first Mellin moments of the polarized quark and anti-quark distributions were found to confirm that the nucleon receives only a small component of its spin from the intrinsic spin of its quark constituents, although the precision of the polarized moments were limited by extrapolation into the unmeasured small-$x$ regime.

Additional constraints on the quark helicity PDFs are provided by polarized semi-inclusive DIS (SIDIS), but require simultaneous knowledge of the relevant fragmentation functions that characterize the emergence of an observed hadron(s) from the struck parton. While {\tt NNPDFpol1.1} did not include light-quark SIDIS data, whose structure functions factorize into a convolution of the helicity PDFs and non-perturbative fragmentation functions, the Jefferson Lab Angular Momentum (JAM) collaboration performed the first simultaneous global analysis of polarized DIS and SIDIS data at NLO~\cite{Ethier:2017zbq}. The analysis, denoted {\tt JAM17} by the authors, leveraged $e^+e^-$ single-inclusive annihilation (SIA) data to provide an independent constraint on the fragmentation functions present in the factorized SIDIS cross sections included from COMPASS~\cite{COMPASS:2009kiy,COMPASS:2010hwr} and HERMES~\cite{HERMES:2004zsh}. The inclusion of the SIDIS data, for example, was found to lead to $\Delta s^+\equiv\Delta s+\Delta\overline{s}$ to become positive for $x\sim0.1$. Recently the {\tt JAM17} analysis was extended to include the latest polarized $W$-production data from STAR at RHIC, while many of the assumed flavor symmetries and PDF positivity constraints were relaxed~\cite{Cocuzza:2022jye}. The resulting PDF set, deemed {\tt JAM22}, supports a nonzero polarized sea asymmetry of the proton as well as a nonzero contribution to its spin from light anti-quarks. Evidently the quark helicity PDFs are less constrained than their unpolarized counterparts, thus warranting continued studies from first-principles lattice QCD.

As DIS, semi-inclusive, and exclusive processes, provide information about the non-perturbative structure of hadrons encoded in PDFs, one might hope that PDFs could also be calculated from first-principles lattice QCD. The Euclidean metric signature of lattice QCD, however, precludes any direct calculation of PDFs. Some of the earliest attempts to bypass this challenge sought to isolate the hadronic tensor~\cite{Liu:1993cv,Liu:1999ak} and forward Compton amplitude~\cite{Detmold:2005gg} directly from suitable Euclidean correlation functions. Within the last decade, renewed interest within the lattice community is generally attributed to the realization that the correlations of space-like separated partons within a boosted hadron, when Fourier transformed to momentum space, share the same infrared physics as the conventional light-cone PDFs~\cite{Ji:2013dva}. The resulting quasi-distributions are amenable to calculation using lattice QCD, while their distinct ultraviolet structures relative to the light-cone PDFs can be computed perturbatively. This forms the basis of Large Momentum Effective Theory (LaMET)~\cite{Ji:2014gla}, a momentum-space factorization, akin to the QCD factorization of hadronic scattering cross sections, that connects the quasi- and light-cone distributions. Determinations of helicity quark PDFs from quasi-distributions computed in lattice QCD include the flavor non-singlet~\cite{Alexandrou:2020qtt,Chen:2016utp,Alexandrou:2018pbm,Lin:2018pvv} and singlet~\cite{Alexandrou:2021oih,Alexandrou:2020uyt} cases, as well as careful accounting for numerous sources of systematic error in such extractions~\cite{Alexandrou:2019lfo}.

Despite the challenges inherent to the extraction of the $x$-dependent PDFs from lattice QCD, there has been considerable progress in the resolution of the flavor-separated quark and gluon contributions to the nucleon spin. The former are the first Mellin moments of the flavor-singlet quark helicity PDFs and are accessible from matrix elements of certain local operators. Advancements in computing infrastructure and algorithm design, including stochastic~\cite{Bhattacharya:2015wna} and truncated solver methods~\cite{Bali:2009hu,Blum:2012uh}, have enabled several recent physical pion mass, high-statistics calculations of the disconnected contributions needed to decompose the proton spin into contributions from its constituents. The first such calculation on a physical pion mass ensemble from the ETM Collaboration found the intrinsic quark spin to the proton to be $\frac{1}{2}\Delta\Sigma_{u+d+s}\simeq0.201$~\cite{Alexandrou:2017oeh}. This result was subsequently verified by the $\chi$QCD collaboration~\cite{Liang:2018pis}, leveraging three $2+1$ domain-wall fermion ensembles of varying pion masses and lattice spacings. 
Both results were found to be consistent with a phenomenological analysis of data from the COMPASS experiment, which found $0.13\lesssim\frac{1}{2}\Delta\Sigma\lesssim0.18$~\cite{COMPASS:2015mhb}. A study of $\frac{1}{2}\Delta\Sigma_{u+d+s}$ by the PNDME collaboration across numerous $2+1+1$ HISQ ensembles, with lattice spacing and pion mass as low as $0.06\text{ fm}$ and $m_\pi\sim135$ MeV, respectively, later found $\frac{1}{2}\Delta\Sigma_{u+d+s}\sim0.143$. This result, entirely within the COMPASS bound, was observed to depend delicately on the chiral-continuum extrapolation of each disconnected contribution. The reader is directed to Ref.~\cite{Liu:2021lke} for a thorough account of recent progress.

As the lattice community nears a decade of intense investigation of the light-cone structure of hadrons, considerable advancements in numerical reconstruction methods and theoretical formalisms, and the approaching Exascale frontier, promise to extend the impact of first-principles lattice QCD into regions largely unknown or poorly constrained from experiment. With this paradigm as the backdrop, it is imperative that all sources of systematic error in such a lattice QCD calculation are suitably accounted for. The reader is directed to Refs.~\cite{Constantinou:2020pek,Cichy:2021lih,Ji:2020ect} for recent contemporary reviews on the state of the art in hadronic structure calculations from lattice QCD.

The introduction of quasi-distributions and LaMET has led to many dedicated calculations of PDFs ~\cite{Liu:1993cv,Liu:1998um,Liu:1999ak,Detmold:2005gg,Chambers:2017dov,Karpie:2018zaz,Detmold:2021uru,Detmold:2003rq,Braun:2007wv,Radyushkin:2017cyf,Ma:2017pxb,Ji:2013dva,Aglietti:1998ur,Musch:2010ka,Xiong:2013bka,Lin:2014yra,Lin:2014zya,Ji:2014gla,Ma:2014jla,Ji:2015jwa,Ji:2015qla,Monahan:2015lha,Alexandrou:2015rja,Li:2016amo,Chen:2016utp,Alexandrou:2016jqi,Monahan:2016bvm,Radyushkin:2016hsy,Constantinou:2017sej,Alexandrou:2017huk,Chen:2017mzz,Orginos:2017kos,Ji:2017rah,Ji:2017oey,Stewart:2017tvs,Ishikawa:2017faj,Lin:2017ani,Hobbs:2017xtq,Jia:2017uul,Bali:2017gfr,Radyushkin:2017lvu,Ishikawa:2019flg,Radyushkin:2018cvn,Zhang:2018ggy,Izubuchi:2018srq,Xu:2018eii,Alexandrou:2018pbm,Chen:2018xof,Chen:2017mie,Chen:2018fwa,Jia:2018qee,Briceno:2018lfj,Alexandrou:2018eet,Liu:2018uuj,Bali:2018spj,Lin:2018qky,Radyushkin:2018nbf,Fan:2018dxu,Zhang:2018diq,Li:2018tpe,Braun:2018brg,Liu:2018hxv,Sufian:2019bol,Karpie:2019eiq,Cichy:2019ebf,Alexandrou:2019lfo,Bali:2019ecy,Hobbs:2019gob,Detmold:2019ghl,Izubuchi:2019lyk,Joo:2019jct,Joo:2019bzr,Ji:2019sxk,Ji:2019ewn,Radyushkin:2019mye,Sufian:2020vzb,Green:2020xco,Chai:2020nxw,Lin:2020ssv,Braun:2020ymy,Joo:2020spy,Bhat:2020ktg,Ji:2020ect,Zhang:2020dkn,Fan:2020nzz,Chen:2020arf,Zhang:2020dbb,Chen:2020iqi,Li:2020xml,Bhattacharya:2020jfj,Chen:2020ody,DelDebbio:2020cbz,DelDebbio:2020rgv,Gao:2020ito,Ji:2020byp,Alexandrou:2020tqq,Fan:2020cpa,Ji:2020brr,Lin:2020rxa,Alexandrou:2020qtt,Lin:2020fsj,Bringewatt:2020ixn,Zhang:2020rsx,Gao:2021hxl,LatticePartonCollaborationLPC:2021xdx,Bhattacharya:2021boh,Bhat:2022zrw,Dodson:2021rdq,Bhattacharya:2021moj,Gao:2021dbh,Gao:2022iex,He:2022lse,LatticeParton:2022xsd,Fan:2022kcb,Egerer:2021ymv,HadStruc:2021qdf,JeffersonLabAngularMomentumJAM:2022aix,HadStruc:2022yaw}, distributions amplitudes (DAs)~\cite{LatticeParton:2022zqc,Gao:2022vyh,Hua:2020gnw,Zhang:2020gaj,Wang:2019msf,Zhang:2017zfe,Zhang:2017bzy}, generalized parton distributions (GPDs)~\cite{Chen:2019lcm,Alexandrou:2020zbe,Alexandrou:2021bbo,Scapellato:2022mai,Bhattacharya:2022aob}, and transverse momentum dependent distributions (TMDs)~\cite{Ebert:2018gzl,Ebert:2019okf,Ebert:2019tvc,Shanahan:2019zcq,Shanahan:2020zxr,Shanahan:2021tst,Schlemmer:2021aij,LatticeParton:2020uhz,LPC:2022ibr,Li:2021wvl,Zhang:2022xuw,Ebert:2020gxr,Schindler:2022eva,Ebert:2022fmh}. An alternative framework~\cite{Radyushkin:2017cyf}, based on a short-distance factorization of the same lattice-calculated matrix elements utilized to construct quasi-distributions, offers complementary information on hadronic structure from lattice QCD. Within this short-distance factorization framework, the computed matrix elements induce a Lorentz decomposition into invariant amplitudes, or pseudo-distributions. The space-like matrix elements computed in lattice QCD are related to their associated light-cone analogs via a factorization theorem derived using the operator product expansion in a short-distance regime. In this manner the pseudo-distributions can be matched to their light-cone counterparts and the PDF obtained. The efficacy of the pseudo-distribution formalism and the distillation spatial smearing algorithm when applied in tandem to such structure studies was demonstrated in Refs.~\cite{HadStruc:2021qdf,Egerer:2021ymv}, wherein the flavor non-singlet unpolarized and transversity quark PDFs were isolated. The goal of this manuscript is to extend this union to include the isovector helicity quark PDFs, for both the CP-even and CP-odd varieties, thereby completing a first determination of the nucleon's leading-twist isovector quark PDFs using distillation and pseudo-distributions. 

The remainder of this manuscript is organized as follows. Section~\ref{sec:theory} begins with a recapitulation of the defining matrix element of the helicity quark PDF, which will expose the leading-twist invariant amplitudes one may hope to calculate using lattice QCD. Presentation of the coordinate space factorization, on which we will depend, will round out this section. The extracted bare nucleon matrix elements sensitive to the helicity distributions, the methodology employed therein, as well as our adopted gauge ensemble are presented in Sec.~\ref{sec:numerics}. The strategy we exercise to isolate the helicity distributions from our computed Ioffe-time pseudo-distributions is then developed in Sec.~\ref{sec:extraction}. Particular emphasis is given to the reduction of systematic bias in our extractions by considering a model average of an array of model ans{\"a}tze. Discussion of our model-averaged results, as well as any phenomenological insight that may be garnered is reserved for Sec.~\ref{sec:results}. Concluding remarks are shared in Sec.~\ref{sec:conclusions}, together with envisioned prospects for the distillation and pseudo-distribution amalgam.

\section{Theoretical construction and the helicity Ioffe-time Distribution\label{sec:theory}}

\subsection{Matrix element defining helicity PDF}
The fundamental object needed to define the leading-twist quark helicity distribution of the nucleon is the following quark correlation
\be
M^{\mu5}\left(p,z\right)=\bra{N\left(p,\lambda\right)}
\overline{\psi}\left(z\right)\gamma^\mu\gamma^5W^{(f)}
\left(z,0\right)\psi\left(0\right)\ket{N\left(p,\lambda\right)},
\label{eq:hel-mat}
\ee
where the quark fields are separated by a distance $z$, $\lambda$ denotes the nucleon helicity, and a Wilson line, in the fundamental representation, $W^{(f)}$, is included to maintain gauge invariance for a generic choice of gauge. Lorentz invariance demands Eq.~\eqref{eq:hel-mat} have the following decomposition into invariant amplitudes~\cite{Musch:2010ka}:
\be
M^{\mu5}\left(p,z\right)=-2m_NS^\mu\mathcal{M}\left(\nu,z^2\right)-2im_Np^\mu\left(z\cdot S\right)\mathcal{N}\left(\nu,z^2\right)+2m_N^3z^\mu\left(z\cdot S\right)\mathcal{R}\left(\nu,z^2\right),
\label{eq:decomp}
\ee
where we will refer to the invariant $\nu\equiv- p\cdot z$ as the Ioffe-time~\cite{Braun:1994jq}, which is related to the original variable of that name in DIS~\cite{Ioffe:1969kf} up to an inverse power of the hadron mass. The nucleon mass is $m_N$, and the polarization vector of the nucleon 
\be
S^\mu\equiv\frac1{2m_N}
\overline{u}\left(p,\lambda\right)\gamma^\mu\gamma^5u\left(p,\lambda\right)
\ee
is normalized to render the amplitudes in Eq.~\eqref{eq:decomp} dimensionless. 

To define the helicity distribution, one should take a light-like
separation, say  $z^\mu=\left(0,z^-,\mathbf{0_\perp}\right)$.  
Note that $M^{\mu5}(p,z)$ is singular on the light cone $z^2=0$,
so one needs to apply some regularization procedure, which we
denote below by ${\rm Reg}_{\mu^2}$, with $\mu^2$ being 
the regularization parameter.   
Writing the nucleon  four-momentum in the light-front coordinates $p^\mu=\left(p^+,p^-,\mathbf{p_\perp}\right)$, we have
\begin{align}
M^{+5}\left(p,z^-\right)_{{\rm Reg}_{\mu^2}} &=-2m_NS^+\left[\mathcal{M}\left(p^+z^-,0\right)+ip^+z^-\mathcal{N}\left(p^+z^-,0\right)\right]_{{\rm Reg}_{\mu^2}} \nonumber \\
&=-2m_NS^+\left[\mathcal{M}\left(\nu,0\right)-i\nu\mathcal{N}\left(\nu,0\right)\right]_{{\rm Reg}_{\mu^2}}\equiv-2m_NS^+\mathcal{I}\left(\nu,\mu^2\right),
\label{eq:LC}
\end{align}
where $\mathcal{I}\left(\nu, \mu^2\right)$ is 
the Ioffe-time distribution (ITD) whose $\nu$-Fourier transform defines the nucleon quark helicity PDF: 
\begin{align}
g_{q/N}\left(x, \mu^2\right)=\int_{-\infty}^\infty\frac{{\rm d}\nu}{2\pi}e^{-ix\nu}\mathcal{I}\left(\nu, \mu^2\right) \ .
\end{align}
According to Eq.~\eqref{eq:LC},  the $x$-dependence of the helicity distribution is dictated by the Ioffe-time dependence of a linear combination of the amplitudes $\mathcal{M}$ and $\mathcal{N}$, for which we make the abbreviation $\mathcal{Y}\left(\nu,z^2\right)\equiv\mathcal{M}\left(\nu,z^2\right)-i\nu\mathcal{N}\left(\nu,z^2\right)$. 
The (light-cone) ITD $\mathcal{I}\left(\nu, \mu^2\right)$
used in Eq.~(\ref{eq:LC}) may be written as $\mathcal{I}\left(\nu, \mu^2\right) =\mathcal{Y}\left(\nu,0\right)_{{\rm Reg}_{ \mu^2}}  $. 

Lorentz invariance implies the $\nu$-dependence of  $\mathcal{Y}\left(\nu,z^2\right)$ can be computed in any frame. This is especially beneficial in first-principles lattice QCD wherein the light-like separations needed to directly realize the matrix element in Eq.~(\ref{eq:LC})  are expressly precluded by the Euclidean metric.
For non-zero quark masses, the axial current is not conserved, 
and  the axial charge $g_A\left(\mu^2\right)$ determining 
the overall normalization of the helicity PDF 
is  not known a priori. Furthermore, it depends on the choice of renormalization scheme and on the chosen input scale $\mu^2$. 
We  defer a calculation of $g_A\left(\mu^2\right)$ to a future work when a high fidelity calculation of the renormalized axial charge is available on our chosen ensemble. 
In the present work,   following  the strategy used in  our recent paper  \cite{HadStruc:2021qdf}, we will calculate the shapes of the nucleon quark helicity PDFs:
\begin{align}
    g_{q_-/N}\left(x,\mu^2\right)/g_A\left(\mu^2\right)&=g_A\left(\mu^2\right)^{-1}\left[g_{q/N}\left(x,\mu^2\right)-g_{\bar{q}/N}\left(x,\mu^2\right)\right] \\
    g_{q_+/N}\left(x,\mu^2\right)/g_A\left(\mu^2\right)&=g_A\left(\mu^2\right)^{-1}\left[g_{q/N}\left(x,\mu^2\right)+g_{\bar{q}/N}\left(x,\mu^2\right)\right].
\end{align}

A suitable frame amenable to lattice QCD is $p^\mu=\left(\mathbf{0}_\perp,p_z,E\left(p_z\right)\right)$ and $z^\mu=\left(\mathbf{0}_\perp,z_3,0\right)$, with each expressed in the Euclidean Cartesian notation. We elect to compute the $\mu=3$ component of $M^{\mu5}\left(p,z\right)$, for which the bare operator does not mix with other operators under renormalization~\cite{Constantinou:2017sej}. With this kinematic setup our bare space-like matrix elements decompose in Minkowski space into a linear combination of Ioffe-time pseudo-distributions (pseudo-ITDs) according to:
\be
M^{35}\left(p,z_3\right)=-2m_NS^3\left[p_z\hat{z}\right]
\left \{ \mathcal{M}\left(\nu,z_3^2\right)-i p_z z_3 \mathcal{N}\left(\nu,z_3^2\right) \right \}  -2m_N^3 z_3^2S^3\left[p_z\hat{z}\right]\mathcal{R}\left(\nu,z_3^2\right)
,
\ee
where the functional dependence of the nucleon polarization vector on the external momentum has been indicated. One can notice that 
the amplitudes $\mathcal{M}$ and $\mathcal{N}$
appear here as $ \mathcal{M}\left(\nu,z_3^2\right)-i p_z z_3 \mathcal{N}\left(\nu,z_3^2\right)=\mathcal{M}\left(\nu,z_3^2\right)-i \nu  \mathcal{N}\left(\nu,z_3^2\right)$, i.e.,    in the same combination $\mathcal{Y}\left(\nu,z_3^2\right)$ as in the 
light-cone projection (\ref{eq:LC}). Thus, we can write 
\be
M^{35}\left(p,z_3\right)= -2m_NS^3\left[p_z\hat{z}\right]\left \{\mathcal{Y}\left(\nu,z_3^2\right)
+m_N^2 z_3^2 \mathcal{R}\left(\nu,z_3^2\right) \right \},
\label{eq:matIntoYAndR}
\ee
and observe that the $M^{35}$ matrix element
is proportional to the combination
\be
 \widetilde{\mathcal{Y}}\left(\nu,z_3^2\right)=
 \mathcal{Y}\left(\nu,z_3^2\right)
+m_N^2 z_3^2 \mathcal{R}\left(\nu,z_3^2\right),
\label{eq:tildeY}
\ee
which, in addition to $ \mathcal{Y}\left(\nu,z_3^2\right)$,
contains a contaminating term $m_N^2 z_3^2 \mathcal{R}\left(\nu,z_3^2\right)$\footnote{The nucleon mass $m_N$ appears
here as consequence of the definition (\ref{eq:decomp}); one can replace it by any scale $\Lambda$, and appropriately re-scale the magnitude of ${\cal R}$.}.
One may try to separate the $\mathcal{Y}$ and $\mathcal{R}$ terms by considering  
the $\mu=4$ component of $M^{\mu5}\left(p,z\right)$, which receives no contribution from the $\mathcal{R}$ term if $z=z_3$. However, in this case
\be
M^{45}\left(p,z_3\right)=-2m_N\left \{ S^4
 \mathcal{M}\left(\nu,z_3^2\right)-i S^3 E(p_z) z_3 \mathcal{N}\left(\nu,z_3^2\right) \right \}  \ .
\label{eq:decomp4}
\ee
One can see that 
the amplitudes $\mathcal{M}$ and $\mathcal{N}$
appear here in a combination quite distinct from the defining combination of $\mathcal{Y}\left(\nu,z_3^2\right)$. Furthermore, such a calculation of $M^{45}\left(p,z_3\right)$
would require an a priori determination of the finite mixing that exists between the $\gamma_4\gamma_5$ operator and other Dirac matrices due to the lattice regularization~\cite{Constantinou:2017sej} -- an added complication we seek to avoid.

For  purely space-like separations, $M^{\mu5}\left(p,z\right)$ acquires additional ultraviolet (UV) divergences~\cite{Polyakov:1980ca,Dotsenko:1979wb,Brandt:1981kf,Craigie:1980qs} that must be regularized and removed before taking the continuum limit. As these additional divergences are  known to renormalize multiplicatively~\cite{Craigie:1980qs,Ishikawa:2017faj,Ji:2017oey,Green:2017xeu}, we elect to remove them by forming an appropriate renormalization group (RG) invariant ratio. Such a prescription not only ensures a finite continuum limit, but also avoids the introduction of additional sources of systematic error stemming from gauge-fixed calculations, such as the RI/MOM scheme 
used in Refs.~\cite{Constantinou:2017sej,Alexandrou:2017huk}.

The  standard  procedure~\cite{Radyushkin:2017cyf} 
within the pseudo-PDF approach
is to divide the original  matrix element $M(p,z)$ by its
$p_z=0$ counterpart $M(p_z=0,z)$. Since the UV renormalization 
factor $Z(z/a)$ is the same for $M(p,z)$ and \mbox{$M(p_z=0,z)$,}
the ratio $M(p,z)/M(p_z=0,z)$ does not contain 
link-related UV divergences and is an RG invariant,  
referred to as   the reduced Ioffe-time 
pseudo-distribution~\cite{Radyushkin:2017cyf,Orginos:2017kos}, or reduced pseudo-ITD.  To enforce the normalization of unity at $z_3=0$, a double ratio is employed -- the RG invariant $M(p,z_3)/M(p_z=0,z_3)$
is divided by its $z_3=0$ magnitude. In analogy  with the ratio utilized  for the unpolarized~\cite{Egerer:2021ymv} and transversity~\cite{HadStruc:2021qdf} quark PDFs of the nucleon, we 
may consider 
\be
\mathfrak{Y} \left(\nu,z_3^2\right)=
\left. 
\left (
\frac{\widetilde{\mathcal{Y}}\left(\nu,z_3^2\right)}
{\widetilde{\mathcal{Y}}\left(0,z_3^2\right)\mid_{p_z=0}}
\right ) 
\right /
\left (
\frac{\widetilde{\mathcal{Y}}\left(\nu,0\right)\mid_{z_3=0}}
{\widetilde{\mathcal{Y}}\left(0,0\right)\mid_{p_z=0,z_3=0}}
\right ) .
\label{eq:subopt-reduced}
\ee
By construction, $\mathfrak{Y}\left(\nu,z_3^2\right)$ does not contain link-related UV divergences and is RG invariant.
However, as we discussed, $\widetilde{\mathcal{Y}}\left(\nu,z^2\right)$
differs from ${\mathcal{Y}}\left(\nu,z^2\right)$ (which is our goal)
by a contamination term $m_N^2 z_3^2 \mathcal{R}\left(\nu,z_3^2\right)$, which was observed to be quite small in Ref.~\cite{Musch:2010ka}. We will attempt to parameterize, and subsequently remove, this additional $z^2$ contamination, which we note is no worse than the corrections to our factorization relationship, through a parametric description of the reduced pseudo-ITD detailed in Sec.~\ref{sec:extraction}.

\subsection{Matching kernel}
The next step is  to relate space-like matrix elements, such as $M^{35}(p,z_3)$, obtained from lattice QCD to ITDs $\mathcal{I}(\nu,\mu^2)$ corresponding to PDFs taken in the $\msbar$ scheme at a scale $\mu^2$.
To this end, 
we need to derive appropriate matching kernels.   
At NLO, such a derivation is based on a calculation in dimensional regularization of the one-loop correction to the relevant bilocal operator. The latter, ignoring the link $W^{(f)}\left(0,z\right)$ for brevity, is $\bar{\psi}(0)\gamma_{5}\gamma^{\mu}\psi(z)$ in the present quark helicity case. The result of our calculations in the $\msbar$ scheme is given by 
\begin{align}
       \langle  \langle \bar{\psi}(0) \gamma_{5} \gamma^{\mu} \psi(z)\rangle \rangle
          \rightarrow 
          &-\frac{\alpha_s C_{F}}{4 \pi} \left(\frac{1}{\epsilon_{U V}}+\ln \left(-\frac{ e^{2\gamma_E+3}}{4\pi} z^{2}\mu^2 \right)\right)   \langle  \langle \bar{\psi} (0) \gamma_{5} \gamma^{\mu} \psi (z) \rangle \rangle  \nonumber 
           \\
           &+\frac{\alpha_s C_{F}}{2 \pi}
                    \left(\frac{1}{\epsilon_{I R}}-\ln \left(-\frac{ e^{2\gamma_E+1}}{4\pi} z^{2}\mu^2\right)          
                    \right)
                     \int_{0}^{1} \mathrm{~d} u \left[\frac{1+u^2}{1-u}\right]_{+}
                    \langle  \langle \bar{\psi} (0) \gamma_{5} \gamma^{\mu} \psi (u z) \rangle \rangle   \nonumber  \\
          &-\frac{\alpha_s C_{F}}{2 \pi} \int_{0}^{1} \mathrm{~d} u
          \left[4 \frac{ \ln (1-u)}{1-u}-2(1-u)\right]_{+}
          \langle  \langle \bar{\psi} (0) \gamma_{5} \gamma^{\mu} \psi (u z) \rangle \rangle 
           \nonumber  \\
          &+\frac{\alpha_s C_{F}}{2 \pi} \left(\frac{2 z^{\mu}  }{z^{2}}\right) \int_{0}^{1} \mathrm{~d} u \,(1-u)\,
          \langle  \langle \bar{\psi} (0) \gamma_{5} \slashed z \psi (u z) \rangle \rangle         \ ,
          \label{eq:one-loop}
          \end{align}
where $\langle\langle\ldots\rangle\rangle$ 
indicates that the relations are valid only 
when the operators are inserted
into a forward matrix element. 
The first line of Eq.~(\ref{eq:one-loop})  contains the ultraviolet divergences, which are
removed by taking the ratio \mbox{$M(p,z)/M(p_z=0,z)$.} The second line contains the evolution logarithm 
$\ln(-z^2 \mu^2)$ accompanied by the flavor non-singlet DGLAP kernel 
\be
\left[\frac{1+u^2}{1-u}\right]_{+}\equiv B(u) \ .
\ee
The plus-prescription is defined in the standard manner,
\be
\int_0^1{\rm d}u\ G\left(u\right)_+f\left(ux\right)\equiv\int_0^1{\rm d}u\ 
G\left(u\right)\left[f\left(ux\right)-f\left(x\right)\right]\ .
\ee
The third and fourth lines of Eq.~(\ref{eq:one-loop})  contain the  ``constant'', or scale-independent, portion of the one-loop 
correction.
Note that the  term in   the 
fourth line contains 
a $z^\mu$ prefactor. 
Hence, it contributes exclusively to the $\mathcal{R}$ amplitude. If $z$ were purely in the third component, the term $z^\mu/z^2$ would vanish for $\mu=4$, and be non-zero for $\mu=3$. This observation explains the well-known difference (see, e.g., Ref. \cite{Izubuchi:2018srq})
between the matching conditions for (pseudo)vector bilocal operators with temporal or spatial indices. In particular, if $z=z_3$ and $\mu=3$, we have $z^\mu \slashed z/z^2=\gamma^3 $, and one can simply add the contribution of the fourth line of Eq.~(\ref{eq:one-loop}) to that of the third line, resulting in the change $-2(1-u) \Rightarrow -4(1-u)$ in the third line.

As a result, 
we obtain the matching relation
\be
{\mathfrak{Y}}\left(\nu,z_3^2\right)=
\frac1{g_A(\mu^2)} \int_0^1{\rm d}u\ \mathcal{C}\left(u,z_3^2\mu^2,\alpha_s\left(\mu^2\right)\right)\mathcal{I}\left(u\nu,\mu^2\right)+\mathcal{O}\left(z_3^2\Lambda_{\rm QCD}^2\right),
\label{eq:factorization}
\ee
in which  $ {\mathfrak{Y}}\left(\nu,z_3^2\right)$ 
is written in terms of the $\msbar$
twist-2 
helicity
ITD
\be 
\mathcal{I}\left(\nu,\mu^2\right)=
\int_{-1}^1{\rm d}x\ e^{i\nu x}g_{q/N}\left(x,\mu^2\right), 
\ee 
with  $\mathcal{O}\left(z_3^2\Lambda_{\rm QCD}^2\right)$ denoting   higher-twist terms. 
The ingredients  producing 
the one-loop matching kernel 
have been discussed above.
In explicit form, the kernel  is given by
\be
\mathcal{C}\left(u,z_3^2\mu^2,\alpha_s\left(\mu^2\right)\right)=\delta\left(1-u\right)-\frac{\alpha_sC_F}{2\pi}\left\lbrace\ln\left(\frac{e^{2\gamma_E+1}}{4}z_3^2\mu^2\right)B(u) +L(u) \right\rbrace,
\label{eq:nlo-kernel}
\ee
where $L(u)$, the  ``constant''  part of the one-loop matching kernel,
comes from the third and fourth lines of Eq.~(\ref{eq:one-loop}),
\be L(u) = \left[4\frac{\ln\left(1-u\right)}{1-u}-4\left(1-u\right)\right]_+ \ . 
\label{eq:L-kernel}
\ee

The convolution~\eqref{eq:factorization}  may be written  
in terms of  the Mellin moments $a_n (\mu^2)$ of the 
normalized $\msbar$ helicity PDF  
\be
{\mathfrak{Y}}\left(\nu,z_3^2\right)=\sum_nc_n\left(z_3^2\mu^2\right)a_{n+1}\left(\mu^2\right)\frac{\left(i\nu\right)^n}{n!}+\mathcal{O}\left(z_3^2\right),
\ee
with $c_n\left(z_3^2\mu^2\right)$ being  the NLO Wilson coefficients of the local OPE:
\be
c_n\left(z_3^2\mu^2\right)\equiv\int_0^1{\rm d}u\ \mathcal{C}\left(u,z_3^2\mu^2,\alpha_s\left(\mu^2\right)\right)u^n=1-\frac{\alpha_sC_F}{2\pi}\left[\gamma_n\ln\left(\frac{e^{2\gamma_E+1}}{4}
z_3^2\mu^2\right)+l_n\right] . 
\label{eq:wilson-coeffs}
\ee
The anomalous dimensions $\gamma_n$ are  the moments of the DGLAP kernel, 
\be
\gamma_n=\int_0^1{\rm d}u\ 
B(u) u^n =-\frac{1}{2}+\frac{1}{(n+1)(n+2)}-2\sum_{k=2}^{n+1}\frac{1}{k} \ , 
\label{eq:scaleDepMoments}
\ee
while $l_n$ is given by 
\be
l_n=\int_0^1{\rm d}u\ 
L(u)
u^n=2\left[\left(\sum_{k=1}^n\frac{1}{k}\right)^2
+\sum_{k=1}^n\frac{1}{k^2} +1-\frac{2}{\left(n+1\right)\left(n+2\right)}\right]
\label{eq:scaleIndepMoments}.
\ee
With this form, it can be seen that the renormalized moments of the PDF, even those which suffer from power divergent mixing on the lattice, can be determined directly from this matrix element~\cite{Karpie:2018zaz}.

\section{Numerical Implementation}
\label{sec:numerics}
We use a single isotropic ensemble of $2\oplus1$ Wilson clover fermions generated by the JLab/W\&M/LANL/MIT collaboration~\cite{jlab-wm-lanl} in this calculation, with the strange quark fixed to its physical value and the sea quarks described by the same Wilson clover action. This ensemble, denoted herein as {\tt a094m358}, is characterized by a $0.094$ fm lattice spacing and $m_\pi=358$ MeV pion mass within a $32^3\times64$ lattice volume. We exploit the same $349$ configuration subset of the {\tt a094m358} ensemble utilized in our earlier unpolarized~\cite{Egerer:2021ymv} and transversity~\cite{HadStruc:2021qdf} quark PDF calculations. Table~\ref{tab:ensem} summarizes aspects of the \ensem\space ensemble relevant to this work, while the interested reader is referred to Refs.~\cite{Yoon:2016dij,Yoon:2016jzj} for further details.
\begin{table}[b]
    \centering
    \begin{tabular}{c|c|c|c|c|c}
    \hline\hline
    ID & $a$ (fm) & $m_\pi$ (MeV) & $\beta$ & $m_\pi L$ & $L^3\times N_T$ \\
    \hline
    \ensem & $0.094(1)$ & $358(3)$ & $6.3$ & $5.4$ & $32^3\times64$ \\
    \hline\hline
    \end{tabular}
    \caption{Characteristics of the ensemble used in this work.}
    \label{tab:ensem}
\end{table}

To isolate the bare matrix elements, Eq.~\eqref{eq:hel-mat}, of the space-like quark bilinear $\overline{\psi}\left(z\right)\Gamma W_{\hat{z}}^{(f)}\left(z,0\right)\psi\left(0\right)$, we require two-point and connected three-point correlation functions. With the kinematic setup presented in Sec.~\ref{sec:theory}, the spectral representations of these correlation functions read
\begin{align}
    C_{2{\rm pt}}\left(p_z\hat{z},T\right)&=\langle\mathcal{N}\left(-p_z\hat{z},T\right)\overline{\mathcal{N}}\left(p_z\hat{z},0\right)\rangle=\sum_n\frac{\left|\mathcal{Z}_n\left(p_z\right)\right|^2}{2E_n\left(p_z\right)}e^{-E_n\left(p_z\right)T}\label{eq:2pt} \\
    C_{3{\rm pt}}^{\left[\gamma_3\gamma_5\right]}\left(p_z\hat{z},T;z_3,\tau\right)&=\langle\mathcal{N}\left(-p_z\hat{z},T\right)\overline{\psi}\left(z_3,\tau\right)\gamma_3\gamma_5W_{\hat{z}}^{(f)}\left(z_3,0\right)\psi\left(0,\tau\right)\overline{\mathcal{N}}\left(p_z\hat{z},0\right)\rangle \nonumber\\
    &=\sum_{n',n}\frac{\mathcal{Z}_{n'}\left(p_z\right)\mathcal{Z}_n^\dagger\left(p_z\right)}{4E_{n'}\left(p_z\right)E_n\left(p_z\right)}\bra{n'}\mathring{\mathcal{O}}^{[\gamma_3\gamma_5]}\left(z_3,\tau\right)\ket{n}e^{-E_{n'}\left(p_z\right)\left(T-\tau\right)}e^{-E_n\left(p_z\right)T},\label{eq:3pt}
\end{align}
with the source and sink interpolating fields $\mathcal{N}$ separated by a Euclidean time $T$, momentum-dependent interpolator-state overlaps given by $\mathcal{Z}_n\left(p_z\right)$, and the unrenormalized quark bilinear, abbreviated by $\mathring{\mathcal{O}}^{[\gamma_3\gamma_5]}\left(z_3\right)$, is introduced between the source and sink interpolators for Euclidean times $0\leq\tau< T$. We consider interpolator separations of $T\in\lbrace4,6,8,10,12,14\rbrace a\simeq\lbrace0.38,0.56,0.75,0.94,1.13,1.32\rbrace{\rm fm}$ to aid in suppressing  excited-state contamination. To map the Ioffe-time dependence of the pseudo-ITD, we utilize Wilson line lengths of $0\leq z/a\leq8$ and project our interpolators onto definite momenta $\vec{p}=\pm p_z\hat{z}$ with $p_z=n_z\frac{2\pi}{aL}\in\lbrace0,0.42,0.83,1.25,1.67,2.08,2.50\rbrace{\rm GeV}$ for $n_z\in \mathbb{Z}$.

The distillation~\cite{HadronSpectrum:2009krc} spatial smearing kernel is employed to increase the interpolator overlaps onto the confinement scale physics we are interested in. A realization~\cite{Egerer:2020hnc} of the momentum smearing algorithm~\cite{Bali:2016lva} is implemented to increase our operator-state overlaps for allowed momenta $\left|n_z\right|\geq4$. This amounts in practice to application of a spatially varying phase
\be
\left\lbrace e^{i\vec{\zeta}\cdot\vec{x}}\quad\middle|\quad\vec{\zeta}=\pm2\cdot\frac{2\pi}{L}\hat{z}\right\rbrace
\ee
onto each eigenvector of the discretized gauge covariant Laplacian that comprise the distillation space. This prescription was found to be sufficient to shift the momentum space overlaps within the distillation space to the larger values of momenta we consider. Details concerning our implementation of distillation within the pseudo-distribution formalism can be found in Ref.~\cite{Egerer:2021ymv}. Of that discussion, some salient features, especially concerning the infinite tower of finite-volume energy eigenstates $\lbrace n',n\rbrace$ in~\eqref{eq:2pt} and~\eqref{eq:3pt}, are worth highlighting here.

The lattice discretization regularizes QCD, but it also breaks the continuum rotational symmetry. As a consequence baryons at rest, typically described by spin, are instead classified according to the finite number of irreducible representations (irreps) of the double-cover octahedral group $O_h^D$. 
In other words, the mass eigenstates of a baryon in the continuum, formerly classified by its $J^P$ quantum numbers, is instead characterized by its pattern of subduction across the irreps $\Lambda$ of $O_h^D$. The mixing of mass eigenstates induced by the reduced rotational symmetry is compounded further when considering non-zero momenta, for which $O_h^D$ is broken further into its little groups, or subgroups, dependent on the {\it star} of $\vec{p}$ - denoted $^*\left(\vec{p}\right)$~\cite{Moore:2005dw}. To address the complication of isolating the ground-state $J^+=\frac{1}{2}^+$ nucleon at rest and in motion in our lattice calculation, we construct nucleon interpolators such that they possess definite transformation properties with respect to $O_h^D$ and its little groups.

Following Refs.~\cite{Edwards:2011jj,Dudek:2012ag}, our interpolators realized using distillation are first constructed at rest in the continuum such that they possess definite flavor and $J^P$ quantum numbers. As consequence, such interpolators will only overlap onto continuum energy eigenstates of the same $J^P$ at rest:
\be
\bra{\vec{p}=\vec{0};J',P',M'}\left[\mathcal{O}^{J,P,M}\left(\vec{p}=\vec{0}\right)\right]^\dagger\ket{0}=Z^{[J]}\delta_{J,J'}\delta_{P,P'}\delta_{M,M'},
\ee
where $Z^{[J]}$ is the numeric overlap factor associated with the continuum interpolator $\mathcal{O}^{J,P,M}(\vec{p}=\vec{0})$. Projecting our interpolators to non-zero momenta then represents a breaking of parity, and energy eigenstates are instead classified according to their value of helicity $\lambda$. Tailoring the algorithm developed in Ref.~\cite{Thomas:2011rh} to the case of baryons, continuum helicity operators are obtained by enacting a basis change on the continuum interpolators boosted to non-zero momenta along a chosen quantization direction, say $\hat{z}$:
\be
\left[\mathbb{O}^{J^P,\lambda}\left(\vec{p}\right)\right]^\dagger=\sum_m\mathcal{D}_{m\lambda}^{(J)}\left(R\right)\left[\mathcal{O}^{J^P,m}\left(\left|\vec{p}\right|\hat{z}\right)\right]^\dagger,
\ee
where $\mathcal{D}_{m\lambda}^{(J)}\left(R\right)$ is a Wigner-$\mathcal{D}$ matrix, dependent on the active rotation $R$, that rotates the vector $\left|\vec{p}\right|\hat{z}$ to $\vec{p}$, thus obtaining a continuum helicity (creation) operator $\mathbb{O}^{J^P,\lambda}\left(\vec{p}\right)$ from boosted canonically quantized continuum interpolators.
In other words, constructing our interpolators to have definite helicity in the continuum forces all operator-state overlaps to be in terms of $\vec{p}$ and $\lambda$.
The manifest breaking of continuum rotational symmetry by the cubic lattice, and further reduction for $\vec{p}\neq\vec{0}$, ensures continuum helicity eigenstates of each value of helicity will subduce into at least one irrep of the double-cover octahedral group $O_h^D$ or its associated little groups. That is, our interpolators constructed with distillation will overlap with finite-volume energy eigenstates $\lbrace n',n\rbrace$ characterized by numerous values of helicity.

Although the distillation paradigm facilitates an especially cheap realization of variational improvement (e.g. Refs.~\cite{Blossier:2009kd,Bulava:2011yz}) within a given symmetry channel, recently applied, for instance, to expose the unpolarized~\cite{HadStruc:2021wmh} and helicity gluon~\cite{Egerer:2022tll} PDFs of the nucleon, we select a single, spatially-local interpolating field, denoted $N^2S_s\frac{1}{2}^+$, to couple to the ground-state nucleon. This choice is motivated by computational expediency and a desire to maximize consistency with our earlier unpolarized and transversity quark PDF calculations~\cite{Egerer:2021ymv,HadStruc:2021qdf}.

\subsection{Matrix Element Isolation}
At asymptotically large temporal separations between operators, namely $0\ll\tau\ll T$, the ratio between the three-point, Eq.~\eqref{eq:3pt}, and two-point, Eq.~\eqref{eq:2pt}, correlation functions will plateau to the desired bare matrix element. Of course such asymptotically large Euclidean times cannot be reached with any useful statistical precision, due to the exponential decay of the signal-to-noise ratio, and one must instead contend with excited-state contamination at short Euclidean times, where data are more precise. Although a multi-state fit to the three-point correlators with explicit dependence on both the insertion time slice and source-sink separation can be used to isolate the bare matrix element, excited-state contamination will afflict the matrix element determination on the order of $\mathcal{O}\left(e^{-\Delta ET/2}\right)$, where $\Delta E$ is the energy gap between the ground state and the lowest lying effective excited state. In an effort to further suppress the contamination from excited states, we elect to extract the bare matrix elements via the summation method~\cite{Maiani:1987by,Capitani:2012gj}
whereby the ratio of the three-point and and two-point correlation functions is summed over the time slice $\tau$ of the inserted bare Wilson line operator $\mathring{\mathcal{O}}^{\left[\gamma_3\gamma_5\right]}\left(z_3\right)$:
\be
R\left(p_z\hat{z},z_3;T\right)\equiv\sum_{\tau/a=1}^{T-1}\frac{C_{3{\rm pt}}\left(p_z\hat{z},T;z_3,\tau\right)}{C_{2{\rm pt}}\left(p_z\hat{z},T\right)}.
\label{eq:ratio}
\ee
Contact terms are expressly excluded from the $R\left(p_z\hat{z},z_3;T\right)$ signal by performing the operator summation for $\tau/a\in\left[1,T-1\right]$. As the resulting geometric series depends linearly on the desired bare matrix element $M\left(p_z\hat{z},z_3\right)$, an appropriate fit function to extract the bare matrix element is
\be
R_{\rm fit}\left(p_z\hat{z},z_3;T\right)=\mathcal{A}+M\left(p_z\hat{z},z_3\right)T+\mathcal{O}\left(e^{-\Delta ET}\right),
\label{eq:linFit}
\ee
where, for brevity, we define $M\left(p,z\right)\equiv M^{35}\left(p,z\right)$.
An important consequence of using the summation method is an exponential suppression of excited-state contamination to the extracted bare matrix element signal relative to multi-state methods for the same source-sink separation $T$ and level of statistics - corrections from excited-state contamination scale as $\mathcal{O}\left(e^{-\Delta ET}\right)=BTe^{-\Delta ET}$ when the summation method is employed.

\begin{figure}[h]
    \centering
    \includegraphics[width=0.24\linewidth]{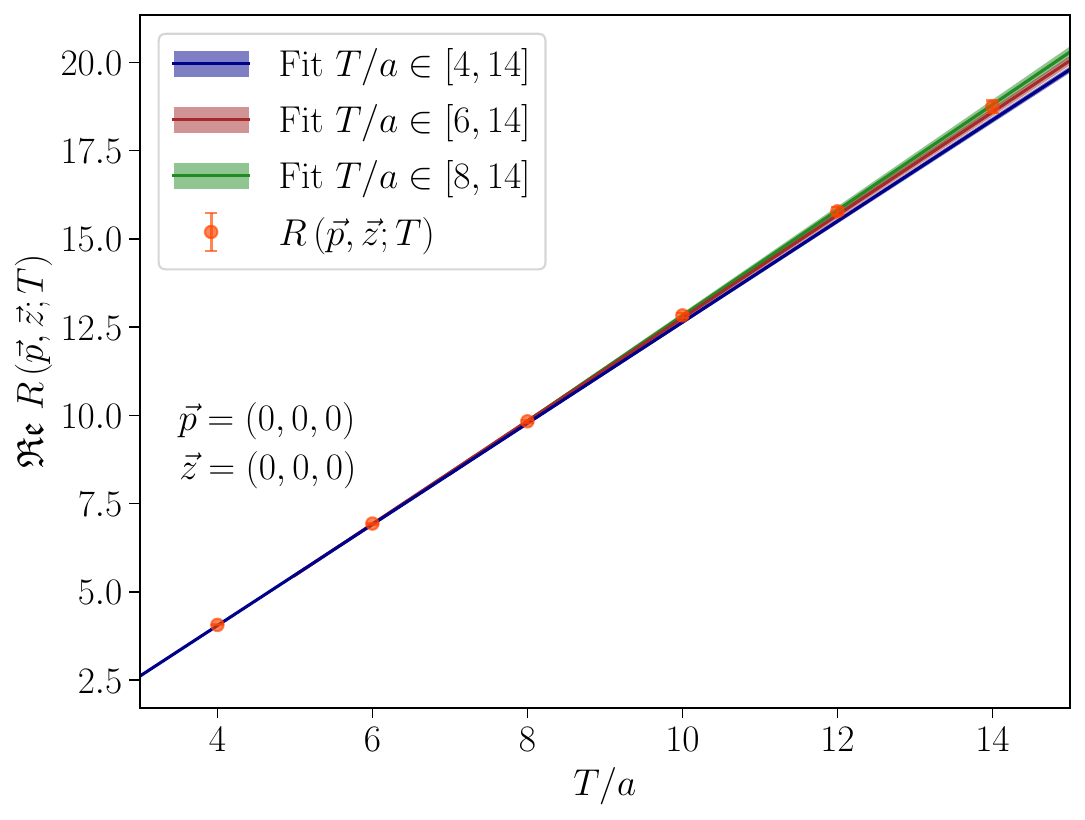}
    \hfill
    \includegraphics[width=0.23\linewidth]{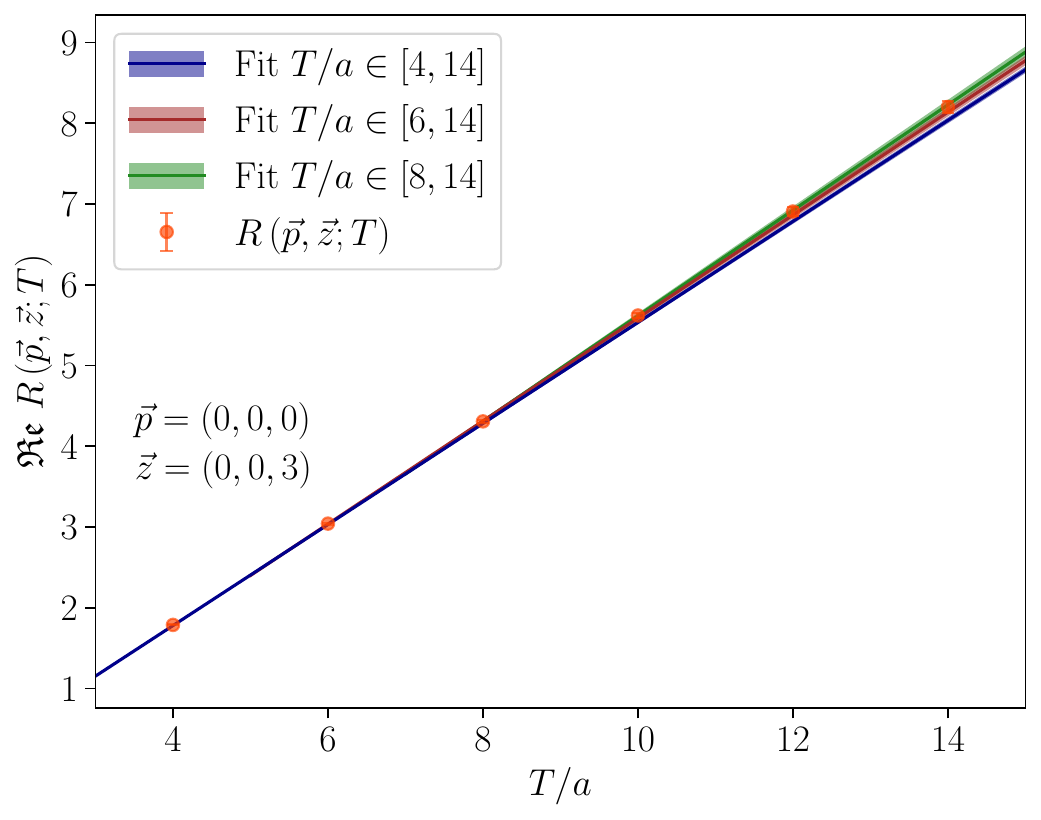}
    \hfill
    \includegraphics[width=0.24\linewidth]{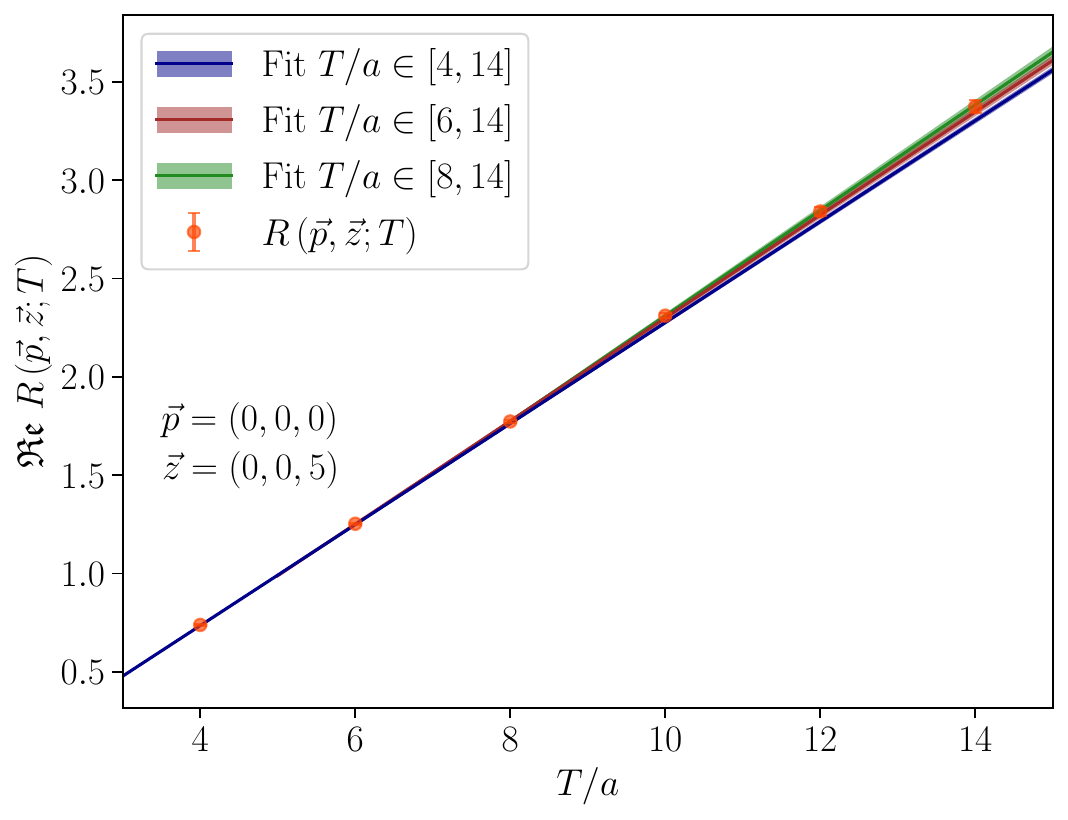}
    \hfill
    \includegraphics[width=0.24\linewidth]{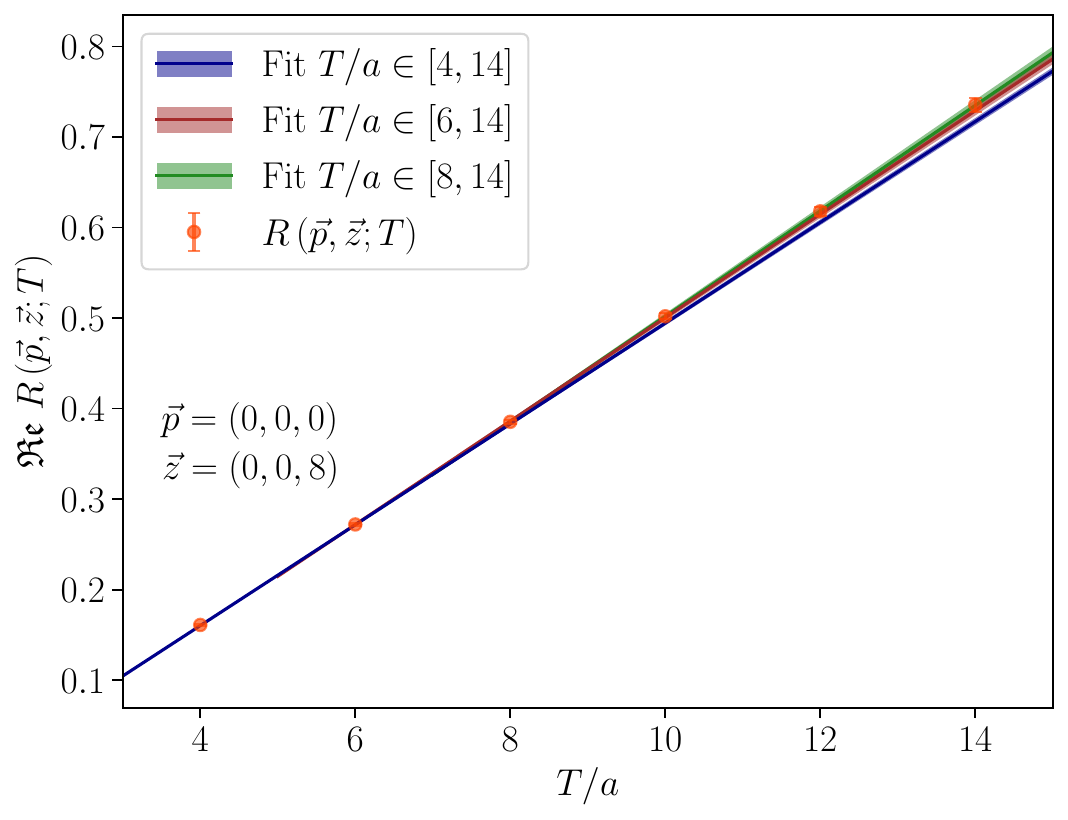}
    \caption{Real component of the summed ratio $R\left(\vec{p},\vec{z};T\right)$ for $\vec{p}=\vec{0}$ together with the linear fit Eq.~\ref{eq:linFit} applied for varying temporal series. The length of the Wilson line is increased from left to right, and given in integer multiples of the lattice spacing. Each panel corresponds to $R\left(\vec{p},\vec{z};T\right)$ determined from a particular {\it subduced} correlation function, which are discussed in Sec.~\ref{sec:pitdFromSVD}.
    % Shown for row1-row1 combination.
    \label{fig:rest_SRs}}
\end{figure}

\begin{figure}[h]
    \centering
    \includegraphics[width=0.32\linewidth]{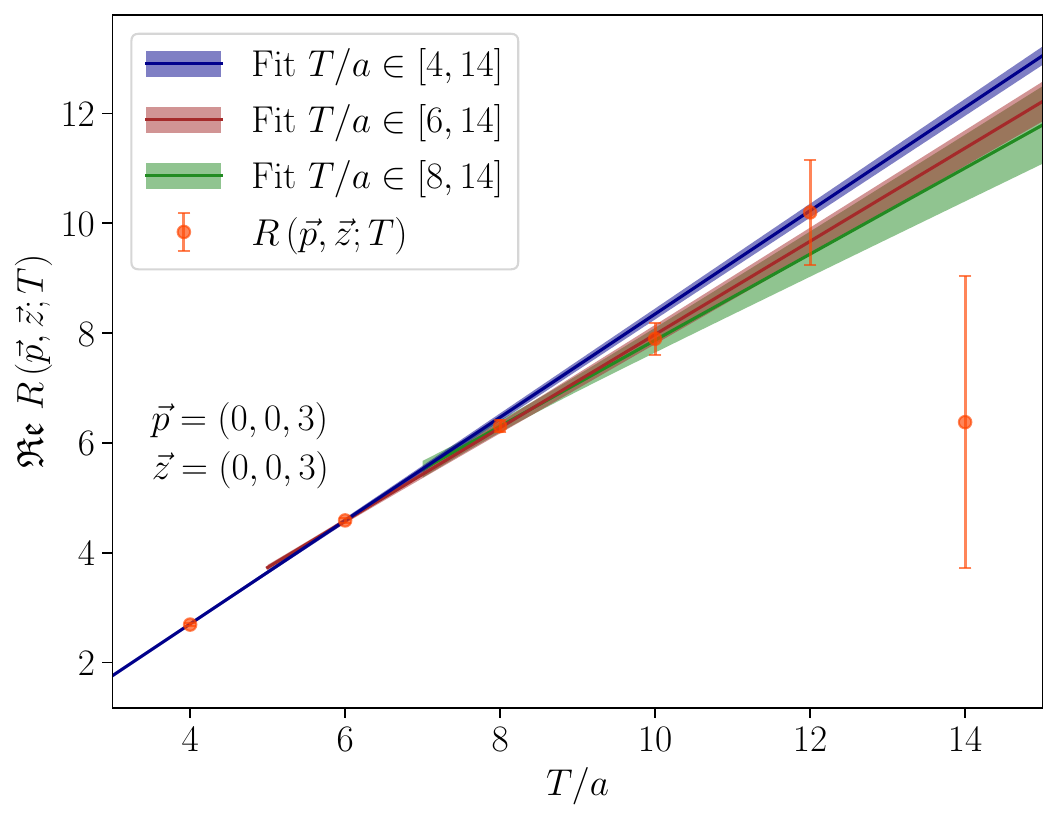}
    \hfill
    \includegraphics[width=0.33\linewidth]{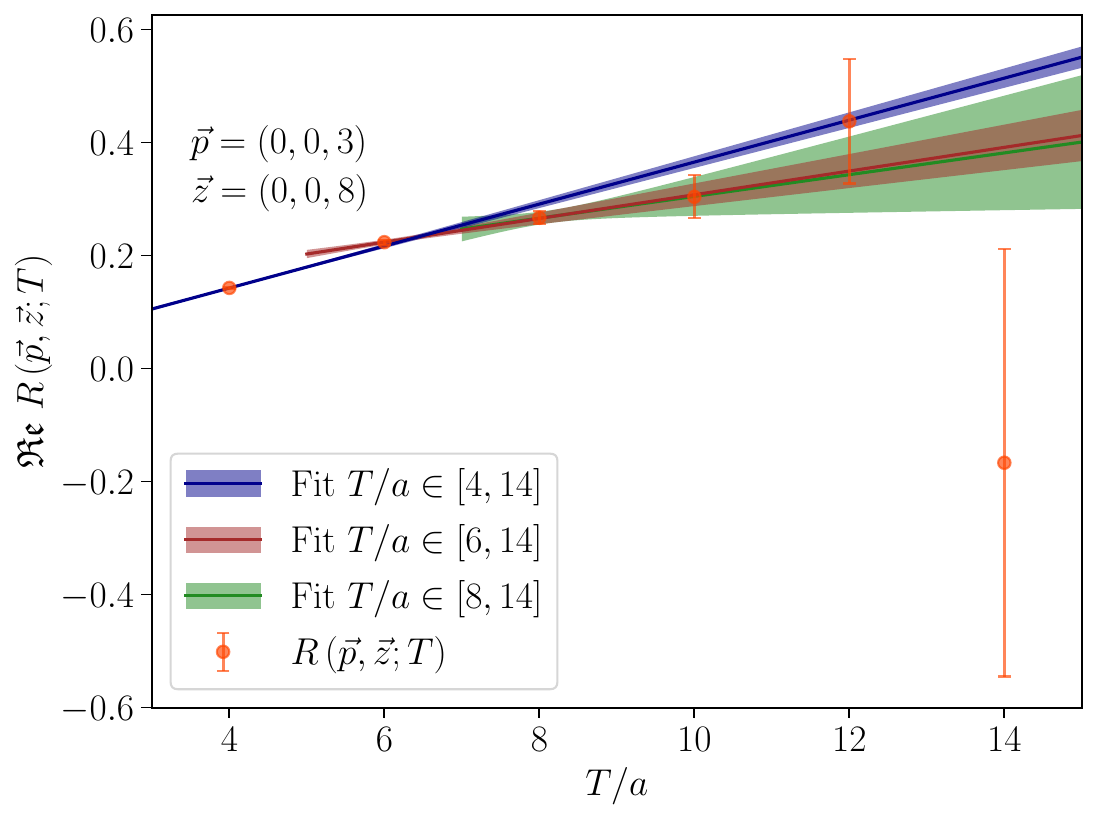}
    \hfill
    \includegraphics[width=0.325\linewidth]{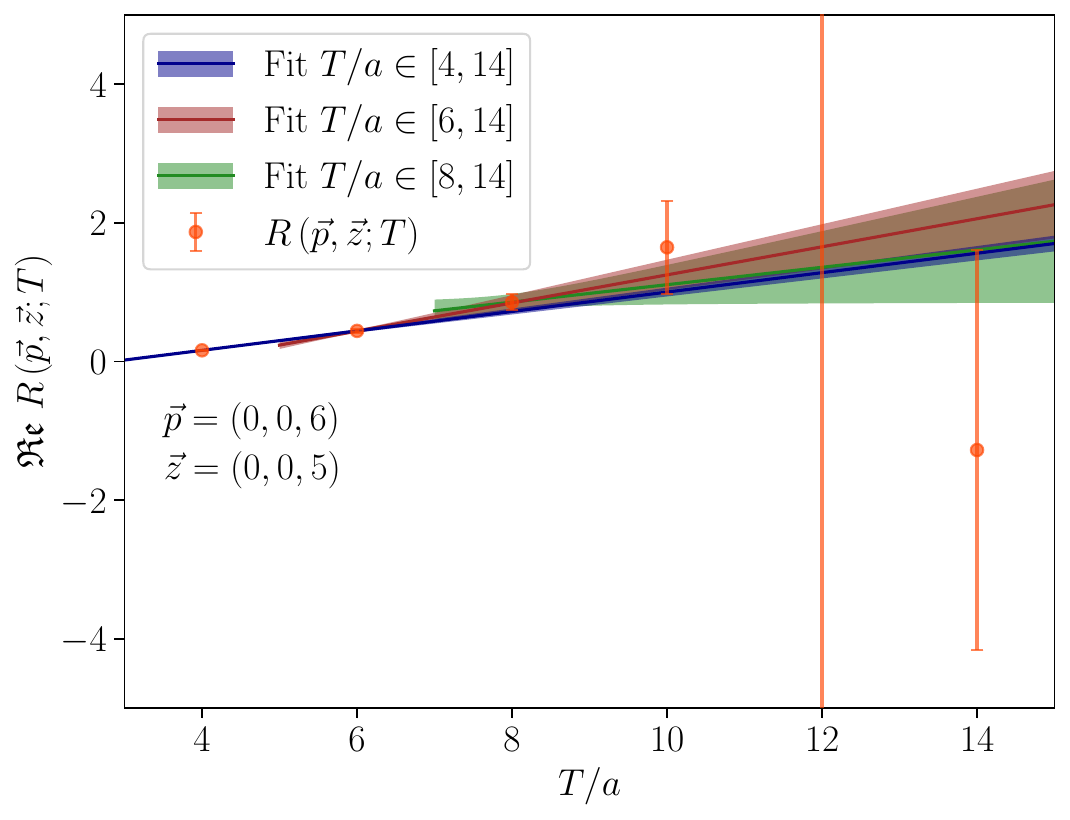}
    \\
    \includegraphics[width=0.32\linewidth]{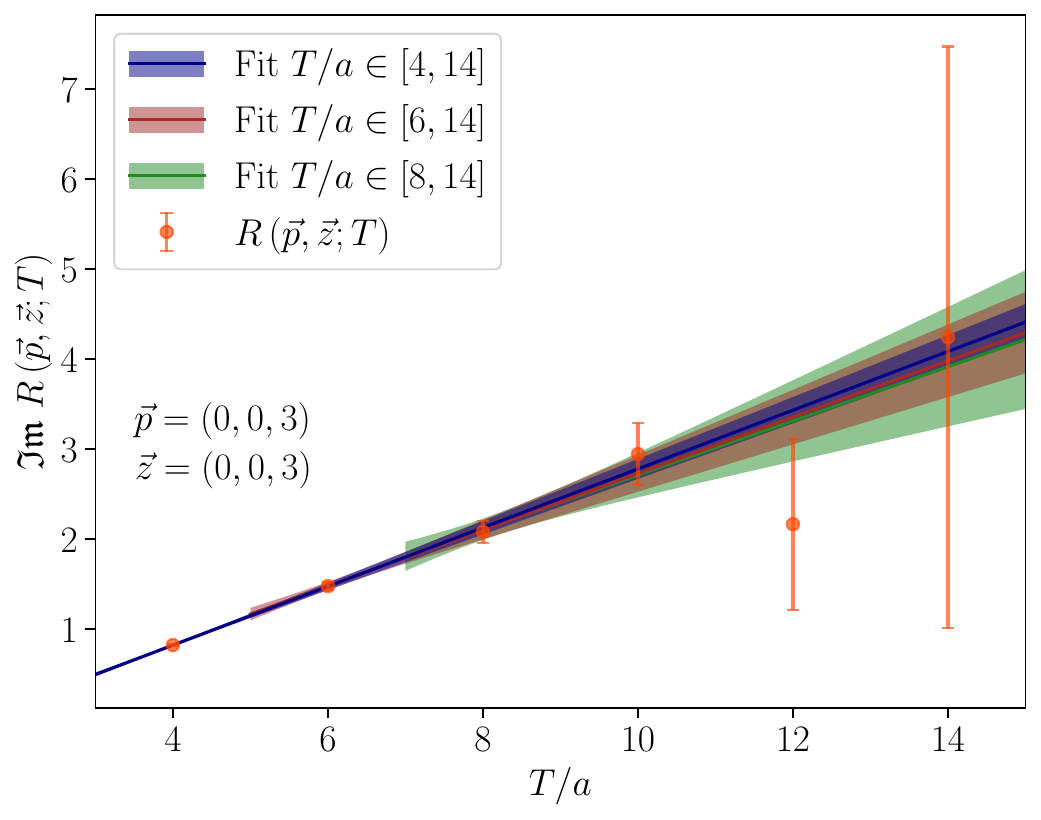}
    \hfill
    \includegraphics[width=0.33\linewidth]{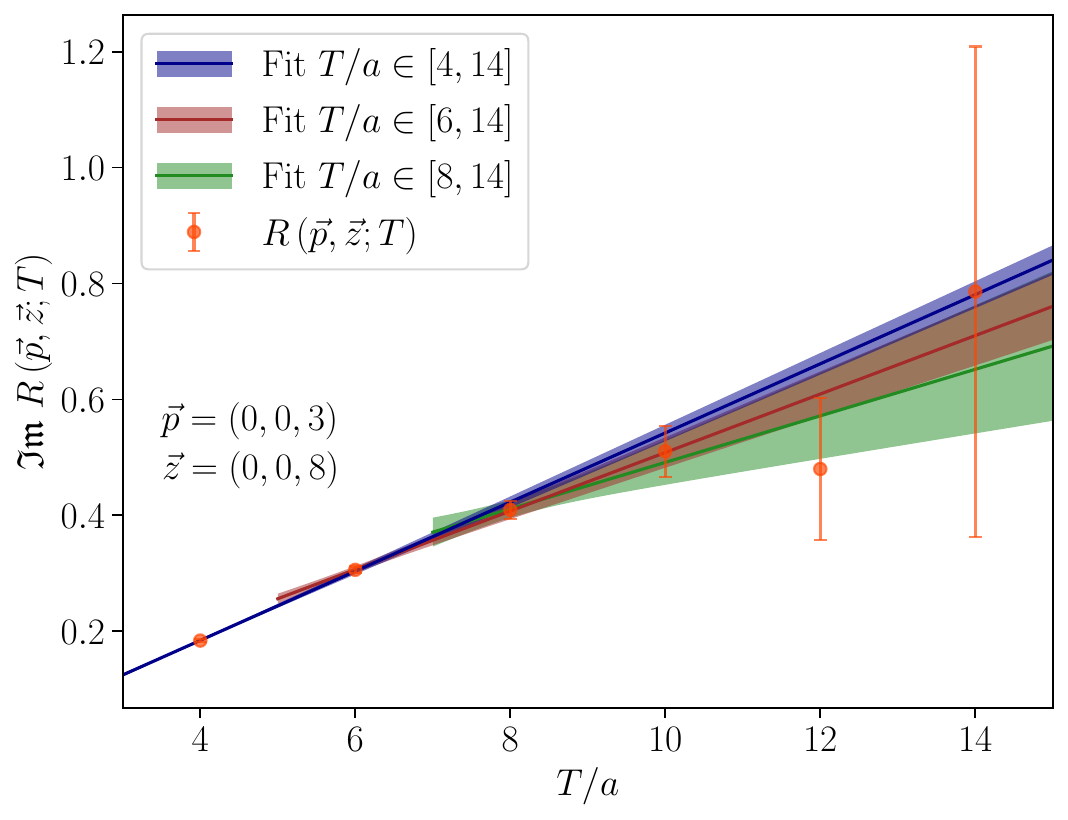}
    \hfill
    \includegraphics[width=0.33\linewidth]{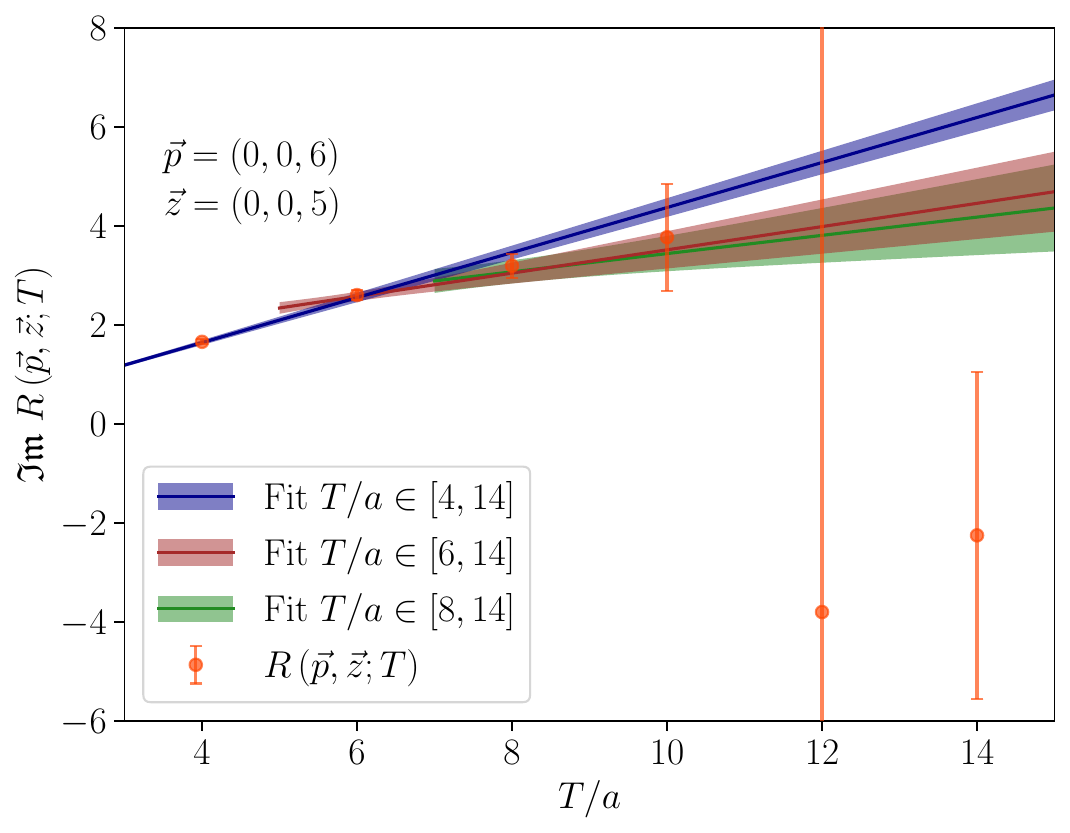}
    \caption{Real (upper row) and imaginary (lower row) components of the summed ratio $R\left(\vec{p},\vec{z};T\right)$ for select lattice momenta $\vec{p}$ together with the linear fit Eq.~\ref{eq:linFit} applied for varying temporal series. The length of the Wilson line is increased from left to right for each momentum, and given in integer multiples of the lattice spacing. Each panel corresponds to $R\left(\vec{p},\vec{z};T\right)$ determined from a particular {\it subduced} correlation function, which are discussed in Sec.~\ref{sec:pitdFromSVD}
    % Shown for row1-row2 combination.
    \label{fig:moving_SRs}}
\end{figure}

\begin{figure}[h]
    \centering
    \includegraphics[width=0.49\linewidth]{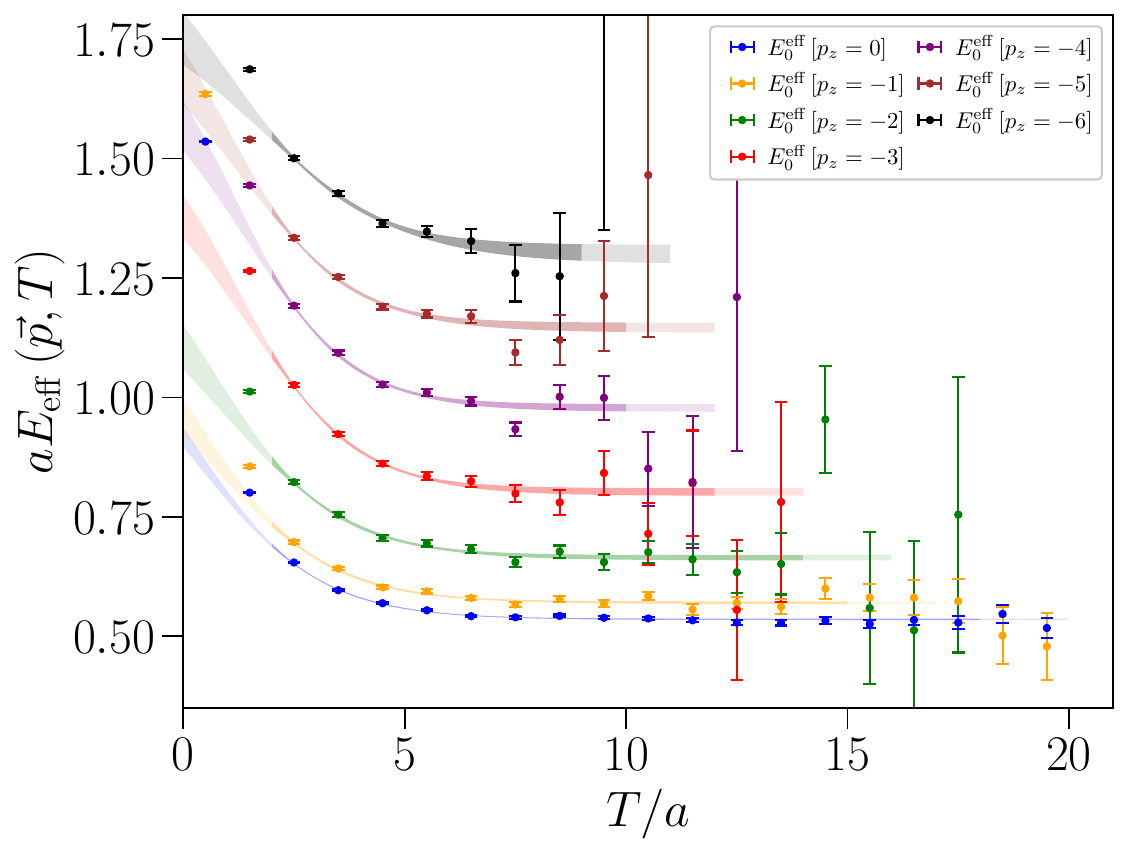}
    \hfill
    \includegraphics[width=0.49\linewidth]{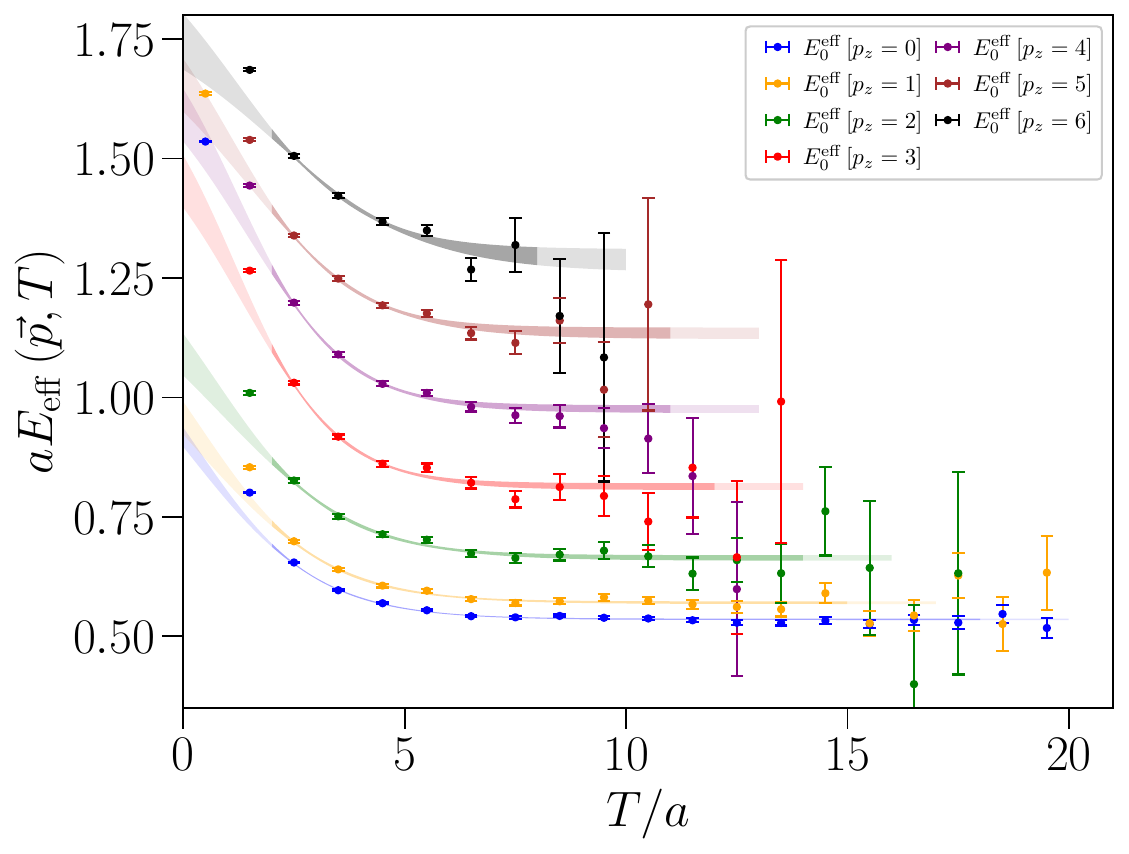}
    \caption{Effective energies for each lattice momenta $p_z^{\rm latt}\leq0$ (left) and $p_z^{\rm latt}\geq0$ (right), together with applied two-state fits. Time series included in each fit are indicated by darkened fit bands, and data are shown for signal-to-noise ratios greater than $2.25$.}  \label{fig:effEnergies}
\end{figure}

A representative subset of summed ratio data $R\left(p_z\hat{z},z_3;T\right)$ and applied linear fits~\eqref{eq:linFit} for source-sink separations $T/a\in\lbrace4-14,6-14,8-14\rbrace$ are presented in Fig.~\ref{fig:rest_SRs} and Fig.~\ref{fig:moving_SRs}. For the real component of the rest-frame matrix elements presented in Fig.~\ref{fig:rest_SRs}, the high quality data for $R\left(p_z\hat{z},z_3;T\right)$ for all Wilson line lengths we consider is seen to lead to a stable determination of the bare matrix elements as the minimum fitted value of $T/a$ is varied.
As the nucleon momentum is increased, the expected degradation of the correlation function signals is encountered. Together with the exponential decay of the bare matrix element signal as $z_3$ is increased, one observes poorer quality $R\left(p_z\hat{z},z_3;T\right)$ data and deviations from linearity for large values of $T$. For example, in Fig~\ref{fig:moving_SRs} we illustrate both the real and imaginary components of $R\left(p_z\hat{z},z_3;T\right)$ data and applied linear fits for nucleon momenta $ap_z=3\times\frac{2\pi}{L}$ and $z_3\in\lbrace3a,8a\rbrace$, as well as for $ap_z=6\times\frac{2\pi}{L}$ and $z_3=5a$. In each case, the summed ratio data $R\left(p_z\hat{z},z_3;T\right)$ are seen to be clearly linear for $T/a\leq8$, with deviations for larger values. As a result, linear fits $R_{\rm fit}\left(p_z\hat{z},z_3;T\right)$ begin to exhibit spread in the fitted bare matrix elements as the minimum fitted value of $T/a$ is increased. This spread is, in part, understood by considering the derived effective energies in Fig.~\ref{fig:effEnergies} of the ground-state nucleon.
The reader is reminded the largest momenta without the use of a phased distillation space are $\left|ap_z\right|=3\times\frac{2\pi}{L}$, shown in red in Fig.~\ref{fig:effEnergies}. For $T/a\gtrsim10$ appreciable variations are observed in the effective energies which, in forming $R\left(p_z\hat{z},z_3;T\right)$, produces the expected spread in the fitted matrix element. Even with the use of phasing (purple, brown, and black in Fig.~\ref{fig:effEnergies}), any statistically meaningful data are lost for $T/a\gtrsim10$.
In the hopes of obtaining a good balance between excited-state control and statistical precision, for the remainder of this manuscript we will thus consider the matrix elements obtained from the $T/a\in\left[6,14\right]$ fits.

\subsection{Inclusion of a Matrix Element Fitting Systematic\label{sec:matelemSystematic}}
The choice to consider the reduced pseudo-ITD $\rpitd\left(\nu,z^2\right)$ obtained from summed ratio $R\left(p_z\hat{z},z_3;T\right)$ fits with $T/a\in\left[6,14\right]$ is a selection that potentially introduces a systematic error into the resulting reduced pseudo-ITD. This systematic arises from uncertainty in the fitted matrix element as distinct time series are considered. For example, in Fig.~\ref{fig:moving_SRs} statistically significant differences are observed between the $T/a\in\left[4,14\right]$ and $T/a\in\left[6,14\right]$ summed ratio fits for many momenta and Wilson line length combinations, a symptom of the combined effect of high-precision points weighting the $T/a\in\left[4,14\right]$ fits and excited-state pollution. We will not consider further the $T/a\in\left[8,14\right]$ summed ratio fits given the low number of usable points in each fit, especially for the highest momenta we consider.

We attempt to estimate this systematic uncertainty and include it in our ensuing analysis by simultaneously considering both $T/a\in\left[4,14\right]$ and $T/a\in\left[6,14\right]$ summed ratio fits. Rather than simultaneously fit both summed ratio datasets, we include the squared difference of $\rpitd\left(\nu,z^2\right)$ independently obtained from both $T/a\in\left[4,14\right]$ and $T/a\in\left[6,14\right]$ summed ratio fits along the diagonal of the $T/a\in\left[6,14\right]$ data covariance:
\be
{\rm Cov}\left[\rpitd\left(\nu,z^2\right)_i,\rpitd\left(\nu,z^2\right)_j\right]={\rm Cov}\left[\rpitd\left(\nu,z^2\right)_i,\rpitd\left(\nu,z^2\right)_j\right]\suchthat{\frac{0}{0}}_{T^{\rm fit}_{\rm min}=6a}+\delta_{ij}\left[\rpitd\left(\nu,z^2\right)\suchthat{\mathfrak{Y}\left(\nu,z^2\right)}_{T^{\rm fit}_{\rm min}=4a}-\rpitd\left(\nu,z^2\right)\suchthat{\mathfrak{Y}\left(\nu,z^2\right)}_{T^{\rm fit}_{\rm min}=6a}\right]^2,
\label{eq:matelem_sys}
\ee
where ${\rm Cov}\left[\rpitd\left(\nu,z^2\right)_i,\rpitd\left(\nu,z^2\right)_j\right]{\Large\mid}_{T^{\rm fit}_{\rm min}=6a}$ is the unaltered reduced pseudo-ITD data covariance with bare matrix elements obtained by fitting over the time series $T/a\in\left[6,14\right]$, and $i,j$ denoting specific $\lbrace p_z,z_3\rbrace$ tuples. This simple procedure is performed at the level of each jackknife fit, thereby retaining correlation within the data and covariance between matrix element fits of distinct momenta and Wilson line lengths. An analogous strategy was conceived in Ref.~\cite{Karpie:2021pap} when resolving the continuum and leading-twist limits of parton pseudo-distributions. Herein the reduced pseudo-ITD $\mathfrak{Y}\left(\nu,z^2\right)$ will be understood to reflect this systematic error estimate.

\subsection{On the Extraction of the Helicity Pseudo-ITD\label{sec:pitdFromSVD}}
The bare matrix element $M^{\mu5}\left(p,z\right)$, once isolated, receives contributions from three invariant amplitudes (see Sec.~\ref{sec:theory}). One of these invariant amplitudes, $\mathcal{R}\left(\nu,z^2\right)$, is absent on the light-cone, but enters as a contamination to the space-like matrix elements we compute. With our choice of kinematics, namely hadron momentum $p_\mu=\left(\mathbf{0}_\perp,p_z,E\left(p_z\right)\right)$ with Wilson line length $z_\mu=\left(\mathbf{0}_\perp,z_3,0\right)$ and gamma matrix Lorentz index $\mu=3$, $\mathcal{R}\left(\nu,z^2\right)$ is unavoidable (cf. Eq.~(\ref{eq:decomp})). Our choice to construct interpolators that transform irreducibly under the double-cover octahedral group $O_h^D$ and its discrete little groups leads to several determinations of $M^{\mu5}\left(p,z\right)$ for a given choice of $\lbrace p_z,z_3\rbrace$ and $\mu=3$. The resulting system can be solved via a singular value decomposition (SVD) for the pseudo-ITD $\widetilde{\mathcal{Y}}\left(\nu,z^2\right)$. Indeed an SVD in this context may seem excessive, however its use establishes a general framework to analyze future off-forward matrix elements relevant for the isolation of GPDs, as well as the hypothetical scenario wherein $\mathcal{R}\left(\nu,z^2\right)$ could be disentangled from the leading pseudo-ITD $\mathcal{Y}\left(\nu,z^2\right)$ were finite mixing not an issue. To provide a cogent foundation for this procedure, we step aside to consider in closer detail the group theoretic construction of our interpolators.

The irrep-based correlation functions we compute of some external current $\mathcal{J}$ expose what are deemed {\it subduced matrix elements} following construction of appropriate ratios of three-point and two-point functions. To then access the canonical matrix element in Eq.~\eqref{eq:hel-mat} we must establish their connection with the subduced matrix elements. The relationship between the subduced matrix elements and invariant amplitudes is given by
\be
\bra{\vec{p},\Lambda,\mu_f}\mathcal{J}^{\Lambda_\Gamma,\mu_\Gamma}\ket{\vec{p},\Lambda,\mu_i}=\sum_l\sum_{\lambda_f,\lambda_\Gamma,\lambda_i}S^{\Lambda,J}_{\mu_f,\lambda_f}\left[S_{\mu_\Gamma,\lambda_\Gamma}^{\Lambda_\Gamma,J_\Gamma}\right]^*\left[S_{\mu_i,\lambda_i}^{\Lambda,J}\right]^*\mathcal{K}_l\left(\lambda_f\left[J,\vec{p}\right];\lambda_i\left[J,\vec{p}\right]\right)\mathcal{A}_l\left(\nu,z^2\right),
\label{eq:subduced-matelem}
\ee
where a subduced matrix element $\bra{\vec{p},\Lambda,\mu_f}\mathcal{J}^{\Lambda_\Gamma,\mu_\Gamma}\ket{\vec{p},\Lambda,\mu_i}$ depends on a linear combination of invariant amplitudes $\mathcal{A}_l$, each of which is weighted by a helicity-dependent kinematic prefactor $\mathcal{K}_l\left(\lambda_f\left[J,\vec{p}\right];\lambda_i\left[J,\vec{p}\right]\right)$ whose irreducibility under the relevant cubic (sub)group is encoded via subduction coefficients\footnote{Subduction coefficients $S^{\Lambda,J}_{\mu,\lambda}$ encode how the different values of helicity $\lambda$ of each object comprising our correlation functions subduce into distinct rows $\mu\in\lbrace1,{\rm dim}\left(\Lambda\right)\rbrace$ of, in general, several irreps $\Lambda$.} $S_{\mu,\lambda}^{\Lambda,J}$. The kinematic prefactors are nothing but the Lorentz covariant structures associated with each invariant amplitude evaluated in a helicity basis, and, as we detail below, each can be straightforwardly isolated. In this notation, we follow the convention of Ref.~\cite{Thomas:2011rh} by treating source interpolators and current insertions as creation operators, whereby the corresponding subduction coefficients are conjugated.

Rather than map the subduced matrix elements back into a canonical form like Eq.~\ref{eq:hel-mat}, thereby abandoning the group symmetries we rely on, we instead re-express the canonical matrix elements appearing in the spectral decomposition of our three-point functions~\eqref{eq:3pt} in terms of the subduced matrix elements. From the discussion concerning operator-state overlaps at the outset of Sec.~\ref{sec:numerics}, it follows the two complete sets of states, introduced between the nucleon interpolators and insertion in $C^{\left[\gamma_3\gamma_5\right]}_{3{\rm pt}}\left(p_z\hat{z},T;z_3,\tau\right)$ to expose its spectral content, are projected by our interpolator-state overlaps onto the infinite tower of continuum helicity eigenstates that subduce into the irrep $\Lambda$ of the little group defined by $^*\left(\vec{p}=p_z\hat{z}\right)$. Despite the presence of excited-states and, in general, negative parity and $J>\tfrac{1}{2}$ states that contribute to the irrep-based correlation functions in motion, in the limit of large Euclidean time $C^{\left[\gamma_3\gamma_5\right]}_{3{\rm pt}}\left(p_z\hat{z},T;z_3,\tau\right)$ will be proportional to the subduced matrix element associated with the continuum $J^P=\tfrac{1}{2}^+$ ground-state nucleon. Evidently the principal step needed to match the canonical matrix elements onto the subduced matrix elements is to evaluate the inner product of spinors representing the initial/final $\frac{1}{2}^+$ subduced states with each Lorentz structure associated with the invariant amplitudes $\mathcal{A}_l$ in Eq.~\eqref{eq:subduced-matelem}.

We now detail an algorithm for obtaining spinors representing the subduced initial/final states. The algorithm begins by considering a standard four-component Dirac spinor at rest and with spin quantized along the $\hat{z}$-axis:
\be
u\left(0,m\right)=\sqrt{E(\vec{p}=\vec{0})+M_N}\quad\left(\vphantom{\frac{1}{2}}\chi\left(m\right)\quad\mathbf{0}\right)^T,
\ee
where $\chi\left(m\right)$ are the standard non-relativistic two-component spinors with $\hat{z}$ component $m$, and $M_N$ the nucleon ground-state mass. To build the relativistic Dirac spinor in a helicity basis, we first perform a Lorentz boost $L_z\left(\left|\vec{p}\right|\right)$ on $u\left(0,m\right)$, such that it carries all the momentum of our desired state along its axis of quantization, followed by an active rotation $\mathfrak{D}\left[R\right]$ of the relativistic spinor to the direction $\vec{p}$:
\be
u\left(\vec{p},\lambda\right)=\mathfrak{D}\left[R\right]u\left(\left|\vec{p}\right|\hat{z},m\right)=\mathfrak{D}\left[R\right]L_z\left(\left|\vec{p}\right|\right)u\left(0,m\right),
\ee
with $\lambda$ the resulting values of helicity. Since the two component spinors $\chi\left(m\right)$ transform under rotations as~\cite{Chung:1971ri}
\be
U\left[R\left(\alpha,\beta,\gamma\right)\right]\chi\left(m\right)=\sum_{m'}\mathcal{D}_{m'm}^{1/2}\left(\alpha,\beta,\gamma\right)\chi\left(m'\right),
\ee
where $\mathcal{D}_{m'm}^{1/2}$ is a Wigner-D matrix and $U\left[R\right]$ a unitary operator encoding the the active rotation $R$, denoted by the Euler angles $\lbrace\alpha,\beta,\gamma\rbrace$ in the $zyz$-convention, it follows
\be
u\left(\vec{p},\lambda\right)=\mathfrak{D}\left[R\right]u\left(\left|\vec{p}\right|\hat{z},m\right)=
\begin{pmatrix}
  U\left[R\left(\alpha,\beta,\gamma\right)\right] & \mathbf{0} \\
  \mathbf{0} & U\left[R\left(\alpha,\beta,\gamma\right)\right]
\end{pmatrix}
u\left(\left|\vec{p}\right|\hat{z},m\right).
\ee
An explicit realization of $U\left[R\right]$ for spin-$1/2$ states:
\be
U\left[R\left(\alpha,\beta,\gamma\right)\right]=\exp\left(-i\frac{\alpha}{2}\sigma_z\right)\exp\left(-i\frac{\beta}{2}\sigma_y\right)\exp\left(-i\frac{\gamma}{2}\sigma_z\right),
\ee
then completes the construction of the relativistic helicity spinor. To then impose the appropriate group symmetry, the same subduction coefficients employed in our irrep-based correlation functions are applied to the relativistic helicity spinor:
\be
u\left(\vec{p},\Lambda,\mu\right)=\sum_\lambda S^{\Lambda J}_{\mu\lambda}u\left(\vec{p},\lambda\right)=\sum_\lambda S^{\Lambda J}_{\mu\lambda}\mathcal{D}\left[R\right]u\left(\left|\vec{p}\right|\hat{z},m\right),
\ee
thereby producing a subduced spinor $u\left(\vec{p},\Lambda,\mu\right)$ that transforms irreducibly within the irrep $\Lambda$ of the appropriate little group with {\it row} $\mu\in\lbrace1,\dim\left(\Lambda\right)\rbrace$. Since in this calculation we consider nucleon momenta $\vec{p}=p_z\hat{z}$ collinear to the space-like Wilson line, the relevant irreps we must consider are $G_{1g}$ and $E_1$. The former is the $O_h^D$ irrep non-relativistic spin-$1/2$ objects subduce into, while the latter is relevant for the subduction of $\lambda=1/2$ objects residing within the order-16 dicyclic (${\rm Dic}_4$) cubic subgroup, or little group - this little group is defined by the set of rotations that leave momenta of the form $\vec{p}=\left(0,0,n\right)\mid_{n\in\mathbb{Z}\setminus\lbrace0\rbrace}$, and its permutations, invariant. Indeed the patterns of subduction for objects with non-zero helicity is quite nuanced. For example, $\lambda=7/2$ objects within ${\rm Dic}_4$ are also represented by the $E_1$ irrep, while for momenta of the form $\vec{p}=\left(0,n,n\right)$ the irrep $E_1$ will also contain information concerning $\lambda=5/2$ states~\cite{Moore:2006ng}.

The $\lambda\neq1/2$ states present within an irrep enter as additional contamination to the correlation functions we compute. Under the supposition of ground-state dominance, we need only consider the pattern of subduction for a $J=1/2$ object at rest and a $\lambda=1/2$ object in motion. The subduction of interpolators at rest with continuum spin $J=0,1$ and $J=1/2,3/2$ are faithfully represented by the $A_1,T_1$ and $G_1,H$ irreps, respectively - the subduction coefficients for baryons at rest with continuum spin $3/2\leq J\leq9/2$ are given in Ref.~\cite{Edwards:2011jj}. The subduction coefficients for baryons in-flight can be obtained following the prescription established in Ref.~\cite{Thomas:2011rh}. Discussion concerning the inserted current's subduction coefficients $S_{\mu_\Gamma,\lambda_\Gamma}^{\Lambda_\Gamma,J_\Gamma}$ is omitted, as the inserted $A_3=\overline{\psi}\gamma_3\gamma_5\psi$ current in the forward limit is faithfully represented by the three-dimensional $T_1$ irrep of the cubic group. Were we to consider an insertion with $J>3/2$ in the forward limit or an off-forward case more generally, the subduction of the inserted current would need to be carefully considered.

\begin{figure}
    \centering
    \includegraphics[width=0.49\linewidth]{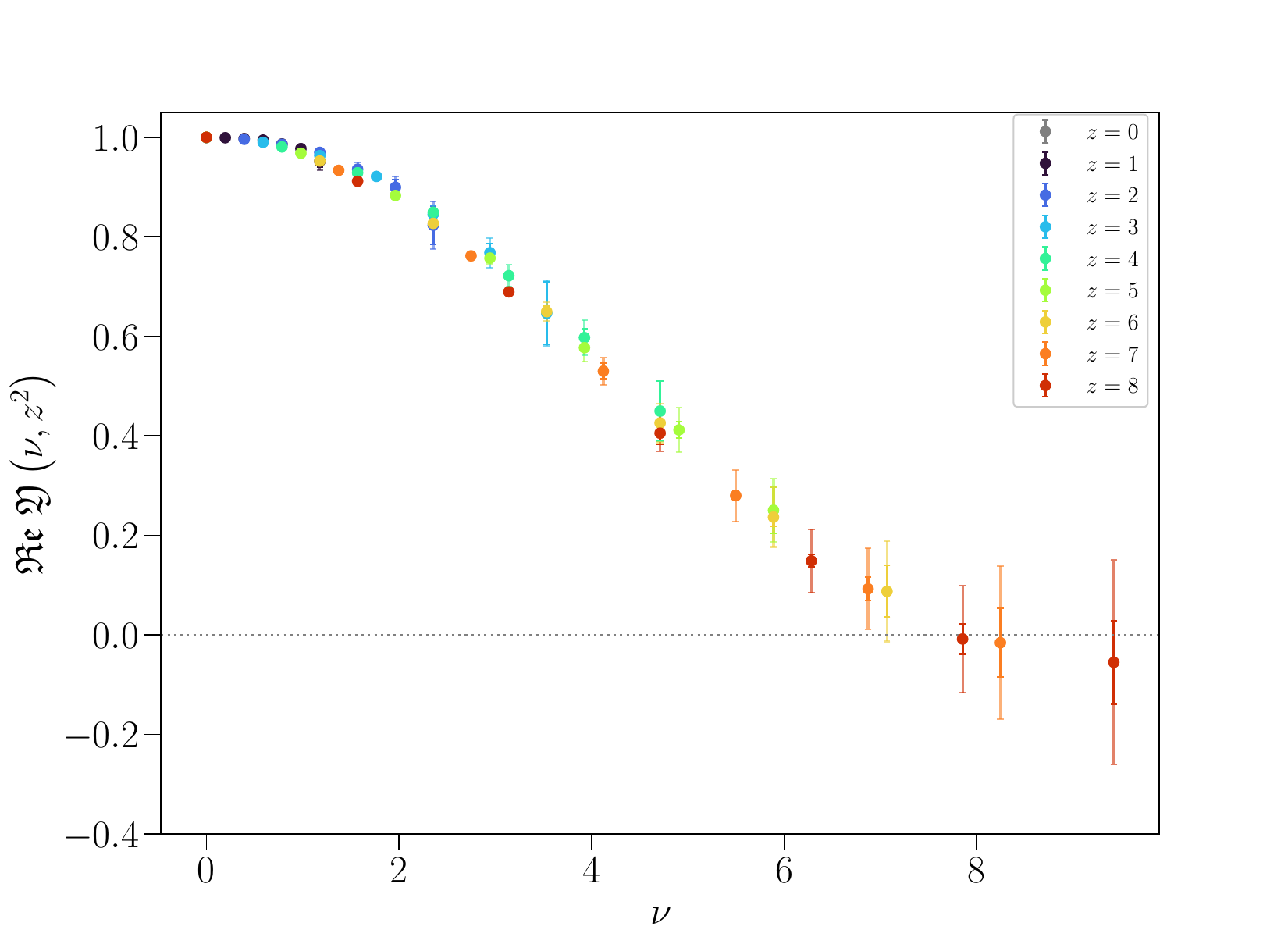}
    \includegraphics[width=0.49\linewidth]{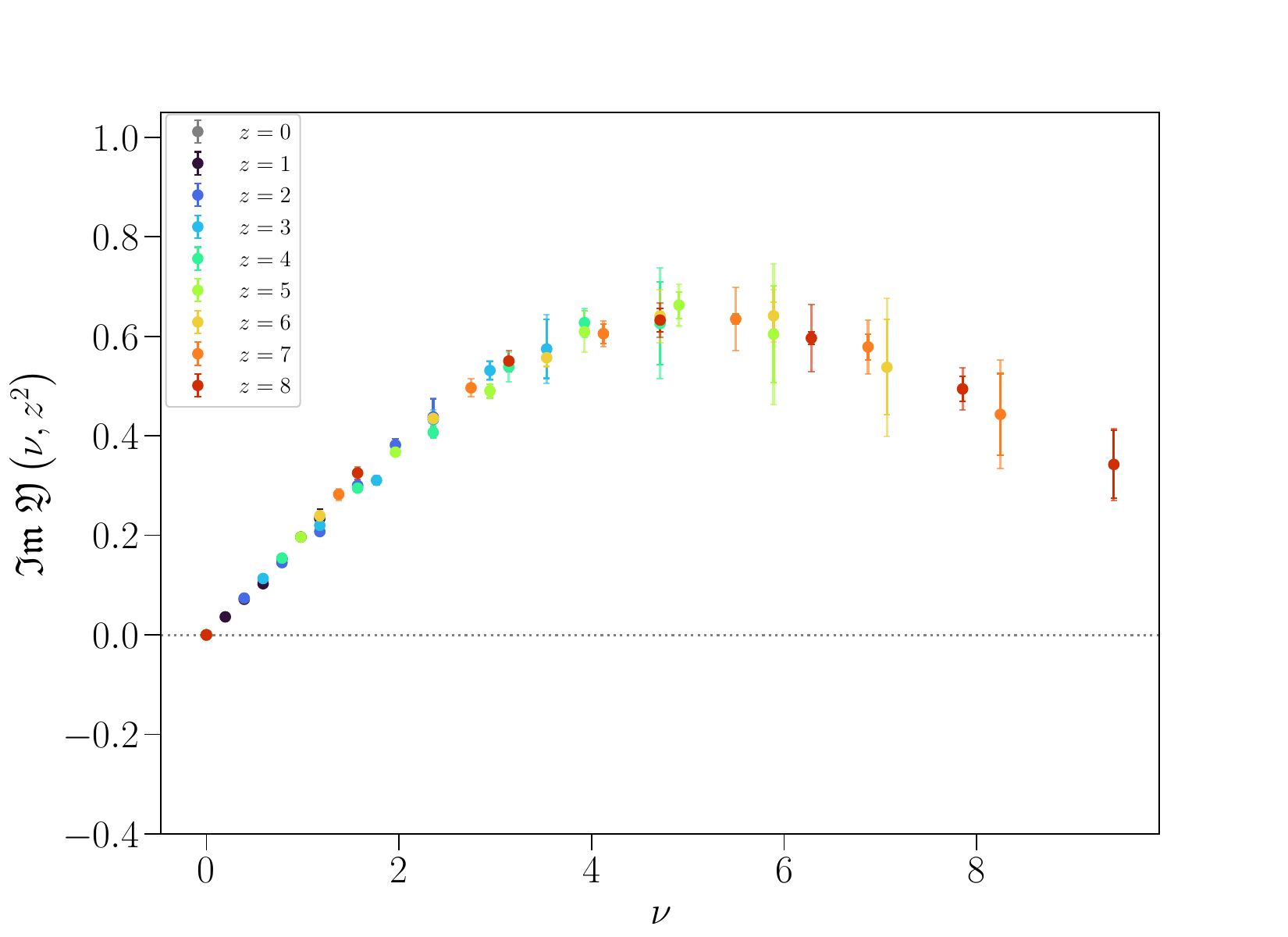}
    \caption{The real (left) and imaginary (right) components of the reduced pseudo-ITD $\rpitd\left(\nu,z^2\right)$ obtained from the SVD applied to the summed ratio $R_{\rm fit}\left(p_z\hat{z},z_3;T\right)$ fits for $T/a\in\left[6,14\right]$ (darkened inner errors), with a matrix element fitting systematic estimated using the $T/a\in\left[4,14\right]$ fits (lightened outer errors).}
    \label{fig:rpitd_with_fitsys}
\end{figure}

The final step in connecting our irrep-based correlation functions with the canonical matrix element in Eq.~\eqref{eq:hel-mat} is to evaluate the contraction of the final/initial state subduced spinors with each Lorentz structure. As $G_{1g}$ and $E_1$ are both two-dimensional irreps, a total of four distinct subduced spinor contractions are realized for each combination of $\lbrace p_z,z_3\rbrace$. This produces what we deem a kinematic matrix, where each column corresponds to the four possible subduced spinor contractions associated with a given Lorentz structure, while each row will encode all contractions corresponding to a specific choice for the initial and final state subduced spinor rows. The kinematic matrix, when right-multiplied by the vector of invariant amplitudes (cf. Eq.~\ref{eq:matIntoYAndR}), can then be directly equated with the fitted subduced matrix elements. The result is the system of equations:
\be
\begin{pmatrix}
  \bra{p_z\hat{z},\Lambda,\mu_f=1}\mathring{\mathcal{O}}^{[\gamma_3\gamma_5]}\left(z_3\right)\ket{p_z\hat{z},\Lambda,\mu_i=1} \\
  \bra{p_z\hat{z},\Lambda,\mu_f=1}\mathring{\mathcal{O}}^{[\gamma_3\gamma_5]}\left(z_3\right)\ket{p_z\hat{z},\Lambda,\mu_i=2} \\
  \bra{p_z\hat{z},\Lambda,\mu_f=2}\mathring{\mathcal{O}}^{[\gamma_3\gamma_5]}\left(z_3\right)\ket{p_z\hat{z},\Lambda,\mu_i=1} \\
  \bra{p_z\hat{z},\Lambda,\mu_f=2}\mathring{\mathcal{O}}^{[\gamma_3\gamma_5]}\left(z_3\right)\ket{p_z\hat{z},\Lambda,\mu_i=2}
\end{pmatrix}
=-2m_N
\begin{pmatrix}
  S_3\left[p_z\hat{z}\right]_{11} & m_N^2z_3^2S_3\left[p_z\hat{z}\right]_{11} \\
  S_3\left[p_z\hat{z}\right]_{12} & m_N^2z_3^2S_3\left[p_z\hat{z}\right]_{12} \\
  S_3\left[p_z\hat{z}\right]_{21} & m_N^2z_3^2S_3\left[p_z\hat{z}\right]_{21} \\
  S_3\left[p_z\hat{z}\right]_{22} & m_N^2z_3^2S_3\left[p_z\hat{z}\right]_{22} \\
\end{pmatrix}
\begin{pmatrix}
  \mathcal{Y}\left(\nu,z^2\right) \\
  \mathcal{R}\left(\nu,z^2\right)
\end{pmatrix},
\ee
where the entries on the left-hand side are the fitted subduced matrix elements obtained from the $R_{\rm fit}\left(p_z\hat{z},z_3;T\right)$ fits~\eqref{eq:linFit}, $\mu_f$ ($\mu_i$) the row of the final (initial) state subduced spinor, and with the functional dependence of the nucleon polarization vector $S_3$ on its momentum explicitly denoted. The subscripts of the polarization vector specify which row of the final/initial state subduced spinors are employed in the contraction. However, as noted, $\mathcal{R}\left(\nu,z^2\right)$ cannot be separated from $\mathcal{Y}\left(\nu,z^2\right)$ with our kinematic setup, so this system of equations instead reads:
\be
\begin{pmatrix}
  \bra{p_z\hat{z},\Lambda,\mu_f=1}\mathring{\mathcal{O}}^{[\gamma_3\gamma_5]}\left(z_3\right)\ket{p_z\hat{z},\Lambda,\mu_i=1} \\
  \bra{p_z\hat{z},\Lambda,\mu_f=1}\mathring{\mathcal{O}}^{[\gamma_3\gamma_5]}\left(z_3\right)\ket{p_z\hat{z},\Lambda,\mu_i=2} \\
  \bra{p_z\hat{z},\Lambda,\mu_f=2}\mathring{\mathcal{O}}^{[\gamma_3\gamma_5]}\left(z_3\right)\ket{p_z\hat{z},\Lambda,\mu_i=1} \\
  \bra{p_z\hat{z},\Lambda,\mu_f=2}\mathring{\mathcal{O}}^{[\gamma_3\gamma_5]}\left(z_3\right)\ket{p_z\hat{z},\Lambda,\mu_i=2}
\end{pmatrix}
=-2m_N
\begin{pmatrix}
  S_3\left[p_z\hat{z}\right]_{11}\quad & 0 \\
  S_3\left[p_z\hat{z}\right]_{12}\quad & 0 \\
  S_3\left[p_z\hat{z}\right]_{21}\quad & 0 \\ 
  S_3\left[p_z\hat{z}\right]_{22}\quad & 0 \\
\end{pmatrix}
\begin{pmatrix}
  \widetilde{\mathcal{Y}}\left(\nu,z^2\right) \\
  \mathcal{R}\left(\nu,z^2\right)
\end{pmatrix},
\label{eq:SVD}
\ee
where the contamination arising from $\mathcal{R}\left(\nu,z^2\right)$ for space-like intervals has been subsumed into the pseudo-ITD $\widetilde{\mathcal{Y}}\left(\nu,z^2\right)$, and the exclusive contribution of $\mathcal{R}\left(\nu,z^2\right)$ to this system has been removed by setting the elements of the rightmost column of the kinematic matrix to null. For each pair $\lbrace p_z,z_3\rbrace$, this system of equations is constructed and solved for the unknown amplitude $\widetilde{\mathcal{Y}}\left(\nu,z^2\right)$ using an SVD. Note the use of an SVD will isolate $\widetilde{\mathcal{Y}}\left(\nu,z^2\right)$ such that it is consistent with the subduced matrix elements at the left-hand side of Eq.~\ref{eq:SVD}. Since the non-trivial subduced matrix elements encode the continuum helicity matrix elements that define the helicity PDFs, the trivial use of SVD in this case, being nothing more than a sum of the non-trivial equations of this system, accomplishes the desired difference of continuum helicity matrix elements that theoretically define the PDF. Following the prescription established in Sec.~\ref{sec:theory}, we then populate the reduced pseudo-ITD $\mathfrak{Y}\left(\nu,z^2\right)$ shown in Fig.~\ref{fig:rpitd_with_fitsys}.

\section{On the Extraction of the Helicity PDF\label{sec:extraction}}
With the helicity reduced pseudo-ITD $\mathfrak{Y}\left(\nu,z^2\right)$ in hand, one is presented with an inverse problem that precludes an unambiguous determination of the helicity PDFs from the one-loop matching relationship~\eqref{eq:factorization}, which originates from the limited range of $\nu$. Indeed matching the reduced pseudo-ITD to a common scale in $\msbar$ alleviates some of these numerical challenges, as the resulting $\msbar$ helicity Ioffe-time distribution $\mathcal{I}\left(\nu,\mu^2\right)$, in principle, involves no residual $z^2$-dependence and directly determines the $x$-dependence of the underlying PDFs via an inverse Fourier transform. Such an evolution/matching step, however, does not quell the ill-posed inverse problem and is furthermore a potential source of additional systematic errors stemming from the interpolation or smooth description of the pseudo-distribution data needed for the evolution/matching procedure. Regardless of whether the matching step is applied explicitly to the reduced pseudo-ITD, it is common for a functional form to be assumed for the PDF and for its convolution with the matching kernel to be fit to the data~\cite{Sufian:2019bol,Sufian:2020vzb,HadStruc:2021wmh,Joo:2019jct,Joo:2019bzr,Joo:2020spy}. This paradigm is not unlike what one encounters in global analyses of experimental data, where a physically-motivated functional form is often assumed for the PDF~\cite{Martin:2009iq,Accardi:2016qay,Harland-Lang:2014zoa,Hou:2019efy} and fit to discrete cross-section measurements over a limited range of $x_B$. Several non-parametric reconstruction techniques have also been explored in the lattice QCD literature including the Backus-Gilbert method~\cite{10.1111/j.1365-246X.1968.tb00216.x}, Bayesian reconstruction~\cite{Karpie:2019eiq}, and an admixture giving rise to a Bayes-Gauss-Fourier transform~\cite{Alexandrou:2020tqq}.

To avoid sullying a high-fidelity determination of the helicity PDFs, we elect to parameterize the $CP$-even and $CP$-odd quark helicity PDFs via model ans{\"a}tze and fit their convolution with the one-loop matching kernel relating the $\lbrace\nu,z^2\rbrace$ dependencies of the reduced pseudo-ITD with the $\lbrace x,\mu^2\rbrace$ dependencies of the PDF:
\begin{align}
{\rm Re}\ \mathfrak{Y}\left(\nu,z^2\right)=g_A\left(\mu^2\right)^{-1}\int_0^1{\rm d}x\ \mathcal{K_-}\left(x\nu,z^2\mu^2,\alpha_s\left(\mu^2\right)\right)g_{q_-/N}\left(x,\mu^2\right)+\mathcal{O}\left(z^2\Lambda_{\rm QCD}^2\right) \nonumber \\
{\rm Im}\ \mathfrak{Y}\left(\nu,z^2\right)=g_A\left(\mu^2\right)^{-1}\int_0^1{\rm d}x\ \mathcal{K_+}\left(x\nu,z^2\mu^2,\alpha_s\left(\mu^2\right)\right)g_{q_+/N}\left(x,\mu^2\right)+\mathcal{O}\left(z^2\Lambda_{\rm QCD}^2\right).
\end{align}
Here the quark helicity PDFs $g_{q_-/N}\left(x,\mu^2\right)$ and $g_{q_+/N}\left(x,\mu^2\right)$ are isolated from the $\mathfrak{Y}\left(\nu,z^2\right)$ signal via cosine and sine transforms of the NLO kernel, defined in Eq.~\eqref{eq:nlo-kernel}, respectively:
\begin{align}
  &\mathcal{K}_-\left(x\nu,z^2\mu^2,\alpha_s\left(\mu^2\right)\right)=\int_0^1du\ \mathcal{C}\left(u,z^2\mu^2,\alpha_s\left(\mu^2\right)\right)\cos\left(u\nu x\right) \\
  &\mathcal{K}_+\left(x\nu,z^2\mu^2,\alpha_s\left(\mu^2\right)\right)=\int_0^1du\ \mathcal{C}\left(u,z^2\mu^2,\alpha_s\left(\mu^2\right)\right)\sin\left(u\nu x\right).
\end{align}
There is considerable flexibility in adopting a functional form to describe the unknown helicity PDFs. Absent a continuum of data over the infinite range $\nu\in[0,\infty)$, any functional choice necessarily introduces model bias into the extraction procedure - we will return to this point in Sec.~\ref{sec:aic}. Following the paradigm established in Refs.~\cite{Karpie:2021pap,Egerer:2021ymv,HadStruc:2021qdf}, we parameterize the unknown quark helicity PDFs using a basis of Jacobi polynomials\footnote{The Jacobi polynomials exploited in this work are referred to as such, however the set of conventional Jacobi polynomials, obtained from our Jacobi polynomials via a change of variables, are orthogonal on the interval $\left[-1,1\right]$.} $\lbrace\Omega_n^{\left(\alpha,\beta\right)}\left(x\right)\rbrace$, the set of which form a complete orthogonal set on the interval $x\in\left[0,1\right]$. The orthogonality of the Jacobi polynomials
\be
\int_0^1{\rm d}x\ x^\alpha\left(1-x\right)^\beta\Omega_n^{\left(\alpha,\beta\right)}\left(x\right)\Omega_m^{\left(\alpha,\beta\right)}\left(x\right)=\frac{1}{2n+\alpha+\beta+1}\frac{\Gamma\left(\alpha+n+1\right)\Gamma\left(\beta+n+1\right)}{n!\Gamma\left(\alpha+\beta+n+1\right)}\delta_{n,m}
\label{eq:orthog}
\ee
is assured provided\footnote{This limitation also ensures a properly normalized helicity PDF: $g_A^{u-d}\left(\mu^2\right)=\int_0^1{\rm d}x\ g_{q_-/N}\left(x,\mu^2\right)$, where $g_A^{u-d}\left(\mu^2\right)$ is the isovector axial charge of the nucleon.} 
$\alpha,\beta>-1$. The helicity PDFs $g_{q_-/N}\left(x,\mu^2\right)$ and $g_{q_+/N}\left(x,\mu^2\right)$ at some scale can therefore be unambiguously expressed as
\be
g_{q_\tau/N}\left(x\right)=x^\alpha\left(1-x\right)^\beta\sum_{n=0}^\infty C_{\tau,n}^{\left(\alpha,\beta\right)}\Omega_n^{\left(\alpha,\beta\right)}\left(x\right)
\label{eq:pdfViaJacobiPolys}
\ee
for arbitrarily chosen $\alpha,\beta>-1$, and $\tau\in\lbrace-,+\rbrace$ indicating either the $CP$-even or $CP$-odd quark helicity PDF. The relationship between the fitted parameters of the PDF and the reduced pseudo-ITD is obtained by considering the expansion of the matching kernels $\mathcal{K}_-\left(x\nu,z^2\mu^2,\alpha_s\left(\mu^2\right)\right)$ and $\mathcal{K}_+\left(x\nu,z^2\mu^2,\alpha_s\left(\mu^2\right)\right)$ in terms of Jacobi polynomials. 

In the interest of self-containment, the contribution of an order-$n$ Jacobi polynomial $\Omega_n^{\left(\alpha,\beta\right)}\left(x\right)$ to $\mathfrak{Y}\left(\nu,z^2\right)$ is given by
\begin{align}
  &\sigma_n^{\left(\alpha,\beta\right)}\left(\nu,z^2\mu^2\right)=\int_0^1dx\ \mathcal{K}_-\left(x\nu,z^2\mu^2\right)x^\alpha\left(1-x\right)^\beta\Omega_n^{\left(\alpha,\beta\right)}\left(x\right) \\
  &\eta_n^{\left(\alpha,\beta\right)}\left(\nu,z^2\mu^2\right)=\int_0^1dx\ \mathcal{K}_+\left(x\nu,z^2\mu^2\right)x^\alpha\left(1-x\right)^\beta\Omega_n^{\left(\alpha,\beta\right)}\left(x\right).
\end{align}
Expanding $\mathcal{K}_{-,+}$ in even/odd powers of Ioffe-time, we define:
\begin{align}
  &\sigma_n^{\left(\alpha,\beta\right)}\left(\nu,z^2\mu^2\right)=\sum_{j=0}^n\sum_{k=0}^\infty\frac{\left(-1\right)^k}{\left(2k\right)!}c_{2k}\left(z^2\mu^2\right)\omega_{n,j}^{\left(\alpha,\beta\right)}B\left(\alpha+2k+j+1,\beta+1\right)\nu^{2k}\label{eq:sigma_n} \\
  &\eta_n^{\left(\alpha,\beta\right)}\left(\nu,z^2\mu^2\right)=\sum_{j=0}^n\sum_{k=0}^\infty\frac{\left(-1\right)^k}{\left(2k+1\right)!}c_{2k+1}\left(z^2\mu^2\right)\omega_{n,j}^{\left(\alpha,\beta\right)}B\left(\alpha+2k+j+2,\beta+1\right)\nu^{2k+1}\label{eq:eta_n},
\end{align}
with $B\left(w,z\right)$ a Beta function, $c_n\left(z^2\mu^2\right)$ the Wilson coefficients defined in Eq.~\ref{eq:wilson-coeffs} and $\omega_{n,j}^{\left(\alpha,\beta\right)}\in\mathbb{R}$ defined as
\be
\omega_{n,j}^{\left(\alpha,\beta\right)}=\frac{\Gamma\left(\alpha+n+1\right)}{n!\Gamma\left(\alpha+\beta+n+1\right)}\binom{n}{j}\frac{\left(-1\right)^j\Gamma\left(\alpha+\beta+n+j+1\right)}{\Gamma\left(\alpha+j+1\right)}.
\ee
The leading-twist component of $\mathfrak{Y}\left(\nu,z^2\right)$ can then be written as the series
\begin{align}
\mathfrak{Re}\ \mathfrak{Y}_{lt}\left(\nu,z^2\right)&=\sum_{n=0}^\infty\sigma_n^{\left(\alpha,\beta\right)}\left(\nu,z^2\mu^2\right)C_{-,n}^{lt\ \left(\alpha,\beta\right)} \\
\mathfrak{Im}\ \mathfrak{Y}_{lt}\left(\nu,z^2\right)&=\sum_{n=0}^\infty\eta_n^{\left(\alpha,\beta\right)}\left(\nu,z^2\mu^2\right)C_{+,n}^{lt\ \left(\alpha,\beta\right)},
\end{align}
where $C_{\tau,n}^{lt\ \left(\alpha,\beta\right)}$ are the Jacobi polynomial expansion coefficients. In the spanning case, where all Jacobi polynomials are considered, any $\alpha,\beta>-1$ sets a suitable basis and the same $\sigma_n^{\left(\alpha,\beta\right)}$ and $\eta_n^{\left(\alpha,\beta\right)}$ can be used to describe known discretization and higher-twist contaminations in $x$-space; the latter includes the additional $z^2$ contamination arising from $\mathcal{R}\left(\nu,z^2\right)$ for space-like intervals. Their contributions in Ioffe-time are given analogously by
\begin{align}
\mathfrak{Re}\ \rpitd_{az}\left(\nu\right)&=\sum_{n=1}^\infty\sigma_{0,n}^{\left(\alpha,\beta\right)}\left(\nu\right)C_{-,n}^{az\ \left(\alpha,\beta\right)}\qquad\mathfrak{Im}\ \rpitd_{az}\left(\nu\right)=\sum_{n=0}^\infty\eta_{0,n}^{\left(\alpha,\beta\right)}\left(\nu\right)C_{+,n}^{az\ \left(\alpha,\beta\right)} \label{eq:discJacobi} \\
\mathfrak{Re}\ \rpitd_{ht}\left(\nu\right)&=\sum_{n=1}^\infty\sigma_{0,n}^{\left(\alpha,\beta\right)}\left(\nu\right)C_{-,n}^{ht\ \left(\alpha,\beta\right)}\qquad\mathfrak{Im}\ \rpitd_{ht}\left(\nu\right)=\sum_{n=0}^\infty\eta_{0,n}^{\left(\alpha,\beta\right)}\left(\nu\right)C_{+,n}^{ht\ \left(\alpha,\beta\right)}, \label{eq:htJacobi}
\end{align}
where $\sigma_{0,n}^{\left(\alpha,\beta\right)}$ and $\eta_{0,n}^{\left(\alpha,\beta\right)}$ are the tree-level components of Eq.~(\ref{eq:sigma_n}) and Eq.~(\ref{eq:eta_n}), and $C_{\tau,n}^{az\ \left(\alpha,\beta\right)}$ and $C_{\tau,n}^{ht\ \left(\alpha,\beta\right)}$ are nuisance parameters associated with $\mathcal{O}\left(a/\left|z_3\right|\right)$ discretization and $\mathcal{O}\left(z^{2k}\Lambda_{\rm QCD}^{2k}\right)$ higher-twist effects. Since $\mathfrak{Re}\ \mathfrak{Y}$ is identically unity for zero Ioffe-time and $\sigma_{0,0}^{\left(\alpha,\beta\right)}\left(\nu=0\right)\neq0$ (see Refs.~\cite{Karpie:2021pap} and~\cite{Egerer:2021ymv}), description of nuisance effects in $\mathfrak{Re}\ \mathfrak{Y}$ by Jacobi polynomials must begin at order $n=1$. Indeed the parameterization of the systematic corrections in this manner neglects additional sources of systematic error, notably those of $\mathcal{O}\left(\ln(z^2)\right)$ which could arise if $\sigma_n^{\left(\alpha,\beta\right)}\left(\nu,z^2\right)$ and $\eta_n^{\left(\alpha,\beta\right)}\left(\nu,z^2\right)$ were considered in Eq.~\eqref{eq:discJacobi} and Eq.~\eqref{eq:htJacobi}. We neglect such effects herein, as each nuisance term we consider is preceded by an ideally small prefactor (e.g. $a/\left|z_3\right|$, etc.) and the additional sources of systematic error arising at NLO would receive an additional suppression of $\mathcal{O}\left(\alpha_s\right)$.

In practice, the infinite series of Jacobi polynomials for the leading-twist and nuisance effects must be truncated, precluding a model-independent determination of the PDF at the expense of potential bias. This truncation is in fact necessary to avoid over-fitting the reduced pseudo-ITD given the finite range of Ioffe-time $\nu\in\left[0,9.42\right]$ within which $\mathfrak{Y}\left(\nu,z^2\right)$ is populated. It is known~\cite{Karpie:2021pap} each Jacobi polynomial, via $\sigma_n^{\left(\alpha,\beta\right)}$ and $\eta_n^{\left(\alpha,\beta\right)}$, contributes appreciably to the pseudo-distribution in a small window of Ioffe-time, and this window shifts to larger values of Ioffe-time as the polynomial order is increased. In this sense, only low order ($n\lesssim5$) Jacobi polynomials are needed to capture the information content of our data, as higher order ($n\gtrsim5$) Jacobi polynomials peak outside the Ioffe-time region for which we have data.
The complete functional form we use to describe the reduced pseudo-ITD and hence extract the leading-twist helicity PDFs is:
\begin{align}
  \mathfrak{Re}\ \rpitd_{\rm fit}\left(\nu,z_3^2\right)&=\sum_{n=0}^{N_{lt}}\sigma_n^{\left(\alpha,\beta\right)}\left(\nu,z_3^2\mu^2\right)C_{-,n}^{lt\ \left(\alpha,\beta\right)}+\frac{a}{\left|z_3\right|}\sum_{n=1}^{N_{az}}\sigma_{0,n}^{\left(\alpha,\beta\right)}\left(\nu\right)C_{-,n}^{az\ \left(\alpha,\beta\right)}\nonumber\\
  &\qquad\qquad\qquad+z_3^2\Lambda_{\rm QCD}^2\sum_{n=1}^{N_{t4}}\sigma_{0,n}^{\left(\alpha,\beta\right)}\left(\nu\right)C_{-,n}^{t4\ \left(\alpha,\beta\right)}+z_3^4\Lambda_{\rm QCD}^4\sum_{n=1}^{N_{t6}}\sigma_{0,n}^{\left(\alpha,\beta\right)}\left(\nu\right)C_{-,n}^{t6\ \left(\alpha,\beta\right)}\label{eq:jacFit-real} \\
  \mathfrak{Im}\ \rpitd_{\rm fit}\left(\nu,z_3^2\right)&=\sum_{n=0}^{N_{lt}}\eta_n^{\left(\alpha,\beta\right)}\left(\nu,z_3^2\mu^2\right)C_{+,n}^{lt\ \left(\alpha,\beta\right)}+\frac{a}{\left|z_3\right|}\sum_{k=0}^{N_{az}}\eta_{0,n}^{\left(\alpha,\beta\right)}\left(\nu\right)C_{+,n}^{az\ \left(\alpha,\beta\right)}\nonumber\\
  &\qquad\qquad\qquad+z_3^2\Lambda_{\rm QCD}^2\sum_{n=0}^{N_{t4}}\eta_{0,n}^{\left(\alpha,\beta\right)}\left(\nu\right)C_{+,n}^{t4\ \left(\alpha,\beta\right)}+z_3^4\Lambda_{\rm QCD}^4\sum_{n=0}^{N_{t6}}\eta_{0,n}^{\left(\alpha,\beta\right)}\left(\nu\right)C_{+,n}^{t6\ \left(\alpha,\beta\right)}\label{eq:jacFit-imag},
\end{align}
where the role of higher-twist effects up to and including $\mathcal{O}\left(z_3^4\Lambda_{\rm QCD}^4\right)$ are considered, and $N_{lt},N_{az},N_{t4},N_{t6}$ denote the truncation orders for the leading-twist and nuisance terms. We take the input scale to be $\mu=2$ GeV, with the three flavor $\msbar$ strong coupling $\alpha_s\left(\mu=2\text{ GeV}\right)=0.303$ and $\Lambda_{\rm QCD}=286$ MeV adopted from {\tt LHAPDF6}~\cite{Buckley:2014ana}.

We would like to determine the most likely set of parameters $\lbrace\theta\rbrace$ given our data $\rpitd\left(\nu,z^2\right)$ and any prior knowledge $I$. This is codified by Bayes' Theorem:
\be
\mathcal{P}\left[\theta\mid\rpitd,I\right]=\frac{\mathcal{P}\left[\rpitd\mid\theta\right]\mathcal{P}\left[\theta\mid I\right]}{\mathcal{P}\left[\rpitd\mid I\right]},
\ee
which states the posterior distribution $\mathcal{P}\left[\theta\mid\rpitd,I\right]$, or probability distribution of a set of parameters $\lbrace\theta\rbrace$ to be correct given the data $\rpitd$ and prior information $I$, depends on the product distribution of $\mathcal{P}\left[\rpitd\mid\theta\right]$, the probability distribution of the data $\rpitd$ given the parameters $\lbrace\theta\rbrace$, and $\mathcal{P}\left[\theta\mid I\right]$, the probability distribution of $\lbrace\theta\rbrace$ given the prior information, normalized by the marginal likelihood $\mathcal{P}\left[\rpitd\mid I\right]$. Although the marginal likelihood, which denotes the probability that the data are correct given the prior information $I$, normalizes the posterior distribution, its impact will be neglected herein as it does not depend on the model parameters $\lbrace\theta\rbrace$. The most likely set of parameters is found by maximizing the posterior distribution, or, equivalently, minimizing its negative logarithm:
\be
L^2\equiv-2\log\left(\mathcal{P}\left[\theta\mid\rpitd,I\right]\right).
\ee
The central limit theorem demands the probability distribution $\mathcal{P}\left[\rpitd\mid\theta\right]$ be given by
\be
\mathcal{P}\left[\rpitd\mid\theta\right]\propto{\rm Exp}\left\lbrace-\frac{\chi^2}{2}\right\rbrace={\rm Exp}\left\lbrace-\frac{1}{2}\sum_{i,j}\left(\rpitd_i-\rpitd_i\left[\theta\right]\right)\mathbf{C}^{-1}_{ij}\left(\rpitd_j-\rpitd_j\left[\theta\right]\right)\right\rbrace,
\ee
where $i,j$ denote a fitted matrix element with momentum $p_z$ and Wilson line length $z_3$,
$\rpitd_k\left[\theta\right]$ a prediction of the data given $\lbrace\theta\rbrace$, and $\mathbf{C}$ the data covariance estimated via a jackknife re-sampling of the data.

We turn now to the prior distributions for each model parameter. Recall our choice in this manuscript to parameterize the PDFs via a basis of Jacobi polynomials on the interval $x\in\left[0,1\right]$ requires $\alpha,\beta>-1$. This prior information on the non-linear parameters $p\in\lbrace\alpha,\beta\rbrace$ can be suitably encoded via shifted log-normal distributions:
\be
\mathcal{P}\left[p\mid I\right]=\frac{1}{\left(p-p_0\right)\sigma_p\sqrt{2\pi}}{\rm Exp}\left\lbrace-\frac{\left[\log\left(p-p_0\right)-\mu_p\right]^2}{2\sigma_p^2}\right\rbrace,
\ee
where $\alpha_0,\beta_0=-1$ enforce the requirement for the Jacobi polynomials to form a complete orthogonal set, and the mean and variance of $\log\left(p-p_0\right)$ are set respectively by $\mu_p$ and $\sigma_p$. Each of $\mu_p$ and $\sigma_p$ can be tuned so as to produce a desired mean and variance of the underlying variate $p$.\footnote{To produce a random variate $X$ with mean $\mu_X$ and variance $\sigma_X^2$, the log-normal parameters should be tuned to $\mu_p=\ln\left(\frac{\left(\mu_X-X_0\right)^2}{\sqrt{\left(\mu_X-X_0\right)^2+\sigma_X^2}}\right)$ and $\sigma_p^2=\ln\left(1+\frac{\sigma_X^2}{\left(\mu_X-X_0\right)^2}\right)$.} To guarantee a convergent PDF we make the requirement $\beta>0$, or equivalently $\beta_0=0$. Physical intuition can help set reasonable prior distributions for each of the linear expansion coefficients $C_{\tau,n}^{*,\left(\alpha,\beta\right)}$ parameterizing the leading-twist and nuisance contributions to the reduced pseudo-ITD signal (Eq.~\ref{eq:jacFit-real} and Eq.~\ref{eq:jacFit-imag}).
The only sensible restriction for each $C_{\tau,n}^{*,\left(\alpha,\beta\right)}$ is to require the associated $x$-dependent distributions to be sub-leading relative to the leading-twist helicity PDF; otherwise, each expansion coefficient may assume an arbitrary value. To reflect this knowledge and allow the reduced pseudo-ITD to best dictate the size of each expansion coefficient, we impose normally distributed priors of the form:
\be
\mathcal{P}\left[C_{\tau,n}^{*,\left(\alpha,\beta\right)}\mid I\right]=\frac{1}{\sqrt{2\pi}\sigma_{c_i}}{\rm Exp}\left\lbrace-\frac{\left(c_i-\mu_i\right)^2}{2\sigma_{c_i}^2}\right\rbrace,
\ee
where the prior and width of each coefficient $C_{\tau,n}^{*,\left(\alpha,\beta\right)}$ are denoted by $\mu_i$ and $\sigma_i$. Given the lack of any a priori knowledge on the size of each $C_{\tau,n}^{*,\left(\alpha,\beta\right)}$, we fix $\mu_i=0$ for each $C_{\tau,n}^{*,\left(\alpha,\beta\right)}$ - larger prior widths are given to the expansion coefficients parameterizing the leading-twist component so as to embody the sub-leading effect of the nuisance terms.

All fits to the reduced pseudo-ITD $\mathfrak{Y}\left(\nu,z^2\right)$ are performed with the aid of the Variable Projection (VarPro) algorithm~\cite{Golub}. In this manner, the potentially large space of fit parameters (e.g. $\alpha$, $\beta$ and any expansion coefficients $C^{*,\left(\alpha,\beta\right)}_{\tau,n}$) is reduced to a non-linear optimization within the two-dimensional space defined by the parameters $\alpha$ and $\beta$, which appear non-linearly in our PDF parameterizations of Eq.~\eqref{eq:jacFit-real} and Eq.~\eqref{eq:jacFit-imag}. Each expansion coefficient $C_{\tau,n}^{*,\left(\alpha,\beta\right)}$ appears linearly and is solved for explicitly in terms of the non-linear basis functions $\lbrace\sigma_n^{\left(\alpha,\beta\right)},\sigma_{0,n}^{\left(\alpha,\beta\right)},\eta_n^{\left(\alpha,\beta\right)},\eta_{0,n}^{\left(\alpha,\beta\right)}\rbrace$.

%%%%%%%%%%%%%%%%%%%%%%%%%%
%%%%%%%% PRIORS %%%%%%%%%%
%%%%%%%%%%%%%%%%%%%%%%%%%%
\begin{table}
    \centering
    \begin{tabular}{|c|c|c|c|c|c|c|}
         Parameter & $\mu_\alpha$ & $\sigma_\alpha$ & $\mu_\beta$ & $\sigma_\beta$ & $\sigma_{C_n^{lt\thinspace(\alpha,\beta)}}$ &
         $\sigma_{C_n^{*\thinspace(\alpha,\beta)}}$ \\
         \hline\hline
         {\rm Default} & 0.0 & 0.4 & 3.0 & 1.0 & 0.5 & 0.1 \\
         {\rm Wide} & 0.0 & 0.8 & 3.0 & 2.0 & 1.0 & 0.2 \\
         {\rm Thin} & 0.0 & 0.2 & 3.0 & 0.5 & 0.25 & 0.05 \\
    \end{tabular}
    \caption{Parameters of each Bayesian prior distribution used in the maximum likelihood fits to the reduced pseudo-ITD $\mathfrak{Y}\left(\nu,z^2\right)$ data. {\it Default} corresponds to fits with prior widths based on reasonable physical intuition, while {\it Wide} ({\it Thin}) represent an increase (decrease) of each prior width by a factor of two. The latter two cases are discussed in Appendix~\ref{sec:fitStabilityWithPriors} and offer a window into the sensitivity of our conclusions on the imposed prior distributions. The prior width of each expansion coefficient parameterizing the leading-twist component of $\mathfrak{Y}\left(\nu,z^2\right)$ is denoted by $\sigma_{C_n}^{lt\ \left(\alpha,\beta\right)}$, while $\sigma_{C_n}^{*\ \left(\alpha,\beta\right)}$ denotes the priors widths of the expansion coefficients parameterizing the $\mathcal{O}\left(a/\left|z\right|\right)$, $\mathcal{O}\left(z^2\Lambda_{\rm QCD}^2\right)$, $\mathcal{O}\left(z^4\Lambda_{\rm QCD}^4\right)$ nuisance terms.\label{tab:priors}}
\end{table}

%%%%%%%%%%%%%%%%%%%%%%%%%%%%%%%%%%%%%%%%%%%%%%%%%%%%%%%%%%%%%%%%%%%%%%
%%%%%%%%***********************************************%%%%%%%%%%%%%%%
%%%%%%%%%%%%%%%%%%%%%%%%%%%%%%%%%%%%%%%%%%%%%%%%%%%%%%%%%%%%%%%%%%%%%%
%%%%%%%%%%%%%%%%%%%%%%%%%%% RESULTS %%%%%%%%%%%%%%%%%%%%%%%%%%%%%%%%%%
%%%%%%%%%%%%%%%%%%%%%%%%%%%%%%%%%%%%%%%%%%%%%%%%%%%%%%%%%%%%%%%%%%%%%%
%%%%%%%%***********************************************%%%%%%%%%%%%%%%
%%%%%%%%%%%%%%%%%%%%%%%%%%%%%%%%%%%%%%%%%%%%%%%%%%%%%%%%%%%%%%%%%%%%%%
\section{Results\label{sec:results}}
\begin{figure}[t]
    \centering
    \includegraphics[width=0.9\linewidth]{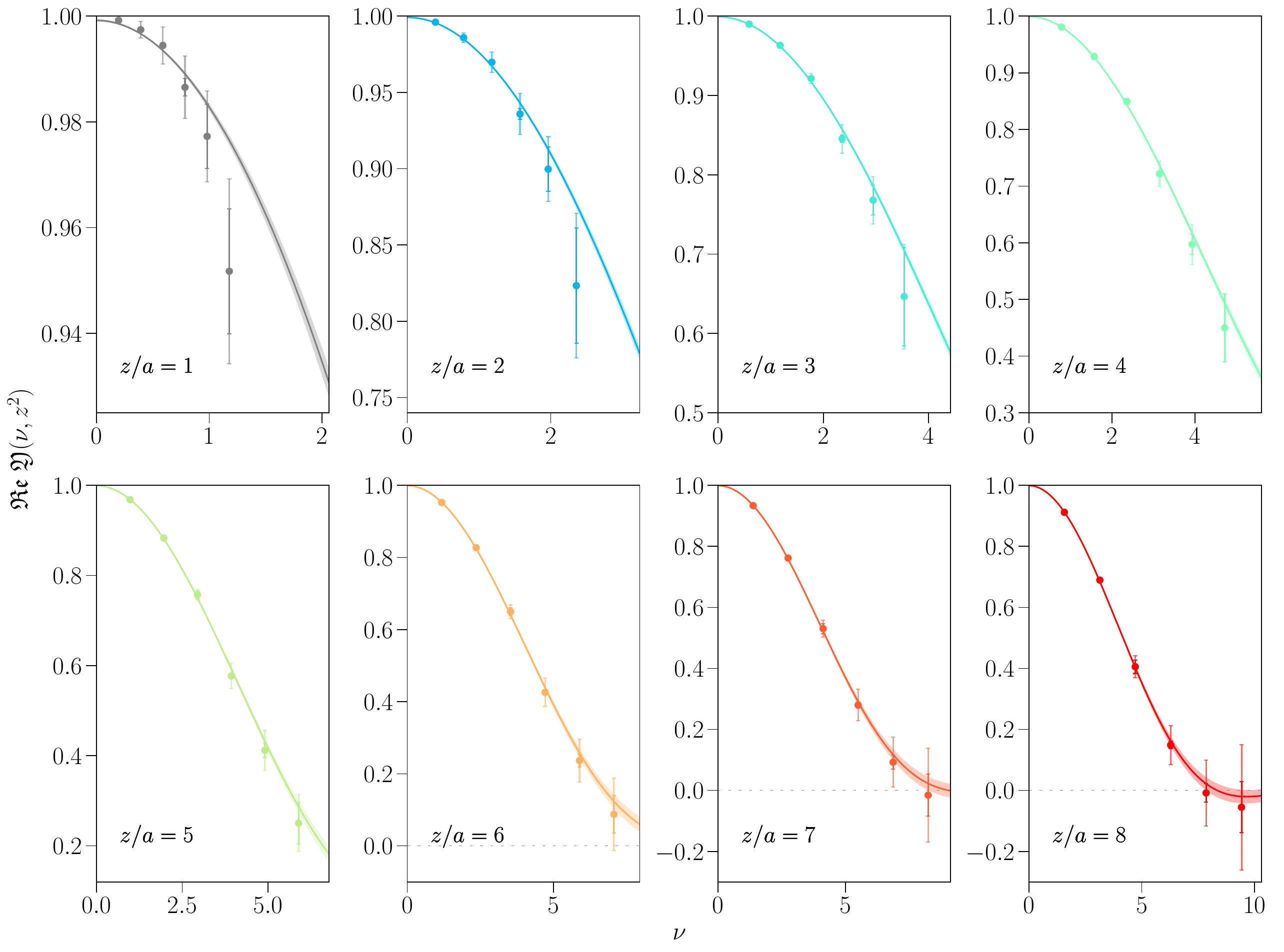}
    \caption{Fit to the real component of the reduced pseudo-ITD $\mathfrak{Y}\left(\nu,z^2\right)$ obtained from summed ratio fits over the time series $T/a\in\left[6,14\right]$ (dark error bars), and where the $T/a\in\left[4,14\right]$ summed ratio fits provide a systematic error estimate (lightened error bars). The leading-twist, discretization, twist-4, and twist-6 corrections have been expanded in Jacobi polynomials up to orders $\left(N_{lt},N_{az},N_{t4},N_{t6}\right)=\left(3,2,2,1\right)$. The data has been cut on $p_{\rm latt}\in\left[1,6\right]$ and $z/a\in\left[2,8\right]$, with data excluded from the fit presented in gray.}
    \label{fig:realPITD_selectFit_p1-6_z2-8_withMatSys_defPriors}
\end{figure}

\begin{figure}[t]
    \centering
    \includegraphics[width=0.49\linewidth]{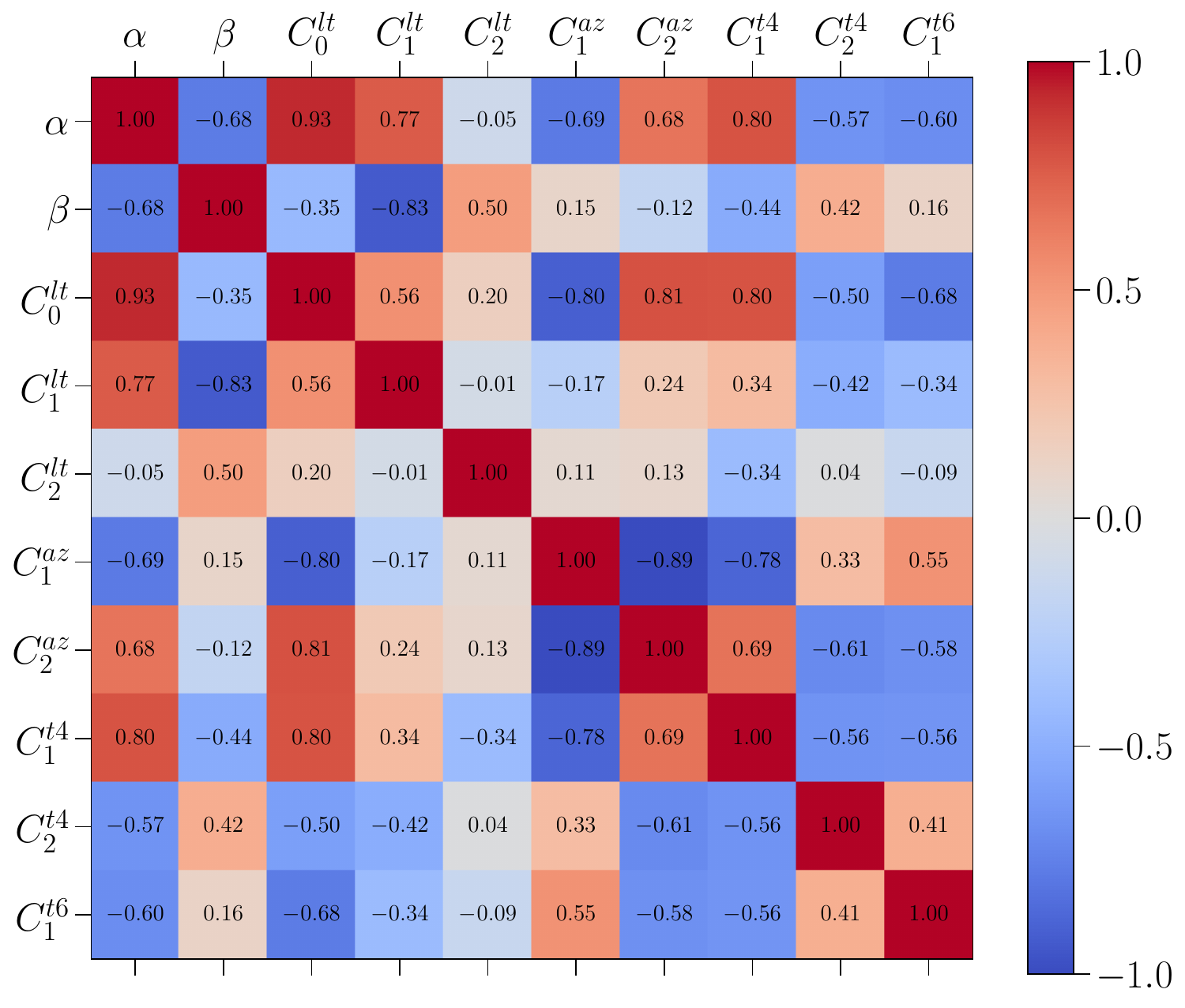}
    \hfill
    \includegraphics[width=0.49\linewidth]{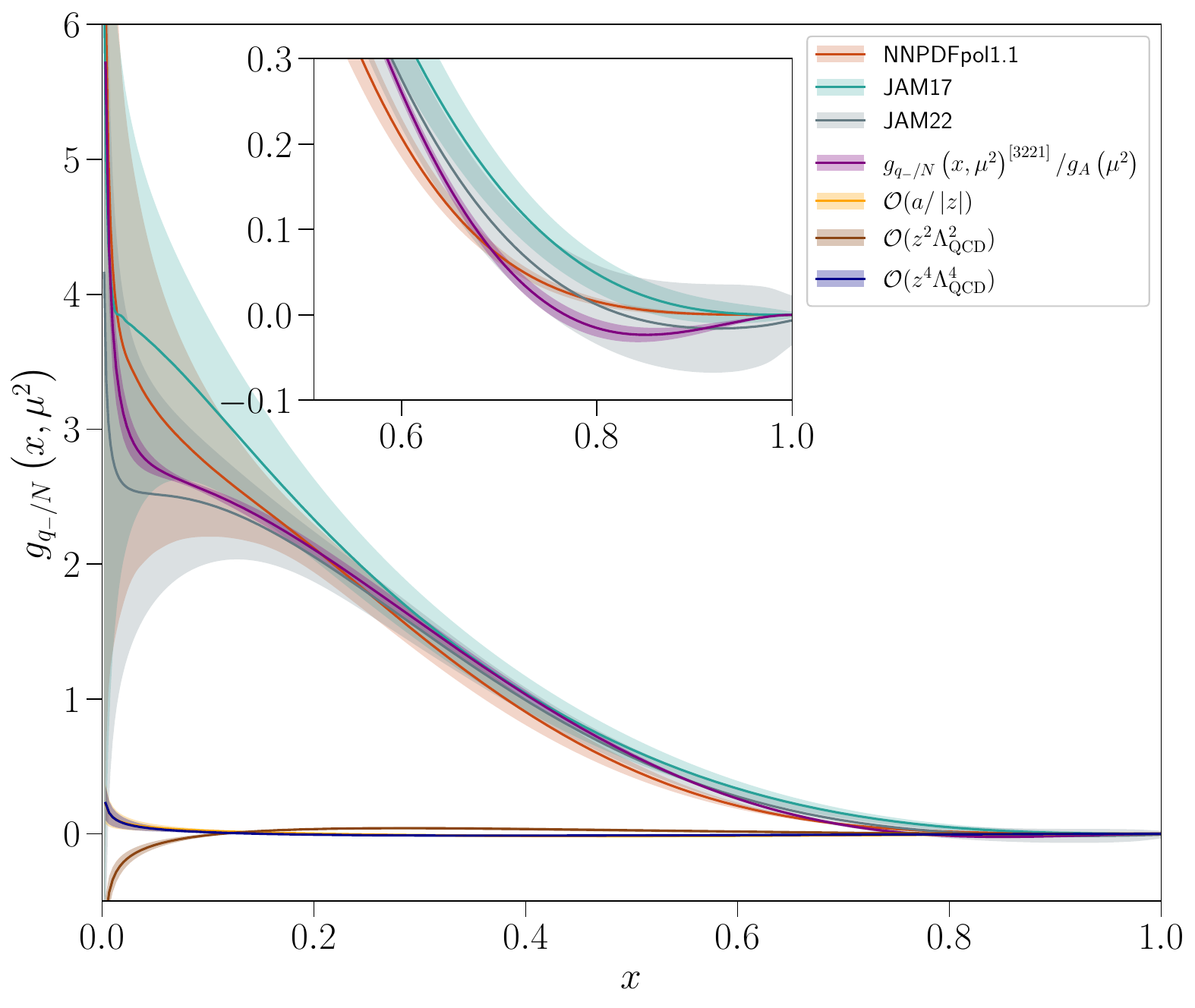}
    \caption{(Left) Parameter correlations obtained from ${\rm Cov}_{ij}/\sqrt{{\rm Cov}_{ii}{\rm Cov}_{jj}}$ of the fit presented in Fig.~\ref{fig:realPITD_selectFit_p1-6_z2-8_withMatSys_defPriors}. (Right) Derived leading-twist valence helicity quark PDF (purple) and $x$-space contaminations compared with the recent global analyses {\tt NNPDFpol1.1}~\cite{Nocera:2014gqa}, {\tt JAM17}~\cite{Ethier:2017zbq}, and {\tt JAM22}~\cite{Cocuzza:2022jye}.}
    \label{fig:paramCovAndPDFQv_selectFit_p1-6_z2-8_withMatSys_defPriors}
\end{figure}

\begin{figure}
    \centering
    \includegraphics[width=0.42\linewidth]{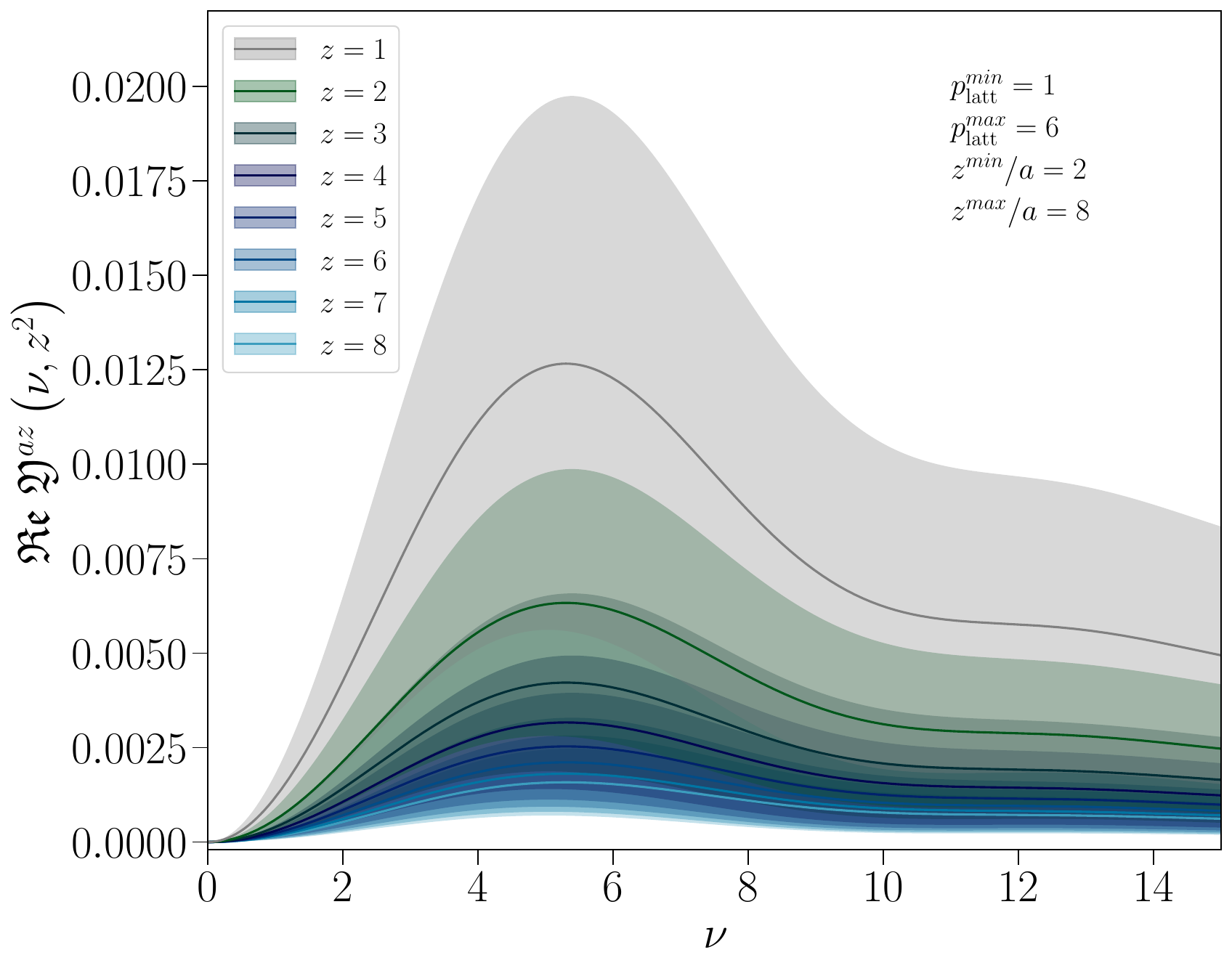}
    \includegraphics[width=0.42\linewidth]{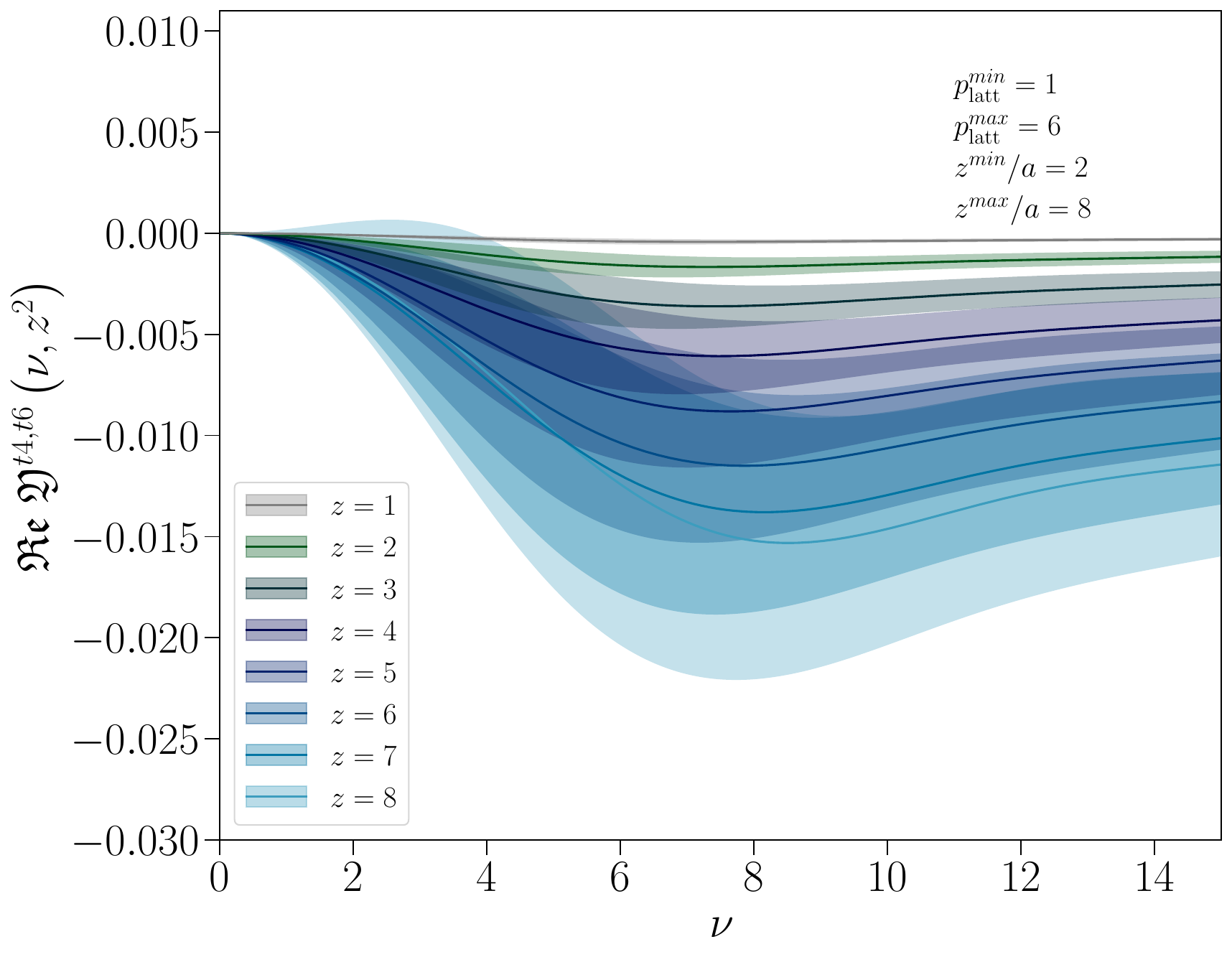} \\
    \includegraphics[width=0.42\linewidth]{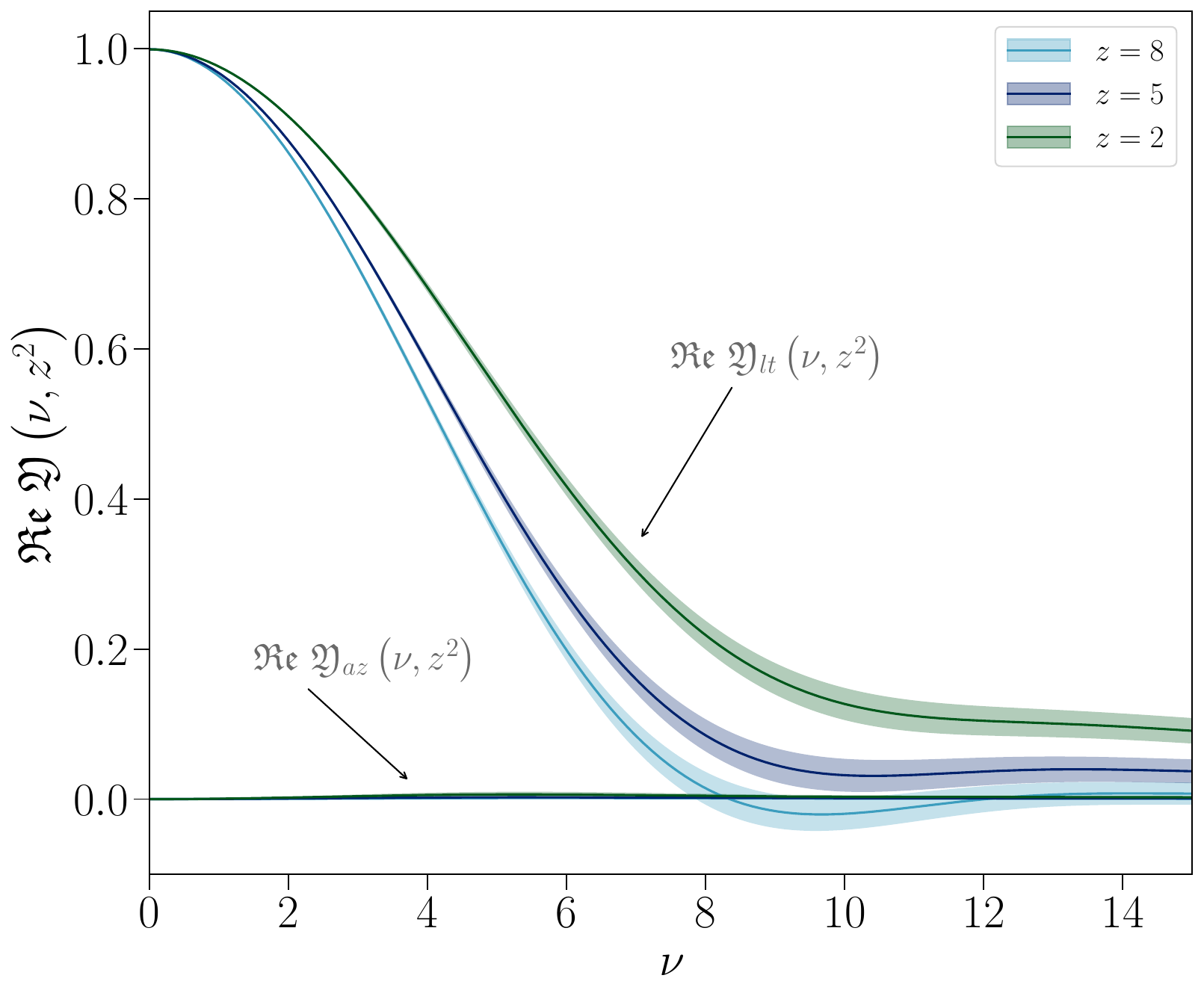}
    \includegraphics[width=0.42\linewidth]{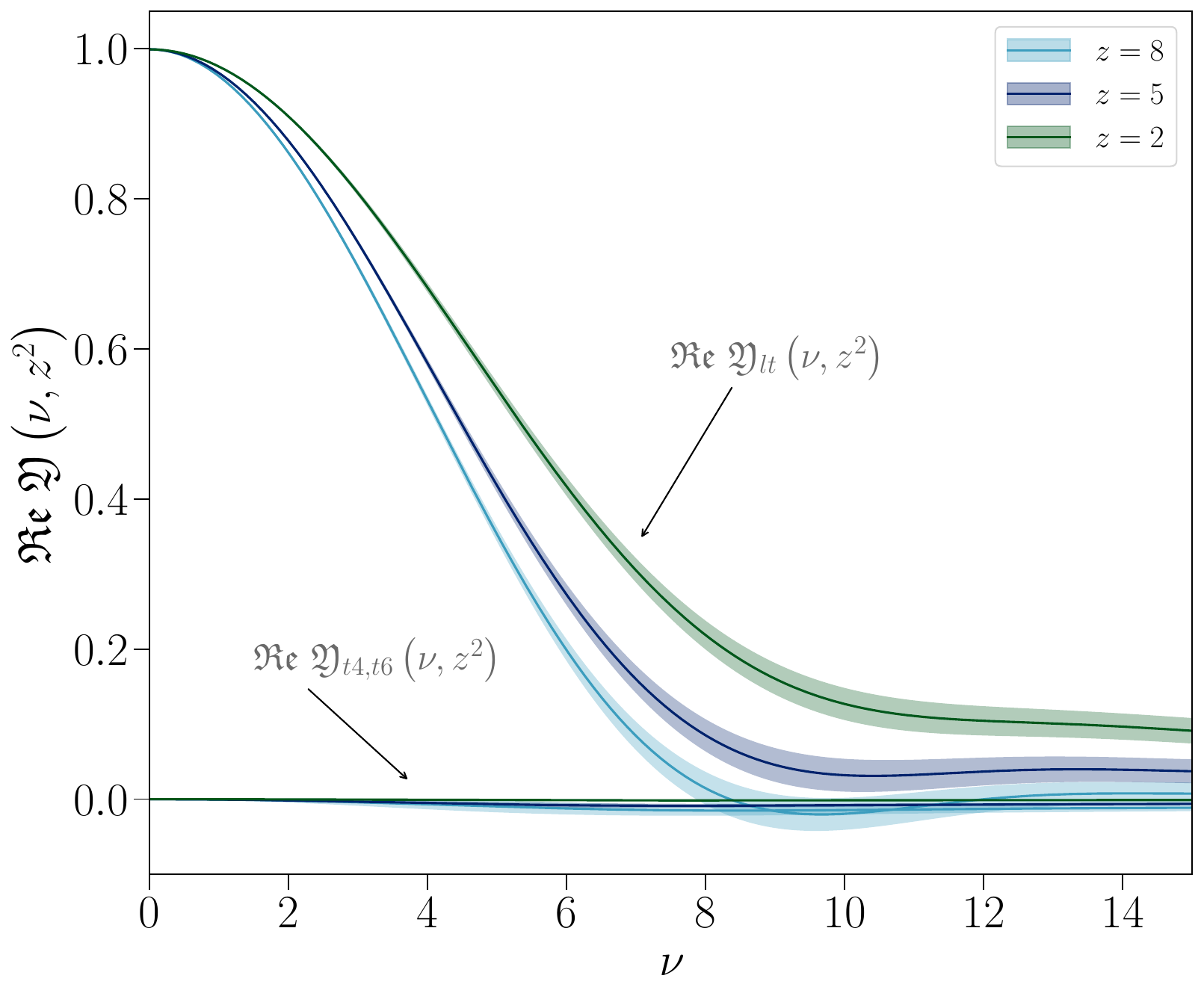}
    \caption{(Upper left) Discretization and (Upper right) net higher-twist nuisance terms determined from the fit in Fig.~\ref{fig:realPITD_selectFit_p1-6_z2-8_withMatSys_defPriors} shown for $z/a\in\left[1,8\right]$. The lower two panels compare each nuisance effect (discretization at the left and higher-twist at the right) on an absolute scale with the parameterized leading-twist reduced pseudo-ITD $\mathfrak{Re}\ \mathfrak{Y}\left(\nu,z^2\right)$. Values of $z/a$ excluded from the fit are presented in gray.}
    \label{fig:realPITD_selectFit_p1-6_z2-8_withMatSys_defPriors_nuisanceViz}
\end{figure}

We begin this section by reporting fit results to the reduced pseudo-ITD $\mathfrak{Y}\left(\nu,z^2\right)$ that includes our matrix element fitting systematic~\eqref{eq:matelem_sys}, where $\mathfrak{Y}\left(\nu,z^2\right)$ has been cut according to $p_{\rm latt}\in\left[1,6\right]$ and $z/a\in\left[2,8\right]$. In Figure~\ref{fig:realPITD_selectFit_p1-6_z2-8_withMatSys_defPriors} the functional form Eq.~\ref{eq:jacFit-real} is fit to $\mathfrak{Re}\ \mathfrak{Y}\left(\nu,z^2\right)$, subject to the aforementioned cuts, where the basis of Jacobi polynomials describing the leading-twist and nuisance terms have been truncated at orders $\left(N_{lt},N_{az},N_{t4},N_{t6}\right)=\left(3,2,2,1\right)$. It is clear all data for a given $z/a\in\left[2,8\right]$ are well described by the chosen functional form, though for $p_{\rm latt}=3$ (middle point of each panel in Fig.~\ref{fig:realPITD_selectFit_p1-6_z2-8_withMatSys_defPriors}) a slight tension of at most 1-$\sigma$ is observed for modest values of $z/a$. Most notable is the presence of the highly precise $p_{\rm latt}=1$ and 2 data for each $z/a\in\left[2,8\right]$. Both are reflections of reasonably precise and extended plateaus in the two-point effective energies (cf. Fig.~\ref{fig:effEnergies}) at $T/a=6$. At fixed cost, the signal-to-noise ratio of lattice correlators decays exponentially with the hadronic energy making these points critical for an efficient high precision analysis. These two data points for each $z/a$ are precise enough to heavily constrain the fits for all jackknife bins, thereby explaining why the resultant fit exhibits rather small statistical fluctuations. Nonetheless, despite removing the entirety of the $z/a=1$ data, it is encouraging to observe the functional dependence on $\nu$ is reproduced reasonably well for the $z/a=1$ data. The resulting parameters corresponding to the fit in Fig.~\ref{fig:realPITD_selectFit_p1-6_z2-8_withMatSys_defPriors} as well as the $L^2$ and $\chi^2$ per degree of freedom are gathered in Tab.~\ref{tab:fitParams_selectFits_p1-6_z2-8_withMatSys_defPriors}. In examining the fit parameter correlations in Fig.~\ref{fig:paramCovAndPDFQv_selectFit_p1-6_z2-8_withMatSys_defPriors}, the leading-twist expansion coefficients $C_{-,n}^{lt}$ appear to strongly correlate with themselves, $\alpha$ and $\beta$. Modest correlation is also observed between $\alpha$, $C^{lt}_{-,0}$, and the systematic error terms. The correlation of the systematic error terms with $C^{lt}_{-,0}$ is not surprising, as $C^{lt}_{-,0}$, though treated as a fitted parameter, fixes the valence quark sum rule and hence is sensitive to overall changes in the leading-twist PDF that arise from variations in the complete functional description of the reduced pseudo-ITD. Otherwise, the systematic errors terms do not exhibit strong correlation with $\beta$ or the remaining leading twist terms. This feature suggests that the systematic error terms are gathering distinct pieces of information that the leading twist terms are unable to capture, and are not generating a large bias. There appears to be a strong anti-correlation between the $C^{az}_{-,n}$ parameters. This feature may imply a cancellation occurring between the two terms leading to the cumulative small effect seen in Fig.~\ref{fig:realPITD_selectFit_p1-6_z2-8_withMatSys_defPriors_nuisanceViz}.

%%%%%%%%%%%%%%%%%%%%%%%%%%%%%%%%%%%%%%%%%%%%%%%%%%%%%%%%%%%%%%%%%%%%%%%%%%%%%%
% REPRESENTATIVE FITS - HIGHEST AICc WEIGHT 
%     --> DEFAULT PRIORS  &&  WITH MATRIX ELEMENT FITTING SYSTEMATIC
%     --> CUTS P\IN[1,6] Z\IN[2,8]
%%%%%%%%%%%%%%%%%%%%%%%%%%%%%%%%%%%%%%%%%%%%%%%%%%%%%%%%%%%%%%%%%%%%%%%%%%%%%%
% % REPRESENTATIVE FIT - REAL RPITD  (3,2,2,1)
%%%%%%%%%%%%%%%%%%%%%%%%%%%%%%%%%%%%%%
% TABLE OF REPRESENTATIVE QV/Q+ FITS
% REPRESENTATIVE FITS - HIGHEST AICc WEIGHT 
%     --> DEFAULT PRIORS  &&  WITH MATRIX ELEMENT FITTING SYSTEMATIC
%     --> CUTS P\IN[1,6] Z\IN[2,8]
%%%%%%%%%%%%%%%%%%%%%%%%%%%%%%%%%%%%%%
\begin{table}[b]
\scriptsize
    \centering
    \begin{tabular}{c|c|c|c|c|c|c|c|c|c|c|c|c|c}
         & $\alpha$ & $\beta$ & $C^{lt}_0$ & $C^{lt}_1$ & $C^{lt}_2$ & $C^{az}_0$ & $C^{az}_1$ & $C^{az}_2$ & $C^{t4}_0$ & $C^{t4}_1$ & $C^{t4}_2$ & $C^{t6}_0$ & $C^{t6}_1$ \\ 
        \hline
        % p\in[1,6] z\in[2,8] WITH MATRIX ELEMENT FITTING SYSTEMATIC
        $q_-$ & $-0.500(11)$ & $1.892(70)$ & $0.918(24)$ & $-0.547(51)$ & $-0.902(78)$ & $-$ & $0.037(20)$ & $-0.015(5)$ & $-$ & $-0.055(22)$ & $-0.030(11)$ & $-$ & $0.027(14)$ \\
        $q_+$ & $-0.547(11)$ & $1.501(63)$ & $0.747(32)$ & $-0.305(47)$ & $-0.762(96)$ & $0.188(5)$ & $-0.094(6)$ & $-$ & $0.014(13)$ & $0.024(11)$ & $-$ & $0.019(9)$ & $-$ \\
        % $q_-$ & $-0.493(12)$ & $1.869(117)$ & $0.934(19)$ & $-0.486(93)$ & $-0.971(99)$ & $-$ & $0.002(18)$ & $-0.007(5)$ & $-$ & $-0.005(24)$ & $-0.038(14)$ & $-$ & $-0.012(21)$ \\
        % $q_+$ & $-0.546(14)$ & $1.639(86)$ & $0.805(40)$ & $-0.309(47)$ & $-0.704(113)$ & $0.163(5)$ & $-0.081(7)$ & $-$ & $-0.042(11)$ & $0.018(14)$ & $-$ & $0.034(14)$ & $-$ \\
    \end{tabular}
    \caption{Fit parameters associated with the representative fits to $\mathfrak{Re}\ \mathfrak{Y}\left(\nu,z^2\right)$ and $\mathfrak{Im}\ \mathfrak{Y}\left(\nu,z^2\right)$ shown in Fig.~\ref{fig:realPITD_selectFit_p1-6_z2-8_withMatSys_defPriors} and Fig.~\ref{fig:imagPITD_selectFit_p1-6_z2-8_withMatSys_defPriors}, respectively. The figures of merit for the fit to $\mathfrak{Re}\ \mathfrak{Y}\left(\nu,z^2\right)$ are $L^2/{\rm d.o.f.}=0.265(131)$ and $\chi^2/{\rm d.o.f}=0.280(126)$, while $L^2/{\rm d.o.f.}=0.756(241)$ and $\chi^2/{\rm d.o.f}=0.659(233)$ for $\mathfrak{Im}\ \mathfrak{Y}\left(\nu,z^2\right)$.
    }
    \label{tab:fitParams_selectFits_p1-6_z2-8_withMatSys_defPriors}
\end{table}

%%%%%%%%%%%%%%%%%%%%%%%%%%%%%%%%%%%%%%%%%%%%%%%%%%%%%%%%%%%%%%%%%%%%%%%
% REPRESENTATIVE Qv FIT - HIGHEST AICc WEIGHT 
%     --> DEFAULT PRIORS  &&  WITH MATRIX ELEMENT FITTING SYSTEMATIC
%     --> CUTS P\IN[1,6] Z\IN[2,8]
%%%%%%%%%%%%%%%%%%%%%%%%%%%%%%%%%%%%%%%%%%%%%%%%%%%%%%%%%%%%%%%%%%%%%%%
Using the fitted values of $\alpha$, $\beta$ and each expansion coefficient $C_{-,n}^{*\ \left(\alpha,\beta\right)}$, we utilize Eq.~\ref{eq:pdfViaJacobiPolys} to map the leading-twist valence quark helicity PDF $g_{q_-/N}\left(x\right)$ and the parameterized $x$-space systematic contaminations in the right panel of Fig.~\ref{fig:paramCovAndPDFQv_selectFit_p1-6_z2-8_withMatSys_defPriors}. The parameterized $g_{q_-/N}\left(x\right)$ exhibits broad statistical consistency with the three global analysis results we consider: {\tt NNPDFpol1.1}~\cite{Nocera:2014gqa}, {\tt JAM17}~\cite{Ethier:2017zbq}, and {\tt JAM22}~\cite{Cocuzza:2022jye}, while for $x\rightarrow1$ the soft approach of the PDF appears to favor the {\tt NNPDFpol1.1} and {\tt JAM22} results. This result, however, represents only one possible solution for $g_{q_-/N}\left(x\right)$ within the space of viable solutions, and thus exhibits an uncertainty that belies the true uncertainty of the PDF. We address this quantitatively in Sec.~\ref{sec:aic} in the context of a model averaging prescription.

Indeed the $x$-space systematic contaminations illustrated in the right panel of Fig.~\ref{fig:paramCovAndPDFQv_selectFit_p1-6_z2-8_withMatSys_defPriors} appear quite small. However, it is more instructive to view the parameterized discretization and higher-twist effects as a function of the two Lorentz invariants of the setup - $\nu$ and $z^2$.
%%%%%%%%%%%%%%%%%%%%%%%%%%%%%%%%%%%%%%%%%%%%%%%%%%%%%%%%%%%%%%%%%%%%%%%
% NUISANCE EFFECTS FROM REPRESENTATIVE Qv FIT - HIGHEST AICc WEIGHT 
%     --> DEFAULT PRIORS  &&  WITH MATRIX ELEMENT FITTING SYSTEMATIC
%     --> CUTS P\IN[1,6] Z\IN[2,8]
%%%%%%%%%%%%%%%%%%%%%%%%%%%%%%%%%%%%%%%%%%%%%%%%%%%%%%%%%%%%%%%%%%%%%%%
In the upper left and upper right panels of Fig.~\ref{fig:realPITD_selectFit_p1-6_z2-8_withMatSys_defPriors_nuisanceViz}, respectively, we visualize the discretization and higher-twist nuisance effects isolated via the fit presented in Fig.~\ref{fig:realPITD_selectFit_p1-6_z2-8_withMatSys_defPriors}. The discretization effect, denoted $\mathfrak{Re}\ \mathfrak{Y}^{az}\left(\nu,z^2\right)$, is seen to be, maximally, on the order of $1-2\%$ for the shortest values of $z/a$ included in the fit - namely, $z/a=2,3$, with the strongest effect for $\nu\sim5-6$. But given that the $z/a=2,3$ data extend up to Ioffe-times of $\nu\sim2-3.5$ for the momenta at our disposal, the fit would suggest a short-distance discretization effect at the sub-percent level (cf. lower left panel of Fig.~\ref{fig:realPITD_selectFit_p1-6_z2-8_withMatSys_defPriors_nuisanceViz}). We do note that despite the apparent smallness of this effect, it is an important one to control and remove, as the effect, if left unaccounted for, would impart a statistically significant shift in the precise low-$p$ data. The net higher-twist effect on the other hand, denoted $\mathfrak{Re}\ \mathfrak{Y}^{t4,t6}\left(\nu,z^2\right)$, shown in the upper right panel of Fig.~\ref{fig:realPITD_selectFit_p1-6_z2-8_withMatSys_defPriors_nuisanceViz} is observed to be small but non-zero across all Ioffe-times for all $2\leq z/a<8$; it is only for $\nu\gtrsim4$ that a non-trivial higher-twist effect is detected by the fit for $z/a\gtrsim2$. By design the ratio in Eq.~\eqref{eq:subopt-reduced} will cancel the leading higher twist contributions in the small $\nu$ limit, so it is reassuring to find the parameterized power corrections in $z^2$ to be numerically small over a broad range of Ioffe-time. However, unlike the discretization effect, the parameterized leading-twist and power correction signals become comparable in precisely the interval for which the $\mathcal{O}\left(z^{2n}\Lambda_{\rm QCD}^{2n}\right)$ nuisance effects are largest. As illustrated in the lower right panel of Fig.~\ref{fig:realPITD_selectFit_p1-6_z2-8_withMatSys_defPriors_nuisanceViz}, for $\nu\gtrsim8$ the leading-twist and $\mathcal{O}\left(z^{2n}\Lambda_{\rm QCD}^{2n}\right)$ effects are indeed of similar magnitude. Since the maximal reach in Ioffe-time of this calculation is for $\nu_{\rm max}\simeq9.42$, we can be assured the $\mathfrak{Re}\ \mathfrak{Y}\left(\nu,z^2\right)$ signal is dominated by the leading-twist contribution we aim to isolate, with the power corrections a relatively small effect that we parameterize and remove. That said, just as different experimental processes are subject to different power corrections when analyzed in a factorization framework, it is of crucial importance to quantify where the power corrections of the reduced pseudo-ITD become appreciable. Despite the na\"ive ab-initio expectations from the size of the scale $z^2$, empirically our data is consistent with the NLO evolution formula with small and effectively zero power corrections.

In the statistical errors of this single fit, a feature common to many previous PDF analyses can be seen. The statistical errors shrink around $x\sim0.1$. The low $x$ region is where the inverse problem is unreliable according to mock data studies~\cite{Karpie:2019eiq}. The individual jackknife samples will have an upward (downward) fluctuation for $x$ above this point and a corresponding downward (upward) fluctuation after this point creating the apparent statistical precision around $x\sim 0.1$. This feature, is created by correlations between the parameters to satisfy the very precise constraints of the data at low Ioffe time. In other words, the precise low $\nu$ data puts a strong constraint on the value of the lowest moment of the PDF. For this model of the PDF to enforce that constraint while fixing the well-controlled large $x$ region, the PDF must have corresponding upward and downward fluctuations above and below $x\sim0.1$. The location of this pinched point will be model dependent. In the subsequent model averaging procedure of Sec.~\ref{sec:aic}, these model dependent features will be seen to average away. This demonstrates the importance of studying many solutions to the inverse problem simultaneously.

%%%%%%%%%%%%%%%%%%%%%%%%%%%%%%%%%%%%%%%%%%%%%%%%%%%%%%%%%%%%%%%%%%%%%%%%%%%%%%
% REPRESENTATIVE FITS - HIGHEST AICc WEIGHT 
%     --> DEFAULT PRIORS  &&  WITH MATRIX ELEMENT FITTING SYSTEMATIC
%     --> CUTS P\IN[1,6] Z\IN[2,8]
%%%%%%%%%%%%%%%%%%%%%%%%%%%%%%%%%%%%%%%%%%%%%%%%%%%%%%%%%%%%%%%%%%%%%%%%%%%%%%
% % REPRESENTATIVE FIT - IMAG RPITD  (3,2,2,1)
\begin{figure}
    \centering
    \includegraphics[width=0.9\linewidth]{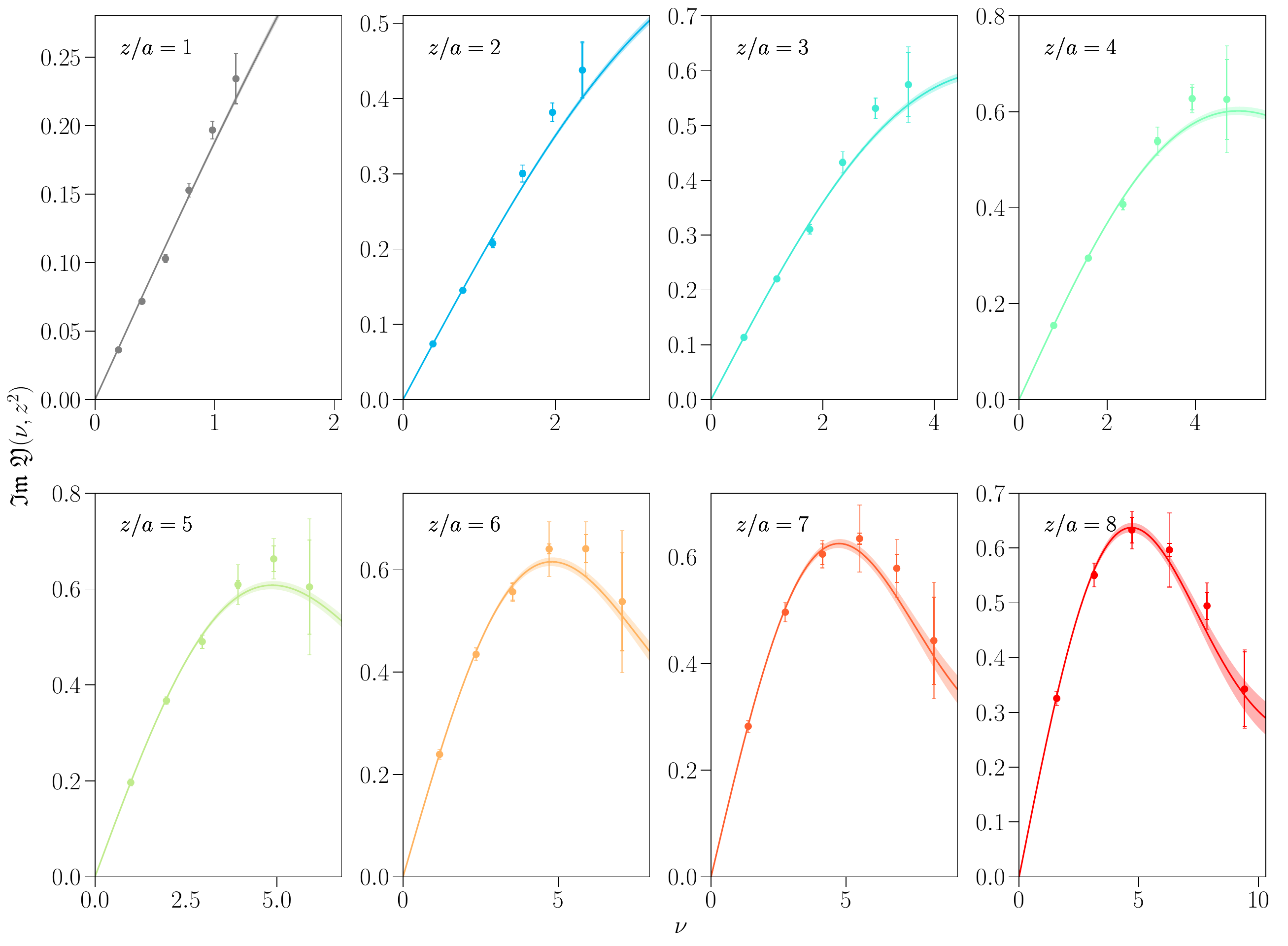}
    \caption{Fit to the imaginary component of the reduced pseudo-ITD $\mathfrak{Y}\left(\nu,z^2\right)$ obtained from summed ratio fits over the time series $T/a\in\left[6,14\right]$ (dark error bars), and where the $T/a\in\left[4,14\right]$ summed ratio fits provide a systematic error estimate (lightened error bars). The leading-twist, discretization, twist-4, and twist-6 corrections have been expanded in Jacobi polynomials up to orders $\left(N_{lt},N_{az},N_{t4},N_{t6}\right)=\left(3,2,2,1\right)$. The data has been cut on
    $p_{\rm latt}\in\left[1,6\right]$ and $z/a\in\left[2,8\right]$, with data excluded from the fit presented in gray.}
    \label{fig:imagPITD_selectFit_p1-6_z2-8_withMatSys_defPriors}
\end{figure}
\begin{figure}
    \centering
    \includegraphics[width=0.49\linewidth]{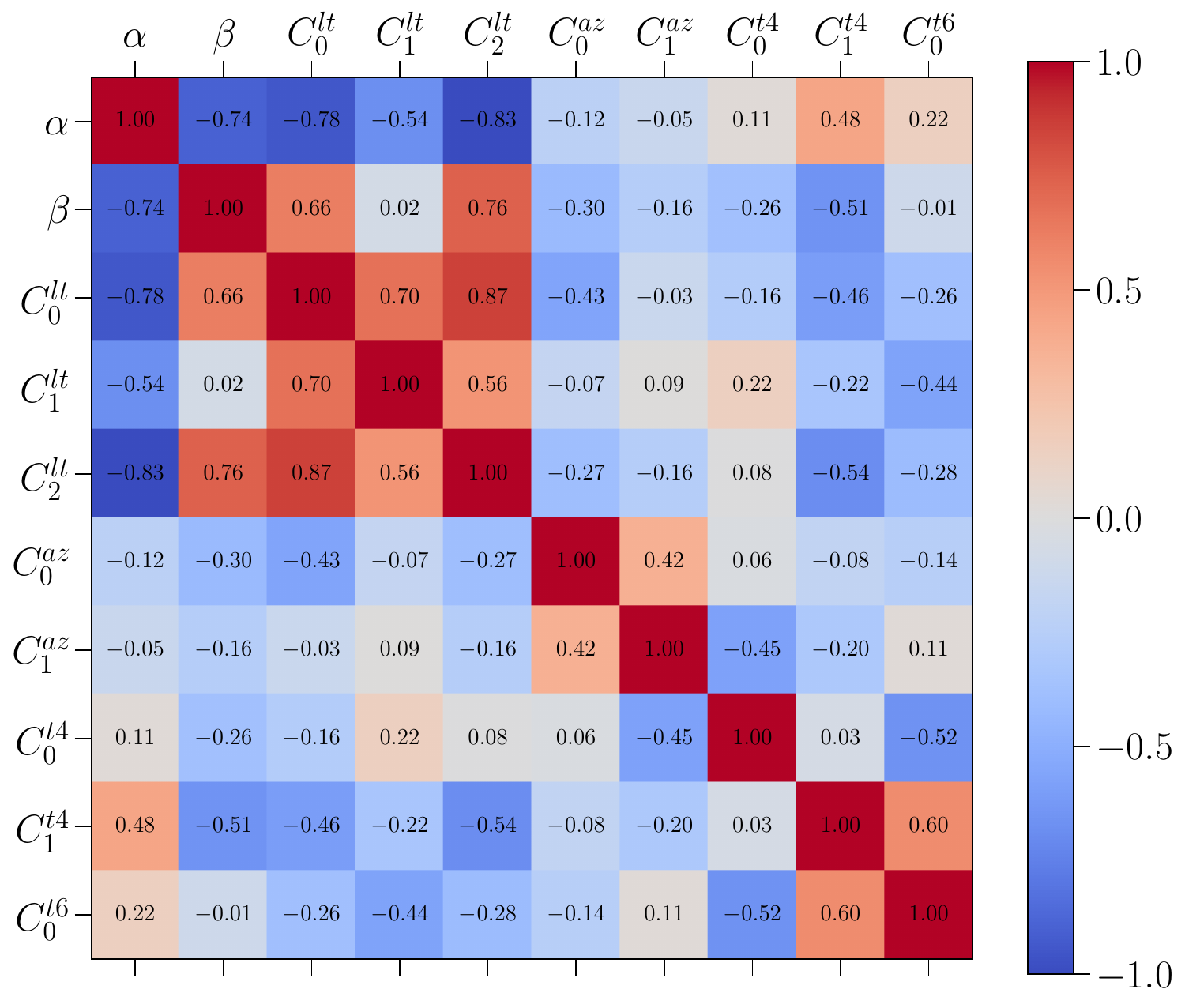}
    \hfill
    \includegraphics[width=0.49\linewidth]{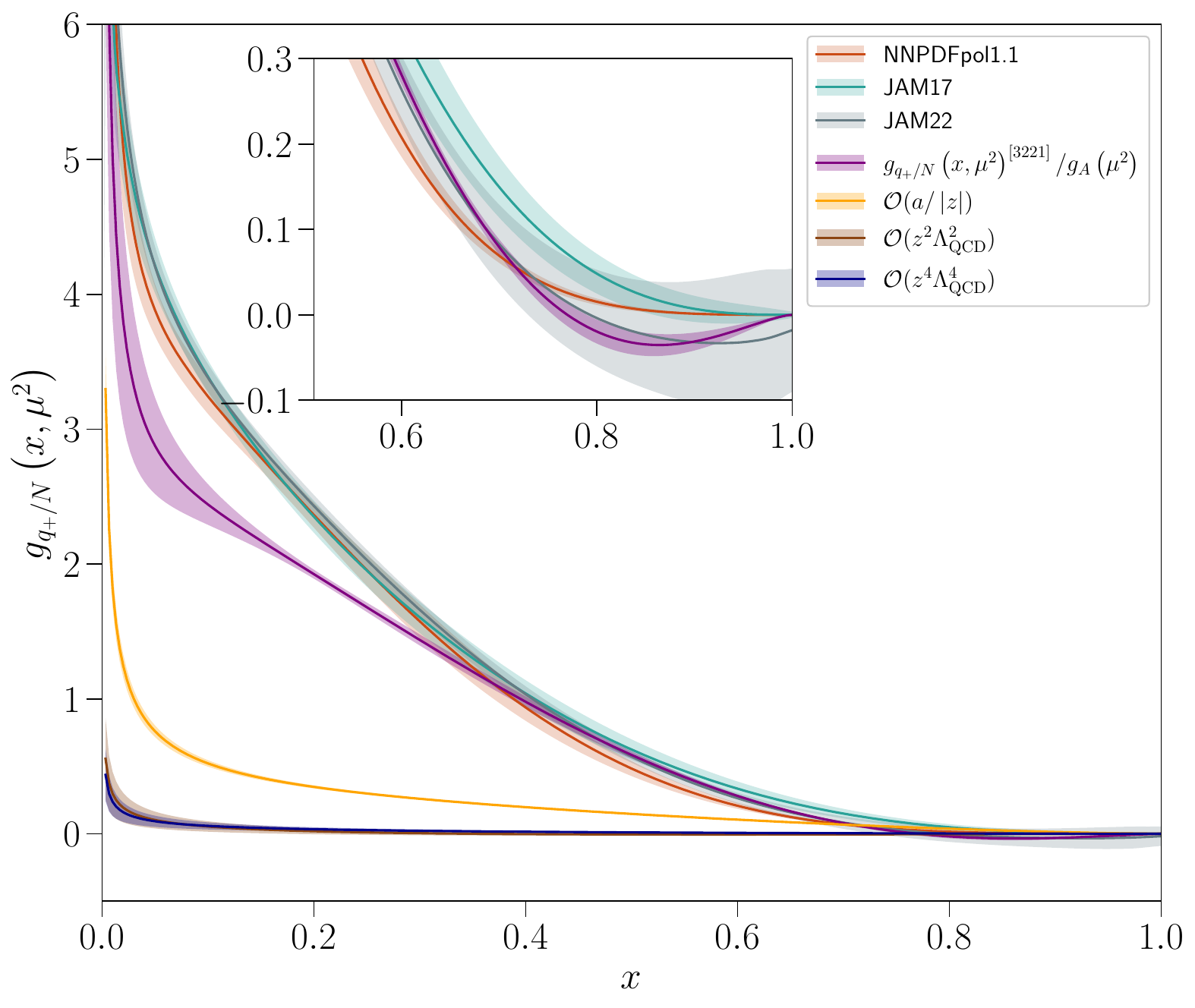}
    \caption{(Left) Parameter correlations obtained from ${\rm Cov}_{ij}/\sqrt{{\rm Cov}_{ii}{\rm Cov}_{jj}}$ of the fit presented in Fig.~\ref{fig:imagPITD_selectFit_p1-6_z2-8_withMatSys_defPriors}. (Right) Derived leading-twist $CP$-odd helicity quark PDF (purple) and $x$-space contaminations compared with the recent global analyses {\tt NNPDFpol1.1}~\cite{Nocera:2014gqa}, {\tt JAM17}~\cite{Ethier:2017zbq}, and {\tt JAM22}~\cite{Cocuzza:2022jye}.}
    \label{fig:paramCovAndPDFQ+_selectFit_p1-6_z2-8_withMatSys_defPriors}
\end{figure}

 In Fig.~\ref{fig:imagPITD_selectFit_p1-6_z2-8_withMatSys_defPriors} a representative fit to $\mathfrak{Im}\ \mathfrak{Y}\left(\nu,z^2\right)$ is illustrated, where the data have again been cut on $p_{\rm latt}\in\left[1,6\right]$ and $z/a\in\left[2,8\right]$, and the basis of Jacobi polynomials describing the leading-twist and nuisance terms have been truncated at orders $\left(N_{lt},N_{az},N_{t4},N_{t6}\right)=\left(3,2,2,1\right)$. As in the fit to $\mathfrak{Re}\ \mathfrak{Y}\left(\nu,z^2\right)$, the most precise $p_{\rm latt}=1$ data for each $z/a\in\left[2,8\right]$ acts as the principal constraint for the candidate model, thereby limiting its statistical fluctuations estimated via jackknife. For this model, there is a tension with the $p_{\rm latt}=3,5$ data points for most $z/a$, where the former is the highest momenta we consider without the use of phasing.  Given that this is only a single model within a large space of models that regularize the inverse problem we face, it is not unreasonable to expect this tension to soften following a model averaging prescription. The parameters of this fit to $\mathfrak{Im}\ \mathfrak{Y}\left(\nu,z^2\right)$ are given in Tab.~\ref{tab:fitParams_selectFits_p1-6_z2-8_withMatSys_defPriors}. Turning to the fit parameter covariance of this fit, shown in Fig.~\ref{fig:paramCovAndPDFQ+_selectFit_p1-6_z2-8_withMatSys_defPriors}, similarly to the real component, there is a strong correlation amongst the leading twist parameters and a weaker correlation between the leading-twist and the systematic correction parameters - albeit the correlations between the leading-twist and systematic correction terms are smaller for $\mathfrak{Im}\ \mathfrak{Y}\left(\nu,z^2\right)$ than for $\mathfrak{Re}\ \mathfrak{Y}\left(\nu,z^2\right)$.

%%%%%%%%%%%%%%%%%%%%%%%%%%%%%%%%%%%%%%%%%%%%%%%%%%%%%%%%%%%%%%%%%%%%%%%
% REPRESENTATIVE Q+ FIT - HIGHEST AICc WEIGHT 
%     --> DEFAULT PRIORS  &&  WITH MATRIX ELEMENT FITTING SYSTEMATIC
%     --> CUTS P\IN[1,6] Z\IN[2,8]
%%%%%%%%%%%%%%%%%%%%%%%%%%%%%%%%%%%%%%%%%%%%%%%%%%%%%%%%%%%%%%%%%%%%%%%

 The right panel of Fig.~\ref{fig:paramCovAndPDFQ+_selectFit_p1-6_z2-8_withMatSys_defPriors} shows the modeled leading-twist $g_{q_+/N}\left(x\right)$ PDF and its systematic contaminating counterparts in $x$-space. Encouraging alignment is seen with the {\tt NNPDFpol1.1}~\cite{Nocera:2014gqa} and {\tt JAM22}~\cite{Cocuzza:2022jye} global analyses for $x\gtrsim0.5$, with deviations apparent for smaller values of $x$, where model biases from the limited range of $\nu$ are known to be largest. As for the $g_{q_-/N}\left(x\right)$ determination above, this particular model does not incorporate the systematic fluctuations induced by the space of viable models capable of describing the $\mathfrak{Im}\ \mathfrak{Y}\left(\nu,z^2\right)$ data and variable prior widths for each fit parameter. The latter is explored in Appendix~\ref{sec:fitStabilityWithPriors}, wherein the PDFs are found to be stable as greater flexibility of the prior distributions is allowed.

\begin{figure}
    \centering
    \includegraphics[width=0.42\linewidth]{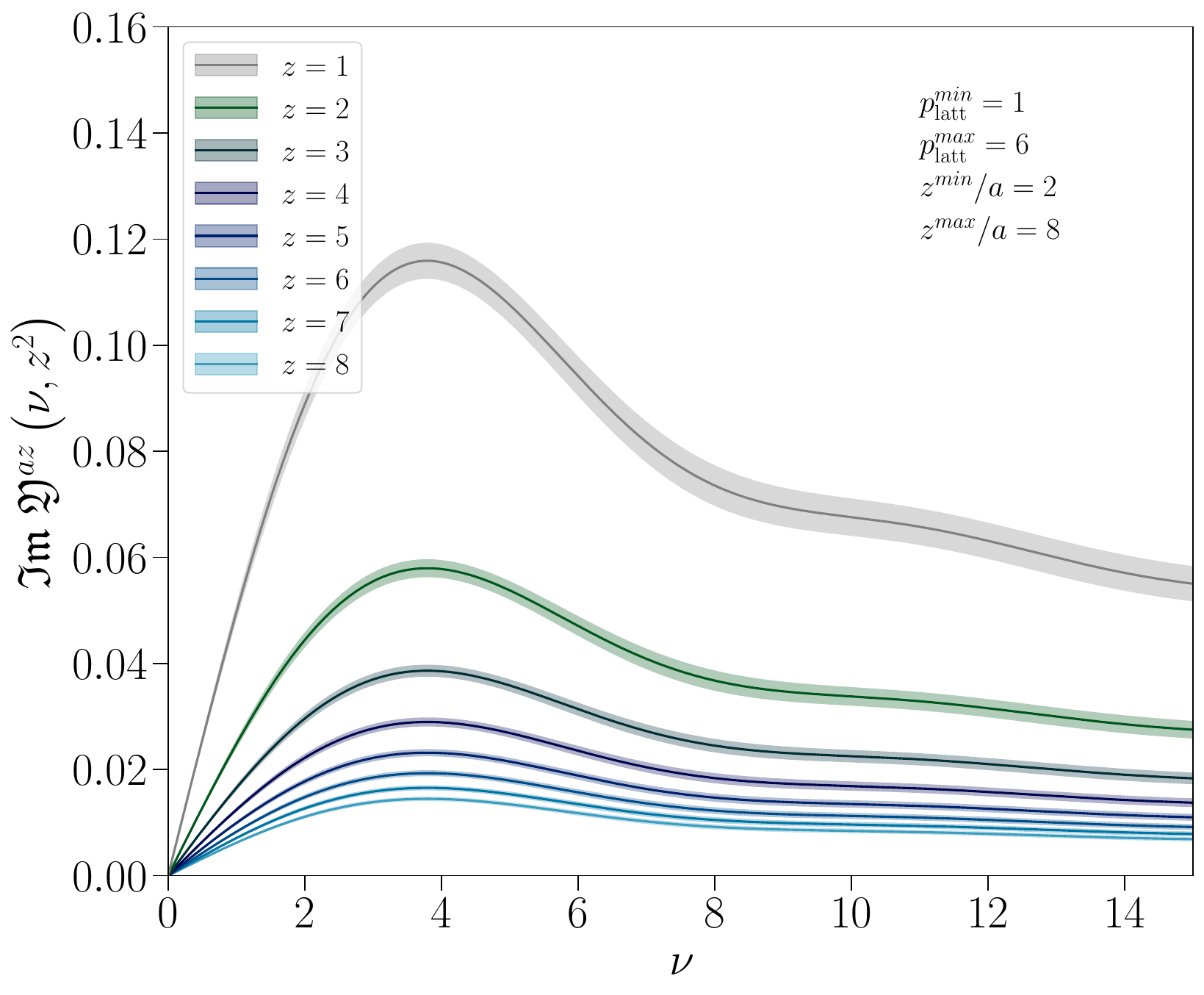}
    \includegraphics[width=0.42\linewidth]{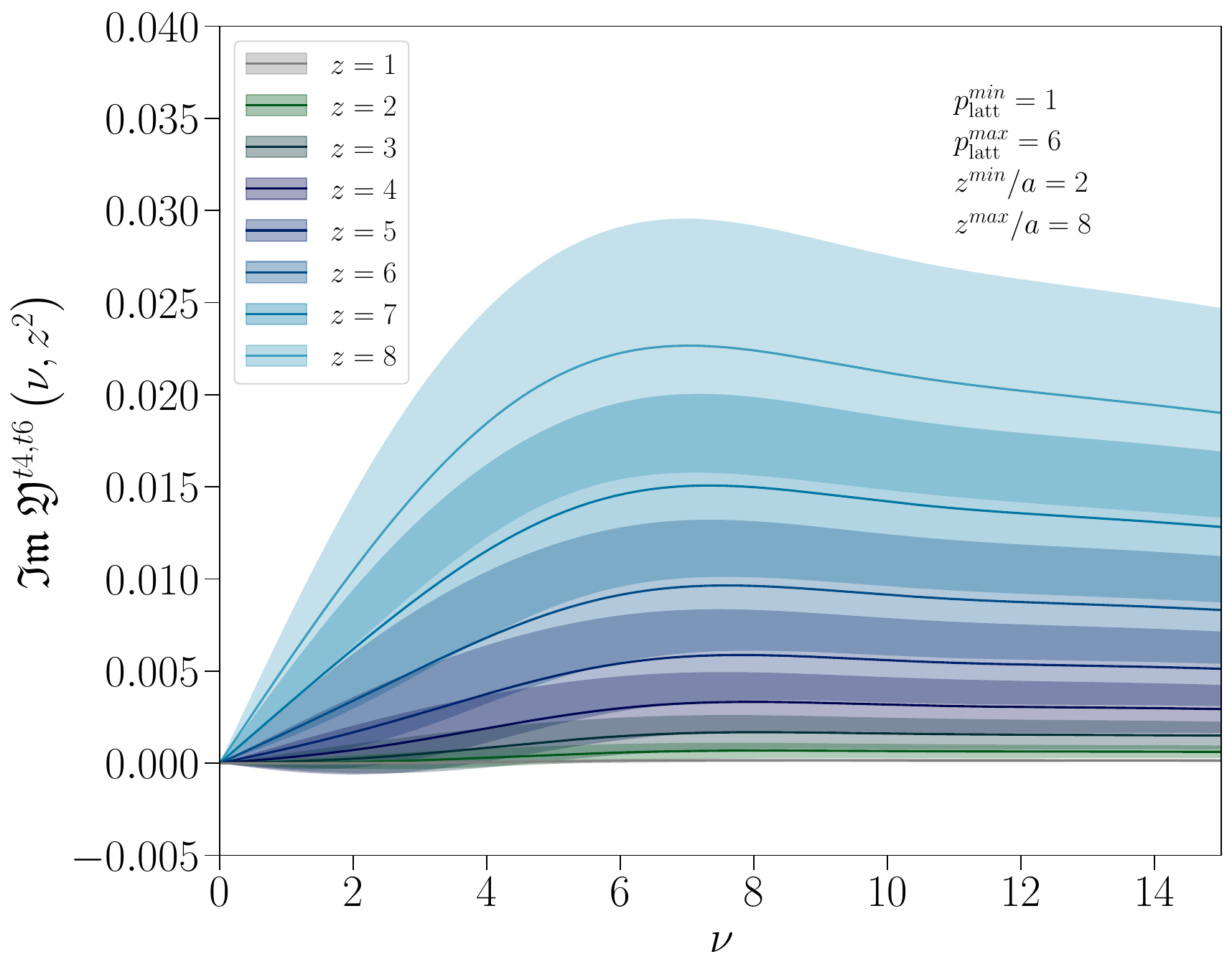} \\
    \includegraphics[width=0.42\linewidth]{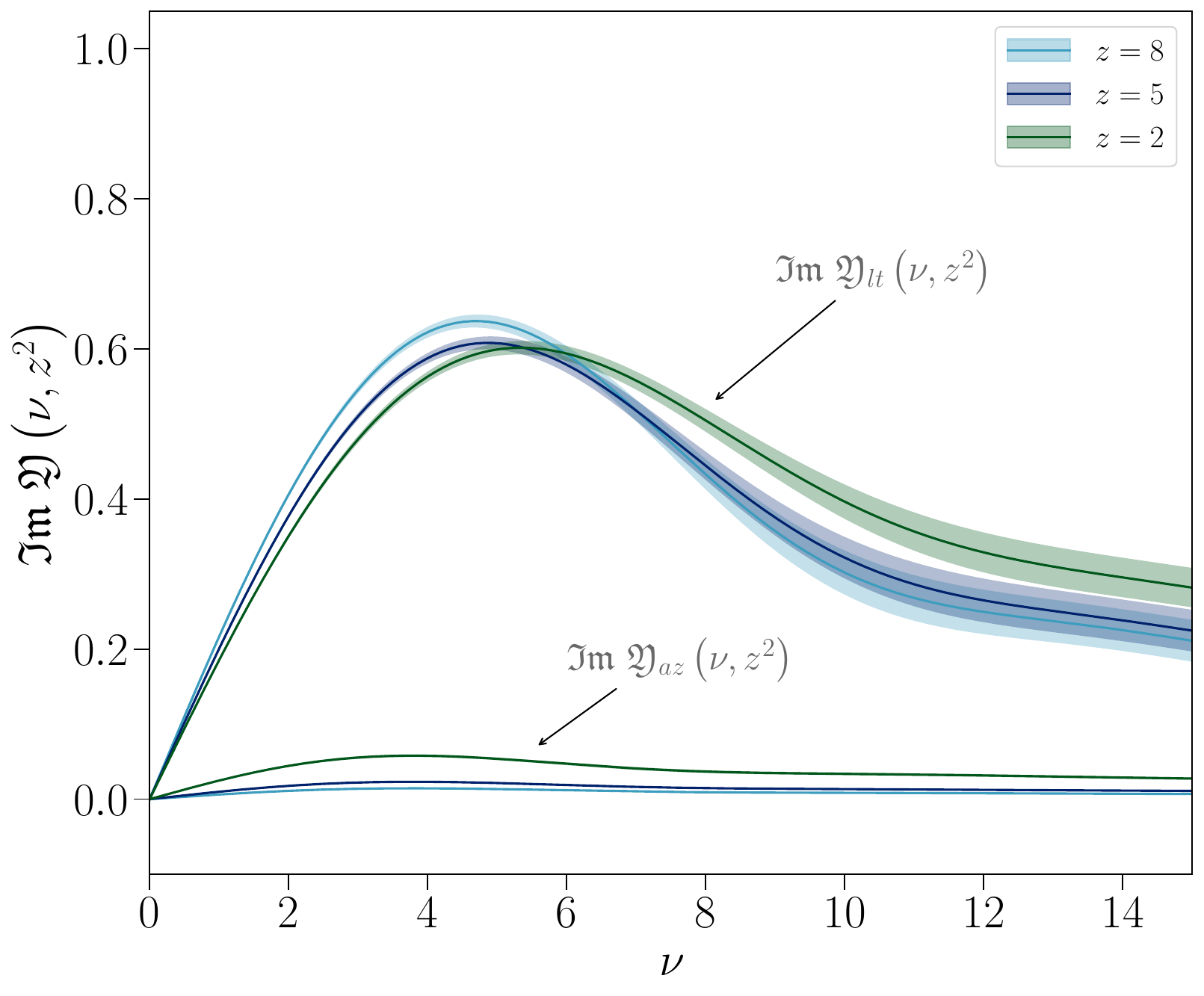}
    \includegraphics[width=0.42\linewidth]{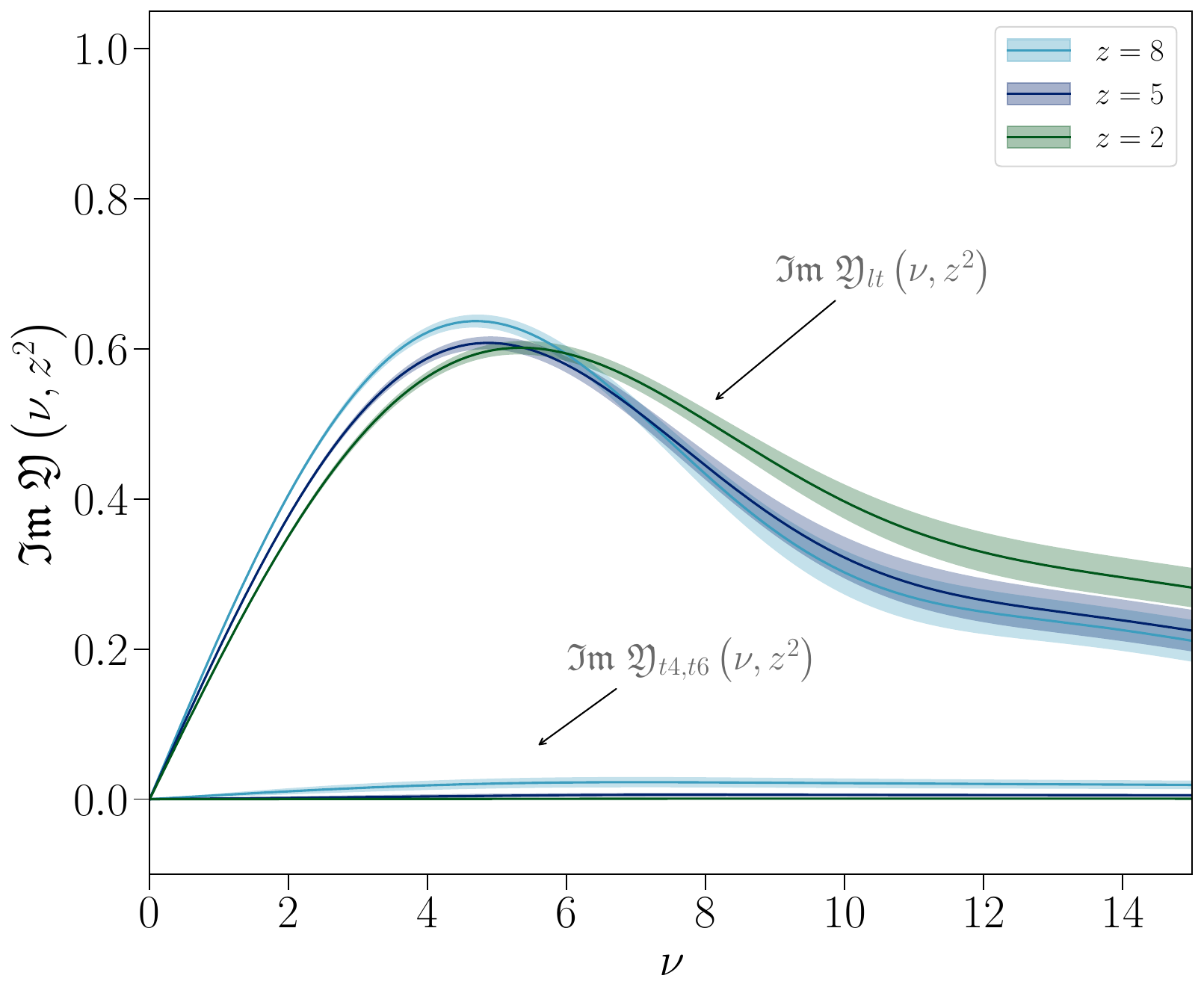}
    \caption{(Upper left) Discretization and (Upper right) net higher-twist nuisance terms determined from the fit in Fig.~\ref{fig:imagPITD_selectFit_p1-6_z2-8_withMatSys_defPriors} shown for $z/a\in\left[1,8\right]$. The lower two panels compare each nuisance effect (discretization at the left and higher-twist at the right) on an absolute scale with the parameterized leading-twist reduced pseudo-ITD $\mathfrak{Im}\ \mathfrak{Y}\left(\nu,z^2\right)$. Values of $z/a$ excluded from the fit are presented in gray.}
    \label{fig:imagPITD_selectFit_p1-6_z2-8_withMatSys_defPriors_nuisanceViz}
\end{figure}

The suggestion by the fit in Fig.~\ref{fig:imagPITD_selectFit_p1-6_z2-8_withMatSys_defPriors} of the presence of $\mathcal{O}\left(a/\left|z\right|\right)$ effects in $\mathfrak{Im}\ \mathfrak{Y}\left(\nu,z^2\right)$ is underscored in the upper left of Fig.~\ref{fig:imagPITD_selectFit_p1-6_z2-8_withMatSys_defPriors_nuisanceViz}, where an appreciable effect is observed for all $\nu\in\left[0,9.42\right]$ for which we have data. The effect is especially pronounced and statistically well-determined for $\nu\sim4$, where, seen in the lower left of Fig.~\ref{fig:imagPITD_selectFit_p1-6_z2-8_withMatSys_defPriors_nuisanceViz}, it enters as a $\sim10\%$ contamination to the leading-twist signal.
Moreover, that this discretization effect, $\mathfrak{Im}\ \mathfrak{Y}^{az}\left(\nu,z^2\right)$, is considerably better resolved than in the real component case is reason alone to understand why the imaginary component appears to be subject to greater systematic contamination. Regardless, our parameterization has proven essential to removing this well-determined, short-distance effect that, if neglected, would have necessarily skewed the $g_{q_+/N}\left(x\right)$ determination. The parameterized higher-twist effects $\mathfrak{Im}\ \mathfrak{Y}^{t4,t6}\left(\nu,z^2\right)$, shown in the upper right of Fig.~\ref{fig:imagPITD_selectFit_p1-6_z2-8_withMatSys_defPriors_nuisanceViz}, are again numerically small and generally consistent with zero for $z/a\lesssim3$, but do pollute $\mathfrak{Im}\ \mathfrak{Y}\left(\nu,z^2\right)$ with an opposite sign compared to the parameterized $\mathfrak{Re}\ \mathfrak{Y}^{t4,t6}\left(\nu,z^2\right)$ effect above. Relative to the leading-twist component, seen in the lower right of Fig.~\ref{fig:imagPITD_selectFit_p1-6_z2-8_withMatSys_defPriors_nuisanceViz}, the parameterized $\mathfrak{Im}\ \mathfrak{Y}^{t4,t6}\left(\nu,z^2\right)$ effect has the potential to spoil the extraction of $g_{q_+/N}\left(x\right)$ within the larger regions of Ioffe-time we isolate (e.g. $\nu\in\left[4,9.42\right]$). However it is within the small-$\nu$ region that the precision of the $\mathfrak{Im}\ \mathfrak{Y}\left(\nu,z^2\right)$ data works to our benefit - the precise small-$p_{\rm latt}$, large-$z$ data help to quantify and remove these $z^2$ effects.
%%%%%%%%%%%%%%%%%%%%%%%%%%%%%%%%%%%%%%%%%%%%%%%%%%%%%%%%%%%%%%%%%%%%%%%
% NUISANCE EFFECTS FROM REPRESENTATIVE Q+ FIT - HIGHEST AICc WEIGHT 
%     --> DEFAULT PRIORS  &&  WITH MATRIX ELEMENT FITTING SYSTEMATIC
%     --> CUTS P\IN[1,6] Z\IN[2,8]
%%%%%%%%%%%%%%%%%%%%%%%%%%%%%%%%%%%%%%%%%%%%%%%%%%%%%%%%%%%%%%%%%%%%%%%

\subsection{Model Selection and Averaging\label{sec:aic}}
The fit to any one model of the PDF likely carries a systematic uncertainty that is undetermined. Despite the description of the reduced pseudo-ITD via orthogonal Jacobi polynomials, the restriction to a truncated set, for both the leading-twist and nuisance terms, exposes a combinatorially large number of distinct functions capable of modeling the data. Selection of any particular model then manifestly carries bias that may skew any quantitative conclusions when considered in isolation. The need for a prescription to average several models together, and in so doing reduce the bias of any individual model, is apparent; this need is reinforced by recognizing that a particular cut on the reduced pseudo-ITD data also introduces bias, as distinct cuts may emphasize contrasting subsets of models.

%%%%%%%%%%%%%%%%%%%%%%%%%%%%%%%%%%%%%%%%%%%%%%%%%%%%%%%%%%%%%%%%%%%%%%%
% Qv AIC RESULTS 
%     --> DEFAULT PRIORS  &&  WITH MATRIX ELEMENT FITTING SYSTEMATIC
%     --> CUTS P\IN[1,6] Z\IN[2,8]
%%%%%%%%%%%%%%%%%%%%%%%%%%%%%%%%%%%%%%%%%%%%%%%%%%%%%%%%%%%%%%%%%%%%%%%
\begin{figure}[t]
    \centering
    \includegraphics[width=0.49\linewidth]{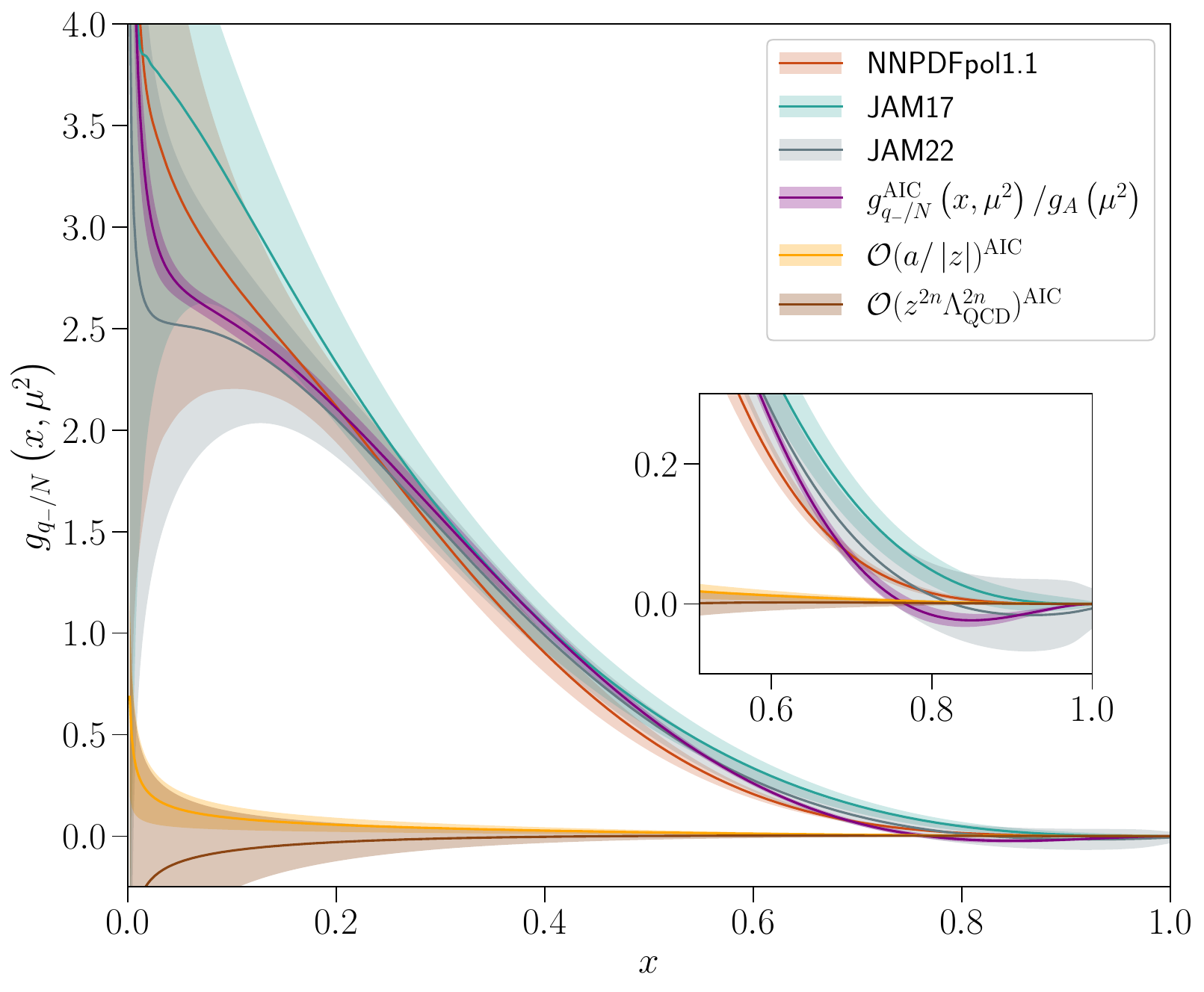}
    \hfill
    \includegraphics[width=0.49\linewidth]{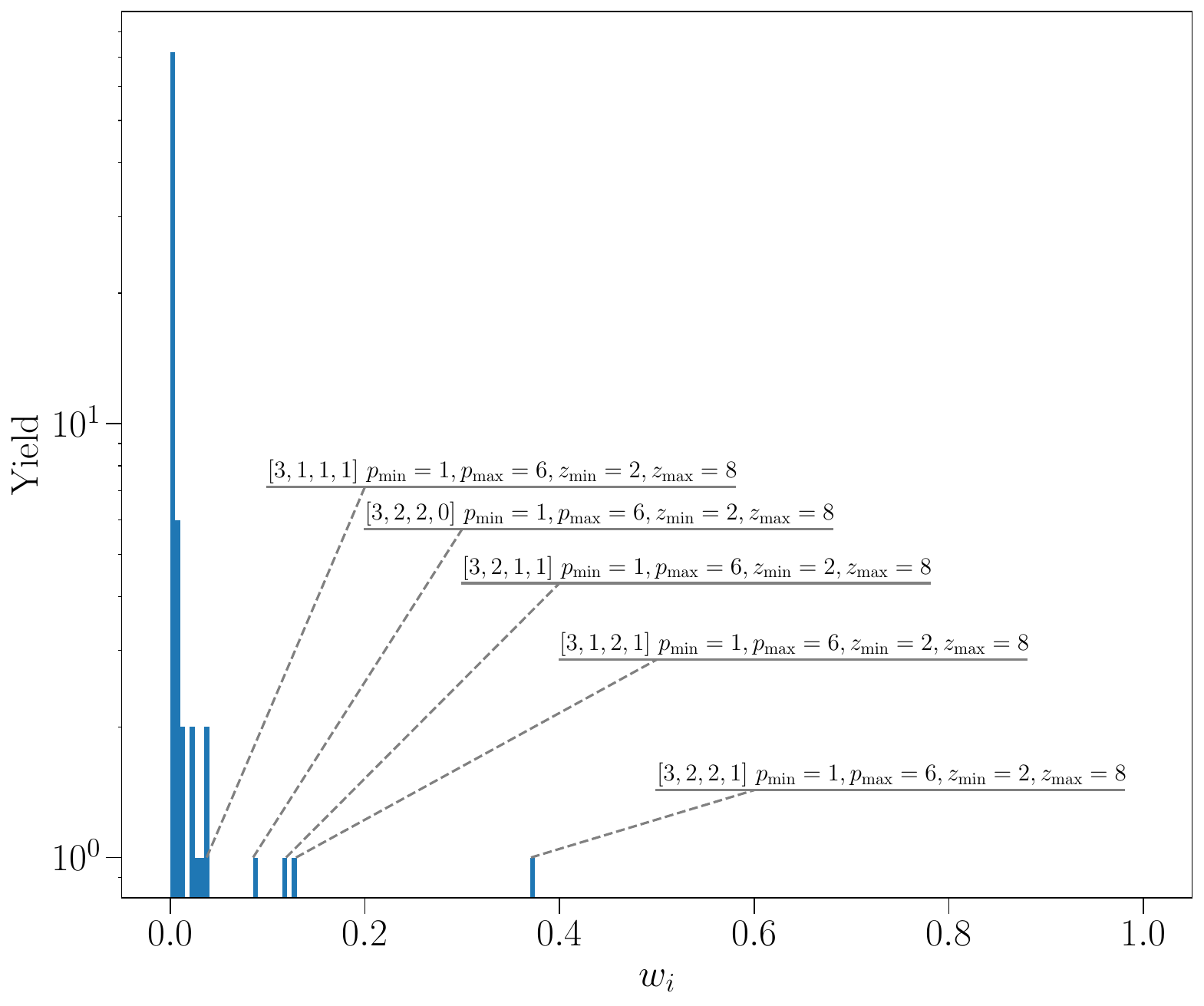}
    \caption{Results of the AICc prescription applied to $\mathfrak{Re}\ \mathfrak{Y}\left(\nu,z^2\right)$ cut on $p_{\rm latt}\in\left[1,6\right]$ and $z/a\in\left[2,8\right]$, where our matrix element fitting systematic~\eqref{eq:matelem_sys} is considered. (Left) The AICc averaged leading-twist valence helicity quark PDF (purple) and model-averaged $x$-space distributions corresponding to an $\mathcal{O}\left(a/\left|z\right|\right)$ discretization (orange) and $\mathcal{O}\left(z^{2n}\Lambda_{\rm QCD}^{2n}\right)$ higher-twist (brown) effects. Comparisons continue to be made with select global analyses. (Right) Histogram of AICc weights associated with all models considered in the data cut.}
    \label{fig:aicQv_p1-6_z2-8_withMatSys_defPriors}
\end{figure}

Among the many prescriptions that can be utilized to create an average model description, one we explore is the Akaike Information Criterion (AIC)~\cite{1100705}. For any given model function $F_i$, the AIC prescription assigns a factor $a_i=2L_i^2+2p_i$, where $L_i^2$ is the negative logarithm of the posterior distribution of the model $F_i$ with $p_i\in\mathbb{Z}^+$ parameters. The factor $a_i$, or ${\rm AIC}\left(F_i\right)$, is then used to assign a weight, or probability, to $F_i$ among the space of all models. In scenarios for which the number of data points $n_i$ fit by a model $F_i$ becomes small or $p_i$ approaches $n_i$, the AIC prescription is known to be biased in its estimate. To account for these scenarios, the corrected AIC~\cite{10.1093/biomet/76.2.297}, or AICc for short, is implemented where $a_i\mapsto A_i=a_i+\frac{2p_i\left(p_i+1\right)}{n_i-p_i-1}$. Using the AICc prescription, which we implement in this work, a model-averaged $F_{\rm AIC}$ is obtained through a weighted sum of each model in consideration:
\be
F_{\rm AIC}=\sum_iw_iF_i,\quad{\rm with}\quad w_i=\cfrac{e^{-A_i/2}}{\left(\sum_{i=1}^Ne^{-A_i/2}\right)}.
\ee
To account for the variation in model choices as well as the variance of a chosen model, the variance of the AICc average is expressed as the weighted sum of the variance of a particular model, ${\rm var}\left(F_i\right)$, plus its squared difference from the AICc model average $\overline{F}_{\rm AIC}$:
\be
{\rm var}\left(F_{\rm AIC}\right)=\sum_iw_i\left[{\rm var}\left(F_i\right)+\left(F_i-\overline{F}_{\rm AIC}\right)^2\right].
\ee
Since the AICc weights $w_i$ depend on the exponential of the AICc value $A_i$, it follows models with the smallest $L^2$ values, that do not over-fit the data, are favored. Although we will consider applying the AICc prescription only to PDFs modeled with Jacobi polynomials, a more robust implementation would consider additional functional forms on the interval $x\in\left[0,1\right]$, including those common from global analyses~\cite{Martin:2009iq,Accardi:2016qay,Harland-Lang:2014zoa,Hou:2019efy} or even neural network parameterizations~\cite{NNPDF:2014otw,NNPDF:2017mvq}. This possibility is reserved for a future work.

%%%%%%%%%%%%%%%%%%%%%%%%%%%%%%%%%%%%%%%%%%%%%%%%%%%%%%%%%%%%%%%%%%%%%%%
% Q+ AIC RESULTS 
%     --> DEFAULT PRIORS  &&  WITH MATRIX ELEMENT FITTING SYSTEMATIC
%     --> CUTS P\IN[1,6] Z\IN[2,8]
%%%%%%%%%%%%%%%%%%%%%%%%%%%%%%%%%%%%%%%%%%%%%%%%%%%%%%%%%%%%%%%%%%%%%%%
\begin{figure}[t]
    \centering
    \includegraphics[width=0.49\linewidth]{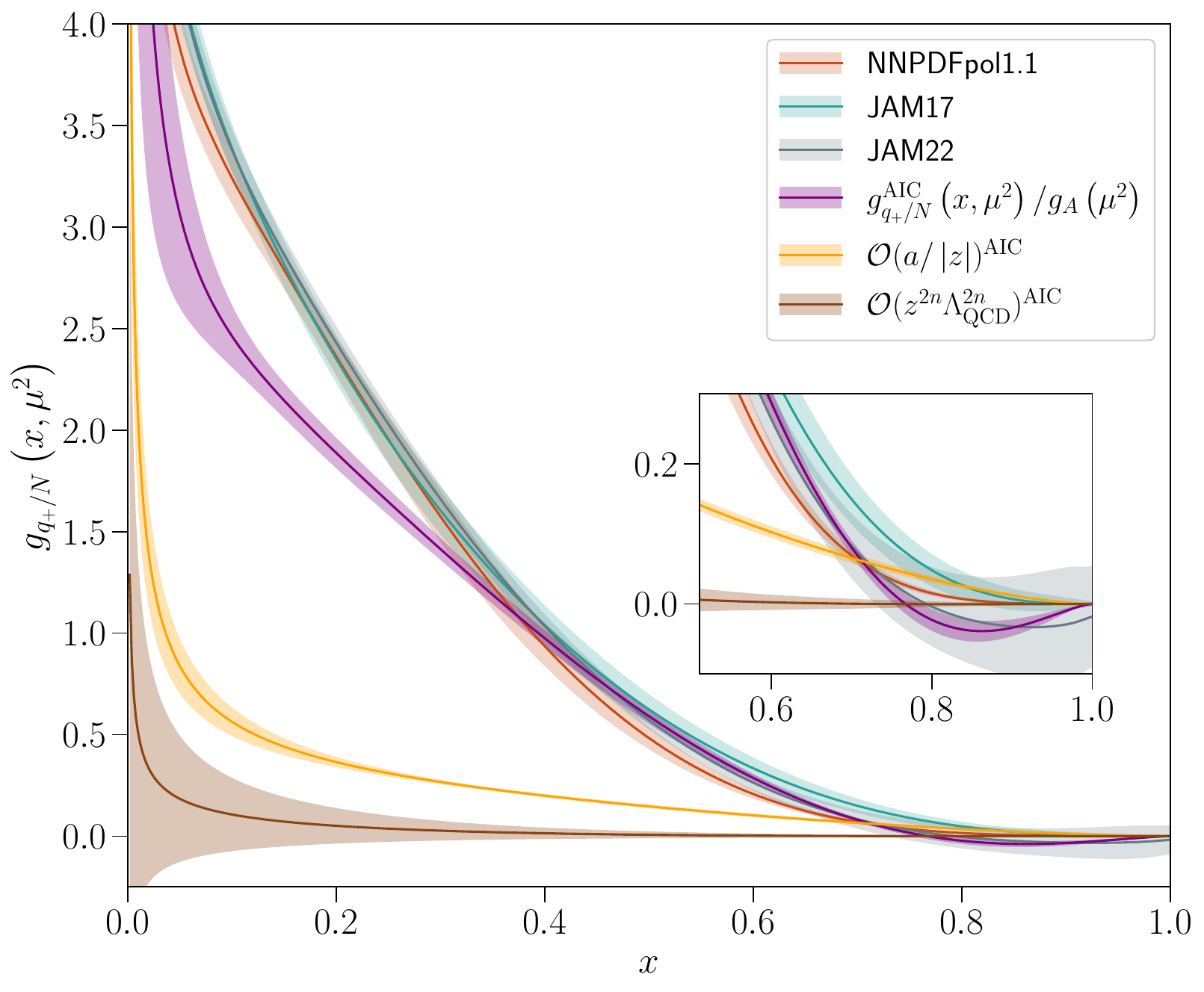}
    \hfill
    \includegraphics[width=0.49\linewidth]{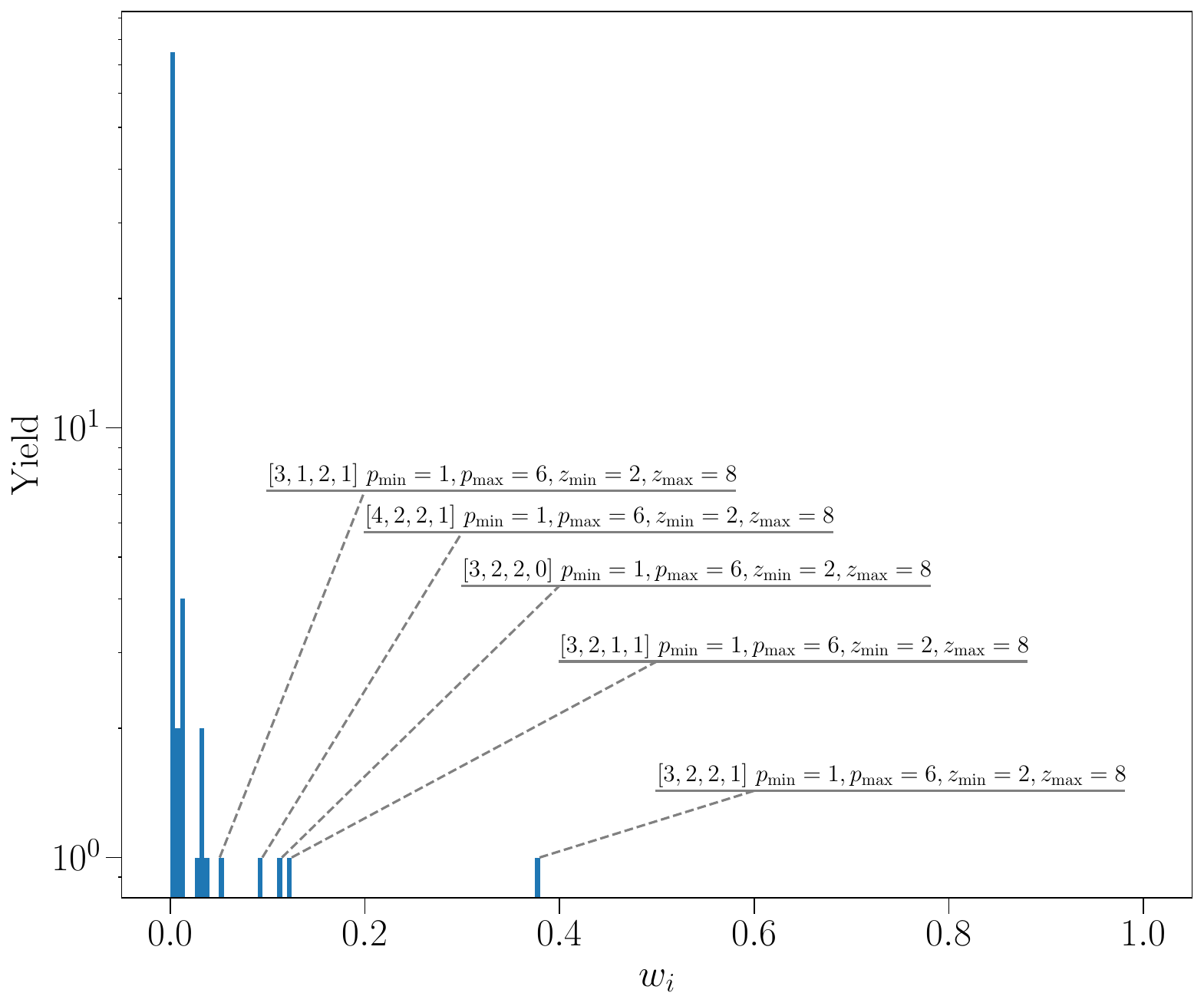}
    \caption{Results of the AICc prescription applied to $\mathfrak{Im}\ \mathfrak{Y}\left(\nu,z^2\right)$ cut on $p_{\rm latt}\in\left[1,6\right]$ and $z/a\in\left[2,8\right]$, where our matrix element fitting systematic~\eqref{eq:matelem_sys} is considered. (Left) The AICc averaged leading-twist $CP$-odd helicity quark PDF (purple) and model-averaged $x$-space distributions corresponding to an $\mathcal{O}\left(a/\left|z\right|\right)$ discretization (orange) and $\mathcal{O}\left(z^{2n}\Lambda_{\rm QCD}^{2n}\right)$ higher-twist (brown) effects. Comparisons continue to be made with select global analyses. (Right) Histogram of AICc weights associated with all models considered in the data cut.}
    \label{fig:aicQ+_p1-6_z2-8_withMatSys_defPriors}
\end{figure}
Within the $p_{\rm latt}=\left[1,6\right]$ and $z/a\in\left[2,8\right]$ data cuts, to construct an AICc model average estimate for both $g_{q_-/N}\left(x\right)$ and $g_{q_+/N}\left(x\right)$ we consider the following variations in the orders of truncation for the Jacobi polynomials: $N_{lt}\in\left[1,5\right]$, $N_{az}\in\left[0,2\right]$, $N_{t4}\in\left[0,2\right]$, and $N_{t6}\in\left[0,1\right]$. The resulting AICc model averaged leading-twist $g_{q_-/N}\left(x\right)$ and $g_{q_+/N}\left(x\right)$ PDFs are shown in the left-hand panels of Fig.~\ref{fig:aicQv_p1-6_z2-8_withMatSys_defPriors} and Fig.~\ref{fig:aicQ+_p1-6_z2-8_withMatSys_defPriors}, respectively, while the right-hand panels depict the histogram of weights determined from the AICc procedure. We observe that only a handful of models contribute appreciably to the AICc averages, while most have negligible impact. In fact, for both $g_{q_-/N}^{\rm AIC}\left(x\right)$ and $g_{q_+/N}^{\rm AIC}\left(x\right)$ the $\left(N_{lt},N_{az},N_{t4},N_{t6}\right)=\left(3,2,2,1\right)$ model was found to dominate the AICc average - hence why this model was presented in Fig.~\ref{fig:realPITD_selectFit_p1-6_z2-8_withMatSys_defPriors} and Fig.~\ref{fig:imagPITD_selectFit_p1-6_z2-8_withMatSys_defPriors}. When comparing the AICc model average $g_{q_-/N}^{\rm AIC}\left(x\right)$ in Fig.~\ref{fig:aicQv_p1-6_z2-8_withMatSys_defPriors} with the selected $g_{q_-/N}\left(x\right)$ fit in Fig.~\ref{fig:paramCovAndPDFQv_selectFit_p1-6_z2-8_withMatSys_defPriors}, it is apparent much of the $x\gtrsim0.5$ information originates from the $\left(N_{lt},N_{az},N_{t4},N_{t6}\right)=\left(3,2,2,1\right)$ model, while the considered models tend to differ in their descriptions of the small-$x$ regime; the latter being evident in the increased uncertainty of $g_{q_-/N}^{\rm AIC}\left(x\right)$ for that region of $x$. The resulting $g_{q_-/N}^{\rm AIC}\left(x\right)$ is found to be in good agreement with the global analysis results we consider, notably the {\tt NNPDFpol1.1}~\cite{Nocera:2014gqa} and {\tt JAM22}~\cite{Cocuzza:2022jye} datasets at small-$x$ and large-$x$, respectively. In an effort to explore the stability of the nuisance effects we attempt to parameterize, the AICc prescription was also applied to the $\mathcal{O}\left(a/\left|z\right|\right)$ and $\mathcal{O}\left(z^{2n}\Lambda_{\rm QCD}^{2n}\right)$ effects.
Seen in Fig.~\ref{fig:aicQv_p1-6_z2-8_withMatSys_defPriors}, it is clear $\mathfrak{Re}\ \mathfrak{Y}\left(\nu,z^2\right)$ is subject to very weak discretization and higher-twist nuisance effects that are generally stable with respect to model variations - albeit the higher-twist effects exhibit a greater variance around zero. As noted earlier in the present section, the increased variability of the leading-twist signal as a result of the AICc prescription causes the pinched errors of the $\left(N_{lt},N_{az},N_{t4},N_{t6}\right)=\left(3,2,2,1\right)$ model PDF around $x\sim0.1$ (cf. Fig.~\ref{fig:paramCovAndPDFQv_selectFit_p1-6_z2-8_withMatSys_defPriors}) to vanish. This highlights the importance of considering as large a space of models as feasible when attempting to extract a PDF, for any one model choice may lead to erroneous conclusions concerning the size and manner of the leading-twist and systematic contamination signals.

Turning attention to the $g_{q_+/N}^{\rm AIC}\left(x\right)$ result in Fig.~\ref{fig:aicQ+_p1-6_z2-8_withMatSys_defPriors}, many structural similarities are observed when compared with the $\left(N_{lt},N_{az},N_{t4},N_{t6}\right)=\left(3,2,2,1\right)$ model considered in isolation (cf. right panel of Fig.~\ref{fig:paramCovAndPDFQ+_selectFit_p1-6_z2-8_withMatSys_defPriors}). Just as for the $g_{q_-/N}^{\rm AIC}\left(x\right)$ PDF, the AICc procedure effectively removes the pinched errors of $g_{q_+/N}\left(x\right)$ seen for $x\sim0.2$ in Fig.~\ref{fig:paramCovAndPDFQ+_selectFit_p1-6_z2-8_withMatSys_defPriors}, and indicates greater variability between any particular model of $g_{q_+/N}\left(x\right)$ for $x\lesssim0.5$; the latter being a reflection of tension between our models and the $p_{\rm latt}=3,5$ data. Furthermore, the model-averaged $g^{\rm AIC}_{q_+/N}\left(x\right)$ is inline with the global analyses we consider for $x\gtrsim0.4$, especially {\tt JAM22}~\cite{Cocuzza:2022jye}. Although any one model of $g_{q_+/N}\left(x\right)$ suggests the presence of $\mathcal{O}\left(z^{2n}\Lambda_{\rm QCD}^{2n}\right)$ nuisance effects, the AICc prescription indicates these effects are largely model-dependent. This model-dependence is seen by the consistency with zero of the AICc higher-twist nuisance effects, while the non-negligible presence in any one model manifests as broad variance around zero in the AICc average. Given the size of the discretization effects parameterized by the fit to $\mathfrak{Im}\ \mathfrak{Y}\left(\nu,z^2\right)$ shown in the left panels of Fig.~\ref{fig:imagPITD_selectFit_p1-6_z2-8_withMatSys_defPriors_nuisanceViz} and that model's dominant weight in the AICc procedure, it is not surprising to find strong support for a short-distance discretization effect after the AICc average. Indeed a more careful study of why the imaginary component of the reduced pseudo-ITD is subject to greater systematic contamination is warranted.

\begin{figure}[t]
    \centering
    \includegraphics[width=0.48\linewidth]{{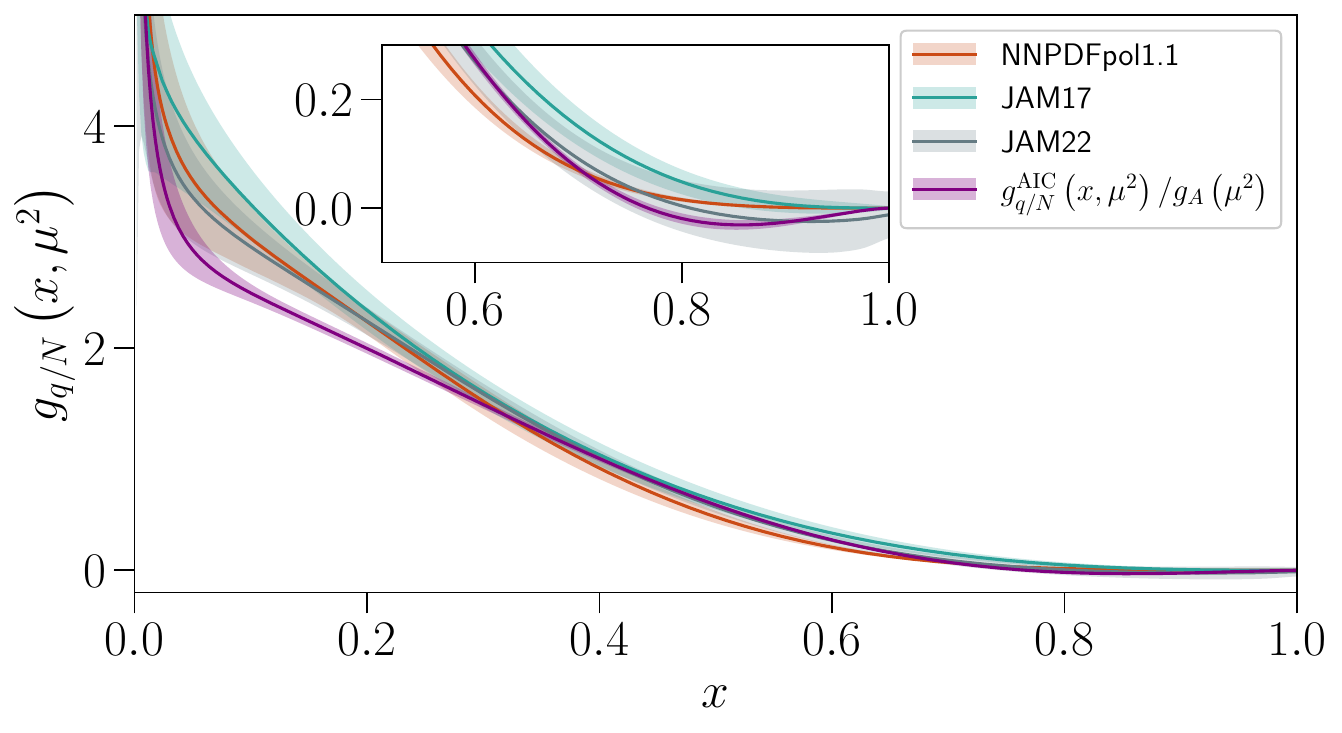}}
    \hfill
    \includegraphics[width=0.5\linewidth]{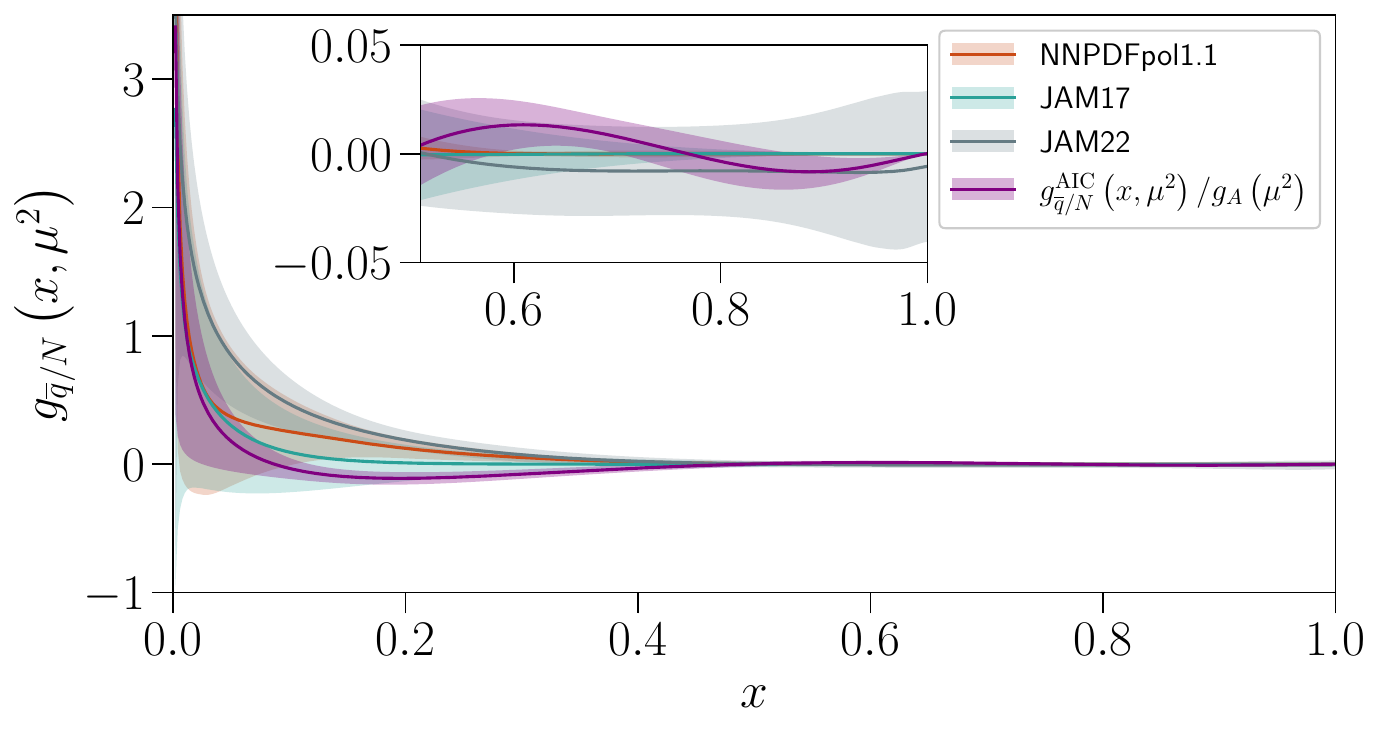}
    \caption{Derived $g_{q/N}\left(x,\mu^2\right)/g_A\left(\mu^2\right)$ (left) and $g_{\overline{q}/N}\left(x,\mu^2\right)/g_A\left(\mu^2\right)$ (right) PDFs obtained from the AICc prescription at an input scale $\mu^2=4\ \text{GeV}^2$ compared with $g_{q/N}\left(x,\mu^2\right)$ and $g_{\overline{q}/N}\left(x,\mu^2\right)$ at the same scale isolated in the global analyses {\tt NNPDFpol1.1}~\cite{Nocera:2014gqa}, {\tt JAM17}~\cite{Ethier:2017zbq}, and {\tt JAM22}~\cite{Cocuzza:2022jye}.}
    \label{fig:Q_QBAR_AIC}
\end{figure}
With the AICc model averaged estimates of $g_{q_-/N}\left(x,\mu^2\right)/g_A\left(\mu^2\right)$ and $g_{q_+/N}\left(x\right)/g_A\left(\mu^2\right)$ in hand, we determine the non-singlet quark and anti-quark helicity PDFs via $g_{q/N}\left(x,\mu^2\right)/g_A\left(\mu^2\right)=\left[g_{q_-/N}\left(x\right)+g_{q_+/N}\left(x,\mu^2\right)\right]/2g_A\left(\mu^2\right)$ and $g_{\bar{q}/N}\left(x,\mu^2\right)/g_A\left(\mu^2\right)=\left[g_{q_+/N}\left(x,\mu^2\right)-g_{q_-/N}\left(x,\mu^2\right)\right]/2g_A\left(\mu^2\right)$, respectively. These distributions are illustrated with the corresponding {\tt NNPDFpol1.1}, {\tt JAM17}, and {\tt JAM22} determinations in Fig.~\ref{fig:Q_QBAR_AIC}. The non-singlet quark helicity PDF is seen to be structurally quite similar to the results from global analyses, albeit with less divergent behavior as $x\rightarrow0$, which is driven by the same behavior in the $g_{q_+/N}^{\rm AIC}\left(x,\mu^2\right)$ result. The non-singlet anti-quark helicity PDF on the other hand, is broadly quite small, favoring two changes in sign for $0.6\lesssim x\leq1$, while being somewhat negative for $0.1\lesssim x\lesssim0.45$. Qualitatively it is straightforward to discern that the first Mellin moment of $g_{\bar{q}/N}^{\rm AIC}\left(x,\mu^2\right)/g_A\left(\mu^2\right)$ will be small, indicating a marginal contribution of the intrinsic light-quark sea to the overall nucleon spin.

\section{Conclusions\label{sec:conclusions}}
The quark helicity PDF, together with the unpolarized and transversity quark PDFs, is essential for a complete characterization at leading-twist of the collinear structure of a hadron involved in an inclusive reaction. Whereas the isovector quark helicity PDFs of the nucleon have been calculated in lattice QCD, we have presented the first such calculation that leverages the pseudo-distribution formalism to perturbatively match renormalization group invariant ratios onto the $\msbar$ helicity PDF. In this work, we have presented our computation of $g_{q_-/N}\left(x,\mu^2\right)/g_A\left(\mu^2\right)$ and $g_{q_+/N}\left(x,\mu^2\right)/g_A\left(\mu^2\right)$ at an input scale $\mu^2=4\text{ GeV}^2$, and found encouraging agreement with an array of global analyses. Moreover, our quoted uncertainty provides for the tantalizing prospect of reducing the uncertainty of the quark helicity PDFs obtained from global analyses when analyzed in a framework akin to Refs.~\cite{JeffersonLabAngularMomentumJAM:2022aix,Bringewatt:2020ixn,Lin:2017stx}. A dedicated, future calculation of the scale-dependent axial charge $g_A\left(\mu^2\right)$ will render the correct overall normalization of our isolated isovector quark helicity PDFs.

We have highlighted the presence of an invariant amplitude that contributes to the space-like matrix elements we have computed, but whose dynamics on the light-cone is absent. This additional amplitude thus enters as an additional source of polynomial $z^2$ corrections that has the potential to sully a reliable determination of the PDF, especially as larger Wilson line lengths are employed to map the Ioffe-time dependence of the leading pseudo-ITD. Through application of the distillation spatial smearing kernel and our construction of nucleon interpolators that transform irreducibly under the double-cover octahedral group $O_h^D$ and its little groups, we proposed a prescription, based on a singular value decomposition (SVD), for disentangling the invariant amplitudes that contribute to matrix elements of the non-local quark bilinear operator we are interested in. Due to an unavoidable finite mixing that exists for the axial vector components that are not collinear to the Wilson line direction, we were unable to establish, in the case of quark helicity, a minimally constrained or over-constrained system of equations to separate the leading pseudo-ITD from the contaminating amplitude (referred to as $\mathcal{R}\left(\nu,z^2\right)$ in Eq.~\ref{eq:decomp}). Regardless, by abandoning the use of correlator projectors and leaning into the group theoretic infrastructure codified by our interpolator construction paradigm via construction of subduced nucleon helicity spinors, we populated a kinematic matrix whose elements are contractions of each subduced spinor with the Lorentz structures associated with the leading and contaminating amplitudes. Since sufficient constraints could not be established (i.e. finite mixing) to separate the leading pseudo-ITD from the contaminating amplitude, we explicitly set to null the kinematic matrix elements that were associated with the contaminating amplitude. By construction, the effect of the contamination was then incorporated into a single amplitude deemed $\widetilde{\mathcal{Y}}\left(\nu,z^2\right)$. Given a kinematic matrix and fitted subduced matrix elements, a trivial application of an SVD solved for $\widetilde{\mathcal{Y}}\left(\nu,z^2\right)$. Our choice to parameterize the derived reduced pseudo-ITD $\mathfrak{Y}\left(\nu,z^2\right)$ in a basis of Jacobi polynomials then afforded the opportunity to account for the additional $z^2$ contamination arising from $\mathcal{R}\left(\nu,z^2\right)$ embedded in $\widetilde{\mathcal{Y}}\left(\nu,z^2\right)$, since this effect is no worse than the corrections to our factorization relationship. Though the prescription we established herein could not rigorously separate $\mathcal{R}\left(\nu,z^2\right)$ from the leading pseudo-ITD, the utility of this approach bears considerable weight for future PDF calculations within the flavor singlet and gluonic sectors, as well as in the off-forward regime relevant for the extraction of generalized parton distributions, when several invariant amplitudes are present.

In line with contemporary {\it HadStruc} calculations of quark and gluon PDFs, we have leveraged a basis of Jacobi polynomials that span the interval $x\in\left[0,1\right]$ to model the functional dependence of the helicity PDFs and any systematic contamination, such as higher-twist and discretization errors. As the expansion of the PDFs and $x$-dependent contaminations in Jacobi polynomials was truncated at a finite order, the potential exists for model bias to afflict any particular extraction treated in isolation. To minimize this bias and allow the information content of the data to drive model selection, we performed a large number of parametric fits of the reduced pseudo-ITD data where the order of truncation for the Jacobi polynomials describing the leading-twist, discretization, and higher-twist effects were varied simultaneously with cuts on the extremal momenta and displacements utilized in this calculation. The results of these fits were combined using the corrected Akaike Information Criterion, or AICc, yielding an appropriate estimate of the statistical and systematic uncertainty in the extracted quark helicity PDFs. The AICc procedure, curiously, found a manifest suppression of higher-twist effects in the computed reduced pseudo-ITD data, yet a strong indication of a short-distance discretization effect. This observation alone warrants continued study of the leading-twist PDFs on lattice ensembles characterized by finer lattice spacings. Finally, the non-singlet quark and anti-quark helicity PDFs were obtained from the AICc averaged $g_{q_-/N}\left(x\right)$ and $g_{q_+/N}\left(x\right)$ PDFs. The non-singlet quark helicity PDF $g_{q/N}\left(x\right)$ was found to be generally consistent with extractions from global analyses, with a less divergent approach to $x\rightarrow0$. The non-singlet anti-quark helicity PDF $g_{\bar{q}/N}\left(x\right)$, on the other hand, was observed to be consistent, or very nearly so, with recent global analysis extractions for all values of the parton momentum fraction.

\subsection*{Acknowledgments}
All members of the HadStruc collaboration are thanked for their support and fruitful and stimulating exchanges. The authors thank J. Green for constructive feedback on an initial version of this manuscript. This material is based upon work supported by the U.S.~Department of Energy, Office of Science, Office of Nuclear Physics under contract DE-AC05-06OR23177. JK is supported
by U.S.~DOE grant \mbox{\#DE-SC0011941}. NK, KO, and RSS are also supported  by U.S.~DOE Grant \mbox{\#DE-FG02-04ER41302.} AR and WM are also supported  by U.S.~DOE Grant \mbox{\#DE-FG02-97ER41028.}  CJM is supported in part by U.S. DOE Grant \mbox{\#DE-AC02-05CH11231.}
We acknowledge the
facilities of the USQCD Collaboration used for this
research in part, which are funded by the Office of
Science of the U.S. Department of Energy. This work used
the Extreme Science and Engineering Discovery
Environment (XSEDE), which is supported by the
National Science Foundation under Grant No. ACI-
1548562~\cite{tacc}. We further acknowledge the Texas
Advanced Computing Center (TACC) at the University
of Texas at Austin for HPC resources on Frontera~\cite{10.1145/3311790.3396656} that have contributed greatly to the results in this work. We
gratefully acknowledge computing cycles provided by
facilities at William and Mary, which were provided by
contributions from the National Science Foundation (MRI
Grant No. PHY-1626177), and the Commonwealth of
Virginia Equipment Trust Fund. The authors acknowledge
William and Mary Research Computing for providing
computational resources and/or technical support that have
contributed to the results reported within this paper.

This work was furthermore made possible using results obtained
at NERSC, a DOE Office of Science User Facility supported by the Office of Science of the U.S.~Department of Energy under Contract No. DE-AC02-05CH11231, as well as resources of the Oak Ridge Leadership Computing Facility (ALCC and INCITE) at the Oak Ridge National Laboratory, which is supported by the Office of Science of the U.S. Department of Energy under Contract No. DE-AC05-00OR22725.
Calculations were performed using the Chroma~\cite{Edwards:2004sx},
QUDA~\cite{Clark:2009wm,Babich:2010mu}, QDP-JIT~\cite{Winter:2014dka}, and QPhiX~\cite{ISC13Phi,qphix} software libraries which were developed with support from the U.S. Department of Energy, Office of Science, Office of
Advanced Scientific Computing Research and Office
of Nuclear Physics, Scientific Discovery through
Advanced Computing (SciDAC) program. This research
was also supported by the Exascale Computing Project
No.~(17-SC-20-SC), a collaborative effort of the U.S.
Department of Energy Office of Science and the
National Nuclear Security Administration.

We acknowledge PRACE (Partnership for Advanced Computing in Europe) for awarding us access to the high performance computing system Marconi100 at CINECA (Consorzio Interuniversitario per il Calcolo Automatico dell’Italia Nord-orientale) under the grant 2021240076. Results were obtained also by using Piz Daint at Centro Svizzero di Calcolo Scientifico (CSCS), via the project with ID s994. We thank the staff of CSCS for access to the computational resources and for their support. This work also benefited from
access to the Jean Zay supercomputer at the Institute for Development and Resources in Intensive Scientific Computing (IDRIS) in Orsay, France under project No. A0080511504.

\appendix
\section{Stability of PDF Results with Variable Prior Widths\label{sec:fitStabilityWithPriors}}
Identifying the most likely set of parameters of some model given data and prior information, as discussed in Sec.~\ref{sec:extraction}, is formalized by maximizing the posterior distribution. The introduction of physically motivated prior information is a delicate procedure that has the potential to unnecessarily constrain the space of viable solutions. In order to gauge the stability of our reported quark helicity PDF results in Sec.~\ref{sec:results} and their sensitivity to the selected {\it Default} priors, we consider again the $L^2$ minimization procedure for the $\left(N_{lt},N_{az},N_{t4},N_{t6}\right)=\left(3,2,2,1\right)$ model of both $\mathfrak{Re}\ \mathfrak{Y}\left(\nu,z^2\right)$ and $\mathfrak{Im}\ \mathfrak{Y}\left(\nu,z^2\right)$, except with distinct prior information.

Since the central values of each {\it Default} prior listed in Tab.~\ref{tab:priors} enforce either the Jacobi polynomial orthogonality or hierarchy between the leading-twist and systematic contaminations, and are otherwise unknown a priori, we elect to maintain each central value. To explore the sensitivity of our results on the chosen priors, we consider, for brevity, two cases in which the prior widths of each model parameter are varied. The first case doubles the {\it Default} prior widths, while the second case halves the {\it Default} prior widths, to which the cases are referred to as {\it Wide} and {\it Thin}, respectively. The adjusted widths are gathered in Tab.~\ref{tab:priors}. To avoid obfuscation of the extracted PDF's dependence on the priors, we again cut on the $\mathfrak{Y}\left(\nu,z^2\right)$ data which includes the matrix element fitting systematic~\eqref{eq:matelem_sys} according to $p_{\rm latt}\in\left[1,6\right]$ and $z/a\in\left[2,8\right]$.

\begin{figure}[t]
    \centering
    \includegraphics[width=0.49\linewidth]{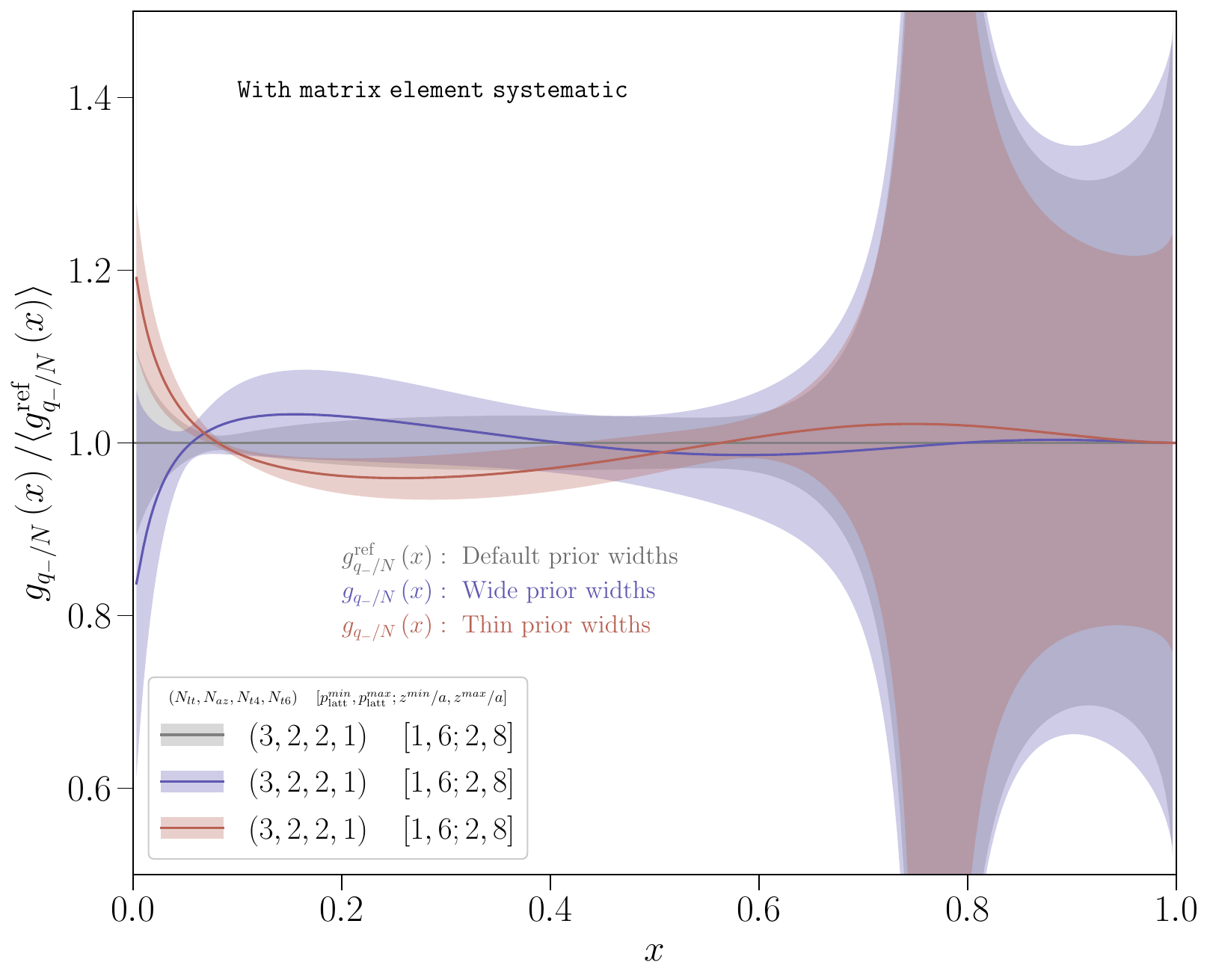}  
    \hfill
    \includegraphics[width=0.49\linewidth]{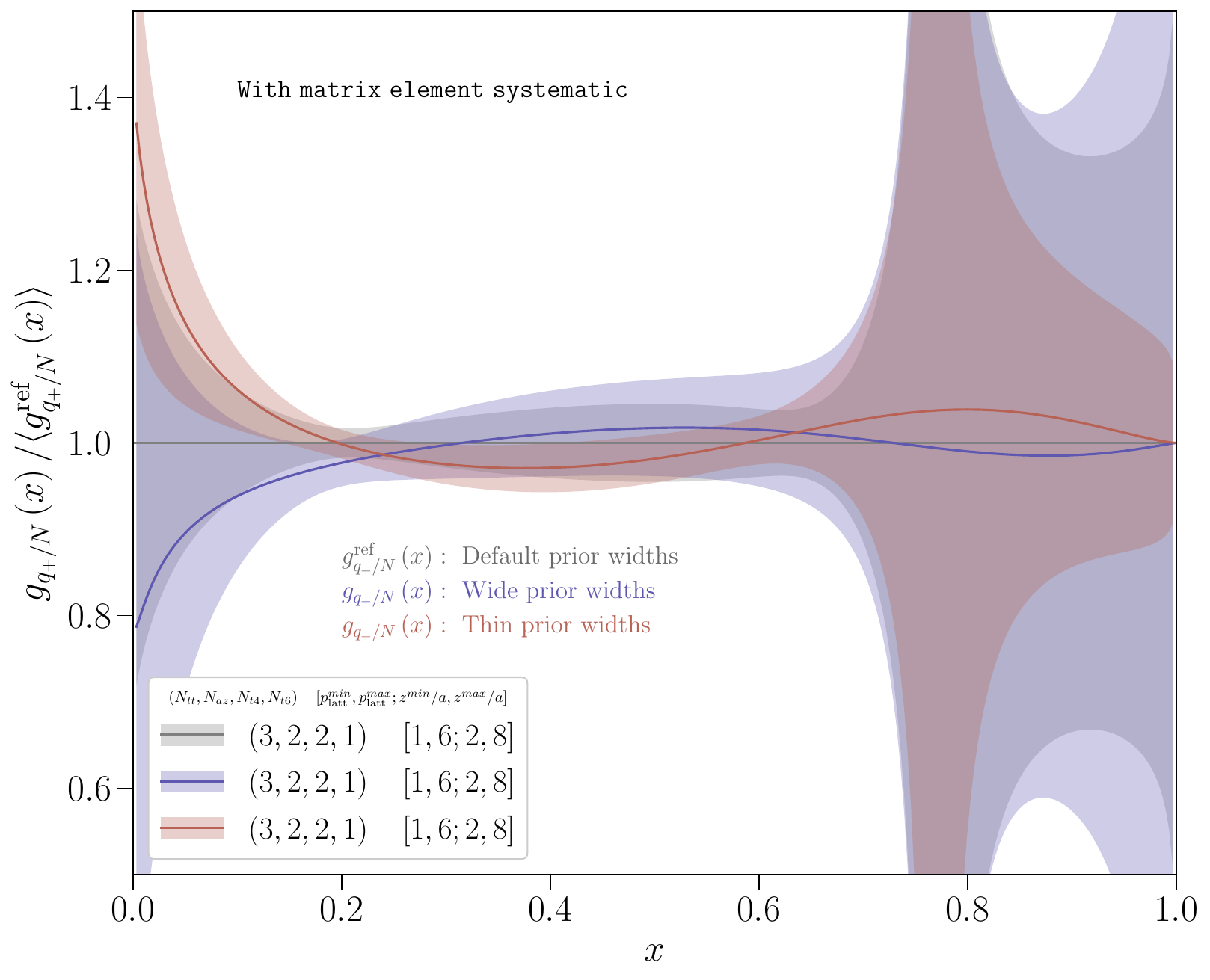}
    \caption{Variability of extracted (left) $g_{q_-/N}\left(x\right)$ and (right) $g_{q_+/N}\left(x\right)$ PDFs when the Bayesian prior widths of the model $\left(N_{lt},N_{az},N_{t4},N_{t6}\right)=\left(3,2,2,1\right)$ fit to $\mathfrak{Y}\left(\nu,z^2\right)$, cut on $p_{\rm latt}\in\left[1,6\right]$ and $z/a\in\left[2,8\right]$, are doubled and halved. The matrix element fitting systematic is included for this comparison.}
    \label{fig:Qv-Q+_variability-with-prior-widths}
\end{figure}
In Fig.~\ref{fig:Qv-Q+_variability-with-prior-widths} the dependence of the extracted $g_{q_-/N}\left(x\right)$ and $g_{q_+/N}\left(x\right)$ PDFs on the re-scaled prior distributions is illustrated. In each case, the $g_{q_-/N}\left(x\right)$ and $g_{q_+/N}\left(x\right)$ PDFs obtained using {\it Wide} and {\it Thin} priors are normalized by the central value of the corresponding PDF obtained using the {\it Default} priors - denoted respectively as $g_{q_-/N}^{\rm ref}\left(x\right)$ and $g_{q_+/N}^{\rm ref}\left(x\right)$. Evidently reducing the widths of each prior distribution by a factor of two yields $g_{q_-/N}\left(x\right)$ and $g_{q_+/N}\left(x\right)$ PDFs (red in Fig.~\ref{fig:Qv-Q+_variability-with-prior-widths}) that are in tension with the {\it Default} PDF determinations across the $x$-space region $0\leq x\lesssim0.6$. The deviations in the central values of the PDFs observed for $x\lesssim0.6$ range between $\sim5-10\%$ and $\sim5-30\%$, respectively, for $g_{q_-/N}\left(x\right)$ and $g_{q_+/N}\left(x\right)$. These findings are unsurprising and highlight the importance of flexible prior distributions, as the amount of variability any model can assume in an $L^2$ optimization process is inextricably linked to the model's prior distributions.

It is especially encouraging to find that the ${q_-/N}\left(x\right)$ and $g_{q_+/N}\left(x\right)$ distributions are stable as the widths of the prior distributions are doubled, shown as blue in Fig.~\ref{fig:Qv-Q+_variability-with-prior-widths}. Indeed the central values of the {\it Wide} prior determined PDFs differ slightly from the {\it Default} determinations for $x\lesssim0.5$ - the deviations being on the order of a few percent for $0.2\lesssim x\lesssim0.5$, and only appreciating in size for $x\lesssim0.2$ where we expect the results from the inverse problem are least reliable anyway. However, unlike the PDFs determined with {\it Thin} priors, the statistical uncertainties of the PDFs obtained with {\it Wide} priors are entirely in line with the central values of the PDFs obtained with the {\it Default} priors. This indicates the {\it Default} prior widths we considered in our initial attempts to extract $g_{q_-/N}\left(x\right)$ and $g_{q_+/N}\left(x\right)$ in Sec.~\ref{sec:results} were not too constraining so as to bias the PDF determinations. This bolsters fidelity in our PDF determinations presented in Sec.~\ref{sec:results}.

\section{Stability of PDF Results with Data Cuts\label{sec:fitStabilityWithCuts}}
In the spirit of exploring the sensitivity of our PDF results on the prior distributions, we use this appendix to investigate any shifts in our PDF determinations as variable cuts on the reduced pseudo-ITD $\mathfrak{Y}\left(\nu,z^2\right)$ are considered. For brevity we consider two cases: no cuts on the data, namely $p_{\rm latt}\in\left[1,6\right]$ and $z/a\in\left[1,8\right]$, and $p_{\rm latt}\in\left[2,6\right]$ and $z/a\in\left[2,8\right]$, which excludes the most precise data from our analysis.
\begin{figure}[t]
    \centering
    \includegraphics[width=0.49\linewidth]{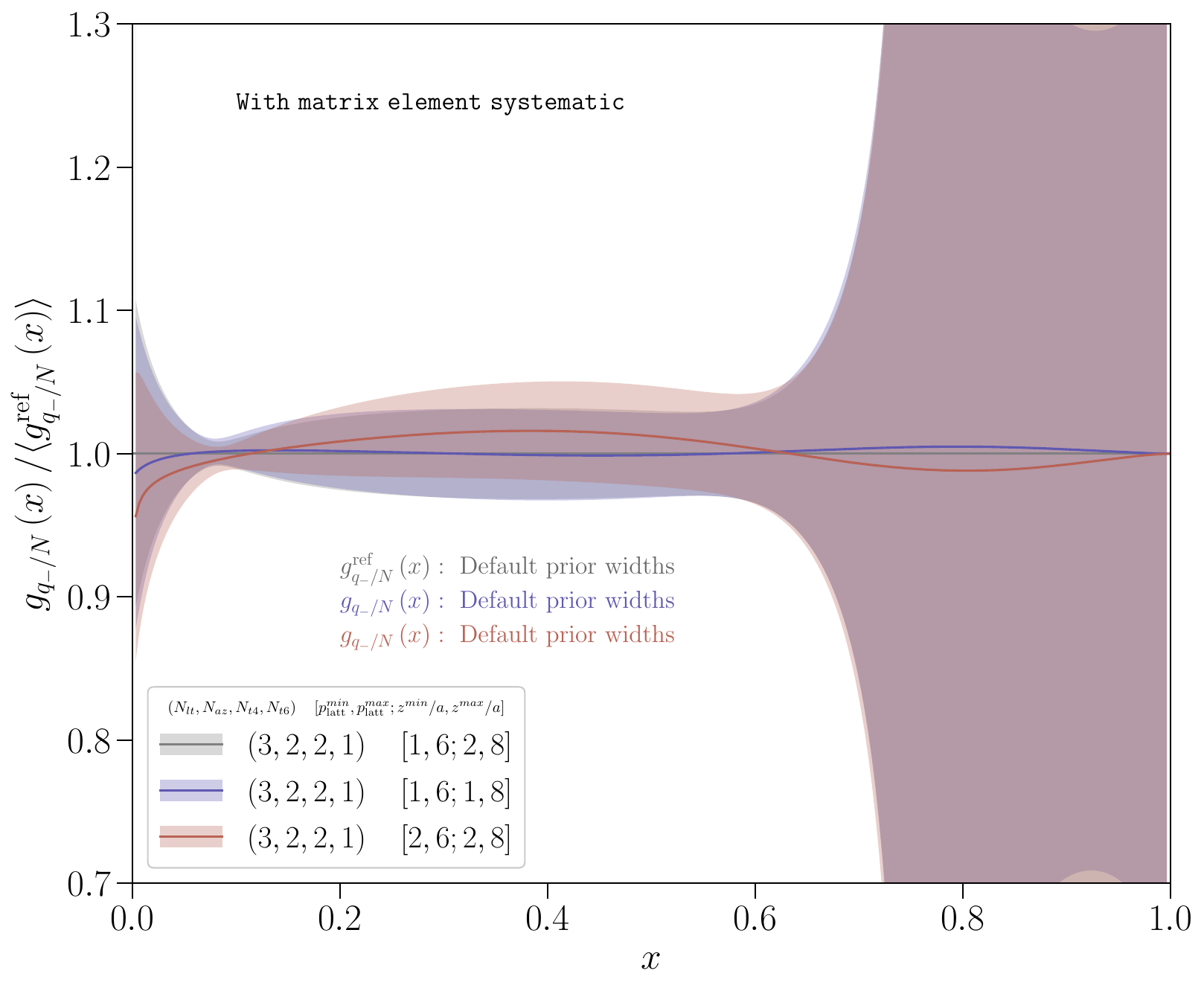}
    \hfill
    \includegraphics[width=0.49\linewidth]{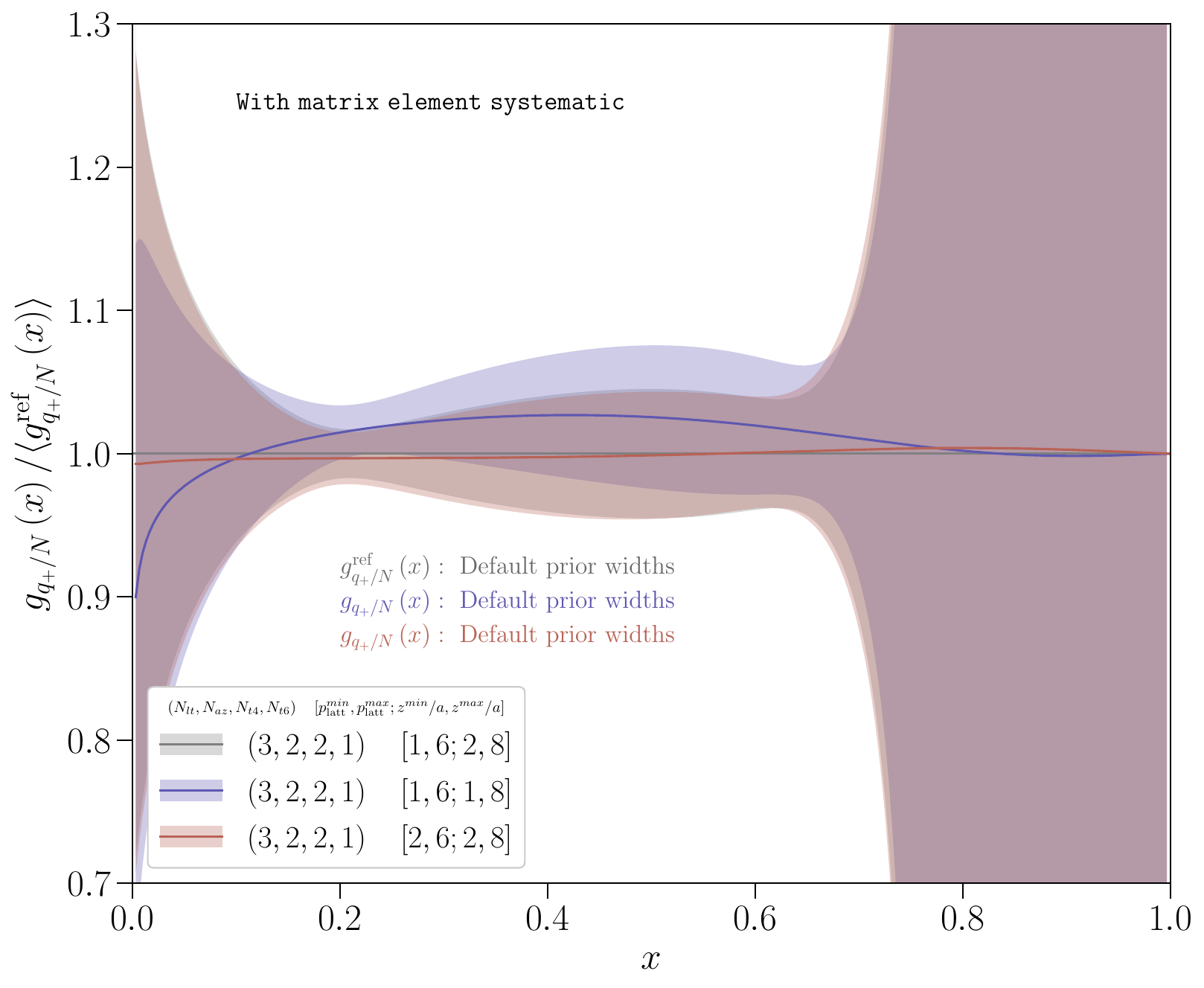}
    \caption{Variability of extracted (left) $g_{q_-/N}\left(x\right)$ and (right) $g_{q_+/N}\left(x\right)$ PDFs when the reduced pseudo-ITD $\mathfrak{Y}\left(\nu,z^2\right)$ is cut on $p_{\rm latt}\in\left[1,6\right]$, $z/a\in\left[1,8\right]$ and $p_{\rm latt}\in\left[2,6\right]$, $z/a\in\left[2,8\right]$. The model $\left(N_{lt},N_{az},N_{t4},N_{t6}\right)=\left(3,2,2,1\right)$ is used for this comparison, and the matrix element fitting systematic remains in consideration.}
    \label{fig:Qv-Q+_variability-with-cuts-on-pmin-zmin}
\end{figure}
Application of the AICc prescription within each of these cuts found that the $\left(N_{lt},N_{az},N_{t4},N_{t6}\right)=\left(3,2,2,1\right)$ model receives the highest weight in determining each cut's model-averaged PDF.

In Fig.~\ref{fig:Qv-Q+_variability-with-cuts-on-pmin-zmin} the dependence of the PDFs $g_{q_-/N}\left(x\right)$ and $g_{q_+/N}\left(x\right)$ on the chosen cuts is illustrated. As in Appendix~\ref{sec:fitStabilityWithPriors}, any deviations from the PDFs reported in Sec.~\ref{sec:results}, denoted again as $g_{q_\pm/N}^{\rm ref}\left(x\right)$, are emphasized through normalization of each PDF by $g_{q_\pm/N}^{\rm ref}\left(x\right)$. It is clear $\mathfrak{Y}\left(\nu,z^2\right)$ that includes the matrix element fitting systematic allows for sufficient flexibility such that the PDFs are essentially invariant under these cuts.

\bibliography{srcs.bib}

\end{document}